\documentclass[journal, 10pt]{IEEEtran}

\usepackage{jdhmacros}

\newcommand{\rk}{{\rm rank}}
\newcommand{\Prob}{{\rm Pr}}

\newcommand{\bZero}{\mathbf 0}

\newcommand{\jdh}[1]{}
\renewcommand{\jdh}[1]{{\color{magenta}{#1}}} % Comment out to leave notes out (final version)

\usepackage[colorlinks]{hyperref}

\begin{document}

%%%%%%%%% TITLE
\title{\huge Identifying Outliers in Large Matrices via\\
Randomized Adaptive Compressive Sampling}

\author{Xingguo Li and Jarvis Haupt\thanks{Submitted June 30, 2014; revised October 12, 2014.  The authors are with the Department of Electrical and Computer Engineering at the University of Minnesota -- Twin Cities. Tel/fax: (612) 625-3300 \ /\ (612) 625-4583. Emails: {\tt \{lixx1661, jdhaupt\}@umn.edu}. (Corresponding author: J. Haupt.) A version of this work was submitted to ICASSP 2015. The authors graciously acknowledge support from the NSF under Award No. CCF-1217751.}}

\maketitle
%\thispagestyle{empty}

%%%%%%%%% ABSTRACT
\begin{abstract}
This paper examines the problem of locating outlier columns in a large, otherwise low-rank, matrix.  We propose a simple two-step adaptive sensing and inference approach and establish theoretical guarantees for its performance; our results show that accurate outlier identification is achievable using very few linear summaries of the original data matrix -- as few as the squared rank of the low-rank component plus the number of outliers, times constant and logarithmic factors.  We demonstrate the performance of our approach experimentally in two stylized applications, one motivated by robust collaborative filtering tasks, and the other by saliency map estimation tasks arising in computer vision and automated surveillance, and also investigate extensions to settings where the data are noisy, or possibly incomplete.
\end{abstract}

%%%%%%%%% BODY TEXT
%-------------------------------------------------------------------------
\section{Introduction}

In this paper we address a matrix \emph{outlier identification} problem.  Suppose $\Mb\in\RR^{n_1\times n_2}$ is a data matrix that admits a decomposition of the form
\begin{equation*}
\Mb = \Lb + \Cb,
\end{equation*}
where $\Lb$ is a low-rank matrix, and $\Cb$ is a matrix of outliers that is nonzero in only a fraction of its columns.  We are ultimately interested in identifying the locations of the nonzero columns of $\Cb$, with a particular focus on settings where $\Mb$ may be very large.  The question we address here is, can we accurately (and efficiently) identify the locations of the outliers from a small number of linear measurements of $\Mb$?

Our investigation is motivated in part by robust collaborative filtering applications, in which the goal may be to identify the locations (or even quantify the number) of corrupted data points or outliers in a large data array.  Such tasks may arise in a number of contemporary applications, for example, when identifying malicious responses in survey data or anomalous patterns in network traffic, to name a few.  Depending on the nature of the outliers, conventional low-rank approximation approaches based on principal component analysis (PCA) \cite{Pearson:01, Jolliffe:05} may be viable options for these tasks, but such approaches become increasingly computationally demanding as the data become very high-dimensional.  Here, our aim is to leverage dimensionality reduction ideas along the lines of those utilized in randomized numerical linear algebra, (see, e.g., \cite{Halko:11, Mahoney:11} and the references therein) and compressed sensing (see, e.g., \cite{Candes:06:Freq, Donoho:06:CS, Candes:06:UES}), in order to reduce the size of the data on which our approach operates.  In so doing, we also reduce the computational burden of the inference approach relative to comparable methods that operate on ``full data.'' 

We are also motivated by an image processing task that arises in many computer vision and surveillance applications -- that of identifying the ``saliency map'' \cite{Koch:87} of a given image, which (ideally) indicates the regions of the image that tend to attract the attention of a human viewer.  Saliency map estimation is a well-studied area, and numerous methods have been proposed for obtaining saliency maps for a given image -- see, for example, \cite{Tsotsos:95:Visual, Itti:98, Harel:06, Liu:07, Rao:10}.   In contrast to these (and other) methods designed to identify saliency map of an image as a ``post processing'' step, our aim here is to estimate the saliency map \emph{directly from compressive samples} -- i.e., without first performing full image reconstruction as an intermediate step.  We address this problem here using a linear subspace-based model of saliency, wherein we interpret an image as a collection of distinct (non-overlapping) patches, so that images may be (equivalently) represented as matrices whose columns are \emph{vectorized} versions of the patches.  Previous efforts have demonstrated that such local patches extracted from natural images may be well \emph{approximated} as vectors in a union of low-dimensional linear subspaces (see, e.g., \cite{Yu:11:Models}). Here, our approach to the saliency map estimation problem is based on an assumption that salient regions in an image may be modeled as outliers from a single common low-dimensional subspace; the efficacy of similar saliency models for visual saliency has been established recently in \cite{Shen:12}. Our approach here may find utility in rapid threat detection in security and surveillance applications in high-dimensional imaging tasks where the goal is not to image the entire scene, but rather to merely identify regions in the image space corresponding to anomalous behavior.  Successful identification of salient regions could comprise a first step in an active vision task, where subsequent imaging is restricted to the identified regions.  

\subsection{Innovations and Our Approach}

We propose a framework that employs dimensionality reduction techniques within the context of a two-step adaptive sampling and inference procedure, and our approach is based on a few key insights.  First, we exploit the fact that the enabling geometry of our problem (to be formalized in the following section) is approximately preserved if we operate not on $\Mb$ directly, but instead on a ``compressed'' version $\bPhi\Mb$ that has potentially many fewer rows.  Next, we use the fact that we can learn the (ostensibly, low-dimensional) linear subspace spanned by the columns of the low rank component of $\bPhi\Mb$ using a small, randomly selected subset of the columns of $\bPhi\Mb$.  Our algorithmic approach for this step utilizes a recently proposed method called \emph{Outlier Pursuit} (OP) \cite{Xu:12} that aims to separate a matrix $\Yb$ into its low-rank and column-sparse components using the convex optimization
\begin{equation}\label{eqn:OP}
\argmin_{\bL,\bC} \ \ \|\bL\|_{*} + \lambda \|\bC\|_{1,2} \ \ \mbox{s.t.} \ \Yb = \bL + \bC
\end{equation}
where $\|\bL\|_*$ denotes the nuclear norm of $\bL$ (the sum of its singular values), $\|\bC\|_{1,2}$ is the sum of the $\ell_2$ norms of the columns of $\bC$, and $\lambda>0$ is a regularization parameter.  Finally, we leverage the fact that correct identification of the subspace spanned by the low-rank component of $\bPhi \Mb$ facilitates (simple) inference of the column outliers.

We analyze two variants of this overall approach. The first (depicted as Algorithm~\ref{alg:main}) is based on the notion that, contingent on correct identification of the subspace spanned by the low-rank component of $\bPhi\Mb$, we may effectively transform the overall outlier identification problem into a compressed sensing problem, using a carefully-designed linear measurement operator whose net effect is to \emph{(i)} reduce the overall $n_1\times n_2$ matrix to a $1\times n_2$ vector whose elements are (nominally) nonzero only at the locations of the outlier columns, and \emph{(ii)} compressively sample the resulting vector.  This reduction enables us to employ well-known theoretical results (e.g., \cite{Candes:08:RIP}) to facilitate our overall analysis. We call this approach Adaptive Compressive Outlier Sensing (ACOS).

\begin{algorithm}[t]
\caption{Adaptive Compressive Outlier Sensing (ACOS)}
\label{alg:main}
\begin{algorithmic}
\STATE \hspace{-1.35em} \textbf{Assume:} $\Mb\in \mathbb{R}^{n_1\times n_2}$ 
\REQUIRE Column sampling Bernoulli parameter $\gamma\in[0,1]$, regularization parameter $\lambda>0$,
Measurement matrices \ \ $\bPhi\in\mathbb{R}^{m\times n_1}$, $\Ab\in\mathbb{R}^{p\times n_2}$, measurement vector $\bphi \in\mathbb{R}^{1\times m}$
\STATE \hspace{-1.35em} \textbf{Initalize:} Column sampling matrix $\Sbb = \Ib_{:,\cS}$, where\\
\STATE  $\cS = \{i:S_i = 1\}$ with $\{S_i\}_{i\in[n_2]}$ i.i.d. Bernoulli$(\gamma)$
\STATE \hspace{-0.9em}\textbf{\uline{Step 1}}  
\STATE Collect Measurements: $\Yb_{(1)} = \bPhi \Mb \Sbb$\\
\STATE Solve: $\{\widehat{\Lb}_{(1)},\widehat{\Cb}_{(1)}\} = \argmin_{\bL,\bC} \|\bL\|_{*} + \lambda \|\bC\|_{1,2}$
\STATE \hspace{10em} $\mbox{ s.t. } \ \Yb_{(1)} = \bL + \bC$\\ 
\STATE Let: $\widehat{\cL}_{(1)}$ be the linear subspace spanned by col's of $\widehat{\Lb}_{(1)}$
\STATE \hspace{-0.9em}\textbf{\uline{Step 2}} 
\STATE Compute: $\Pb_{\widehat{\mathcal{L}}_{(1)}}$, the orthogonal  projector onto $\widehat{\cL}_{(1)}$
\STATE Set: $\Pb_{\widehat{\cL}_{(1)}^{\perp}} \triangleq \Ib - \Pb_{\widehat{\mathcal{L}}_{(1)}}$
\STATE Collect Measurements: $\yb_{(2)} =\bphi \ \Pb_{\widehat{\cL}_{(1)}^{\perp}} \bPhi \Mb \Ab^{T} $
\STATE Solve: $\widehat{\cbb} = \argmin_{\cbb} \ \ \|\cbb\|_{1}  \ \ \mbox{s.t.} \ \yb_{(2)} = \cbb\Ab^{T}$ 
\ENSURE $\widehat{\cI}_{\Cb} = \{i: \widehat{\rm c}_{i} \neq 0\}$
\end{algorithmic}
\end{algorithm}

The second approach, which we call Simplified ACOS (SACOS) and summarize as Algorithm~\ref{alg:simple}, foregoes the additional dimensionality reduction in the second step and identifies as outliers those columns of $\bPhi \Mb$ having a nonzero component orthogonal to the subspace spanned by the low-rank component of $\bPhi\Mb$. The simplified approach has a (perhaps significantly) higher sample complexity than ACOS, but (as we will see in Section~\ref{sec:exp}) benefits from an ability to identify a larger number of outlier columns relative to the ACOS method. In effect, this provides a trade-off between detection performance and sample complexity for the two methods.

\begin{algorithm}[t]
\caption{Simplified ACOS (SACOS)}
\label{alg:simple}
\begin{algorithmic}
\STATE \hspace{-1.35em} \textbf{Assume:} $\Mb\in \mathbb{R}^{n_1\times n_2}$ 
\REQUIRE Column sampling Bernoulli parameter $\gamma\in[0,1]$, regularization parameter $\lambda>0$,
Measurement matrices \ \ $\bPhi\in\mathbb{R}^{m\times n_1}$, $\Ab\in\mathbb{R}^{p\times n_2}$, measurement vector $\bphi \in\mathbb{R}^{1\times m}$
\STATE \hspace{-1.35em} \textbf{Initalize:} Column sampling matrix $\Sbb = \Ib_{:,\cS}$, where\\
\STATE  $\cS = \{i:S_i = 1\}$ with $\{S_i\}_{i\in[n_2]}$ i.i.d. Bernoulli$(\gamma)$
\STATE \hspace{-0.9em}\textbf{\uline{Step 1}}  
%\STATE Collect Measurements: $\Yb_{(1)} = \bPhi \Mb \Sbb$\\
\STATE Collect Measurements: $\Yb = \bPhi \Mb$
\STATE Form: $\Yb_{(1)} = \Yb \Sbb$
\STATE Solve: $\{\widehat{\Lb}_{(1)},\widehat{\Cb}_{(1)}\} = \argmin_{\bL,\bC} \|\bL\|_{*} + \lambda \|\bC\|_{1,2}$
\STATE \hspace{10em} $\mbox{ s.t. } \ \Yb_{(1)} = \bL + \bC$\\ 
\STATE Let: $\widehat{\cL}_{(1)}$ be the linear subspace spanned by col's of $\widehat{\Lb}_{(1)}$\\
\STATE \hspace{-0.9em}\textbf{\uline{Step 2}} 
\STATE Compute: $\Pb_{\widehat{\mathcal{L}}_{(1)}}$, the orthogonal  projector onto $\widehat{\cL}_{(1)}$
\STATE Set: $\Pb_{\widehat{\cL}_{(1)}^{\perp}} \triangleq \Ib - \Pb_{\widehat{\mathcal{L}}_{(1)}}$
%\STATE Collect Measurements: $\Yb_{(2)} =\Pb_{\widehat{\cL}_{(1)}^{\perp}} \bPhi \Mb$
\STATE Form: $\Yb_{(2)} =\Pb_{\widehat{\cL}_{(1)}^{\perp}} \Yb$
\STATE Form: $\widehat{\cbb}$ with $\widehat{c}_i = \|(\Yb_{(2)})_{:,i}\|_2$ for all $i\in[n_2]$ 
\ENSURE $\widehat{\cI}_{\Cb} = \{i: \widehat{\rm c}_{i} \neq 0\}$
\end{algorithmic}
\end{algorithm}

\subsection{Related Work}

Our effort here leverages results from Compressive Sensing (CS), where parsimony in the object or signal being acquired, in the form of \emph{sparsity}, is exploited to devise efficient procedures for acquiring and reconstructing high-dimensional objects \cite{Candes:06:Freq, Donoho:06:CS, Candes:06:UES, Candes:08:RIP}.  The sequential and adaptive nature of our proposed approach is inspired by numerous recent works in the burgeoning area of adaptive sensing and adaptive CS (see, for example, \cite{Ji:08:BCS, Bashan:08, Haupt:09:CBS, Haupt:11:DS, Bashan:11, Indyk:11, Iwen:12, Malloy:12, Singh:12, Castro:12, Price:12, Malloy:12:NOACS, Davenport:12:CBS, Singh:13, Arias-Castro:13, Wei:13, Krishnamurthy:13, Soni:14} 
as well as the summary article \cite{Haupt:11:Chapter} and the references therein).  The column subsampling inherent in the first step of our approaches is also reminiscent of the data partitioning strategy of the \emph{divide-and-conquer} parallelization approach of \cite{Mackey:11} (though our approach only utilizes one small partition of the data for the first inference step).

Our efforts here utilize a generalization of the notion of sparsity, formalized in terms of a low-rank plus outlier matrix model.  In this sense, our efforts here are related to earlier work in Robust PCA \cite{Chandrasekaran:11:Rank, Candes:11:PCA} that seek to identify low-rank matrices in the presence of sparse impulsive outliers, and their extensions to settings where the outliers present as entire columns of an otherwise low-rank matrix \cite{Xu:12, Chen:11, McCoy:11, Hardt:13, Lerman:14}.  In fact, the computational approach and theoretical analysis of the first step of our approach make direct utilization of the results of \cite{Xu:12}.  

We also note a related work \cite{Wright:13}, which seeks to decompose matrices exhibiting some simple structure (e.g., low-rank plus sparse, etc.) into their constituent components from compressive observations.  Our work differs from that approach in both the measurement model and scope.  Namely, our linear measurements take the form of row and column operations on the matrix and our overall approach is adaptive in nature, in contrast to the non-adaptive ``global'' compressive measurements acquired in \cite{Wright:13}, each of which is essentially a linear combination of all of the matrix entries.  Further, the goal of \cite{Wright:13} was to exactly recover the constituent components, while our aim is only to identify the locations of the outliers.  We discuss some further connections with \cite{Wright:13} in Section~\ref{sec:disc}.

A component of our numerical evaluation here entails assessing the performance of our approach in a stylized image processing task of saliency map estimation.  We note that several recent works have utilized techniques from the sparse representation literature in salient region identification, and in compressive imaging scenarios.  A seminal effort in this direction was \cite{Olshausen:97:Sparse}, which proposed a model for feature identification via the human visual cortex based on parsimonious (sparse) representations.  More recently, \cite{Yan:10} applied techniques from \emph{dictionary learning} \cite{Olshausen:97:Sparse, Aharon:06:KSVD} and low-rank-plus-sparse matrix decomposition \cite{Chandrasekaran:11:Rank, Candes:11:PCA} in a procedure to identify salient regions of an image from (uncompressed) measurements.  Similar sparse representation techniques for salient feature identification were also examined in \cite{Li:09}.  An adaptive compressive imaging procedure driven by a saliency ``map'' obtained via low-resolution discrete cosine transform (DCT) measurements was demonstrated in \cite{Yu:10}. Here, unlike in \cite{Li:09, Yan:10}, we consider salient feature identification based on compressive samples, and while our approach is similar in spirit to the problem examined in \cite{Yu:10}, here we provide theoretical guarantees for the performance of our approach.  Finally, we note several recent works \cite{Aksoylar:13, Haupt:13} that propose methods for identifying salient elements in a data set using compressive samples.

\subsection{Outline}

The remainder of the paper is organized as follows.  In Section~\ref{sec:main} we formalize our problem, state relevant assumptions, and state our main theoretical results that establish performance guarantees for the Adaptive Compressive Outlier Sensing (ACOS) approach of Algorithm~\ref{alg:main} and the Simplified ACOS approach of Algorithm~\ref{alg:simple}.  In Section~\ref{sec:proof} we outline the proofs of our main results. Section~\ref{sec:exp} contains the results of a comprehensive experimental evaluation of our approach on synthetic data, as well as in a stylized image processing application of saliency map estimation. In section~\ref{extension} we empirically investigate several extensions of our methods to noisy and ``missing data'' scenarios. In Section~\ref{sec:disc} we provide a brief discussion of the computational complexity of our approach, and discuss a few potential future directions. We relegate proofs and other auxiliary material to the appendix.

\subsection{A Note on Notation}

We use bold-face upper-case letters ($\Mb, \Lb, \Cb, \bPhi, \bL, \bC, \bI$ etc.) to denote matrices, and use the {\tt MATLAB}-inspired notation $\Ib_{:,\cS}$ to denote the sub matrix formed by extracting columns of $\Ib$ indexed by $i\in\cS$.  We typically use bold-face lower-case letters ($\xb, \vb, \bc, \bphi$, etc.) to denote vectors, with an exception along the lines of the indexing notation above -- i.e., that $\Cb_{:,i}$ denotes the $i$-th column of $\Cb$.  Note that we employ both ``block'' and ``math'' type notation (e.g., $ \Lb, \bL$), where the latter are used to denote variables in the optimization tasks that arise throughout our exposition.  Non-bold letters are used to denote scalar parameters or constants; the usage will be made explicit, or will be clear from context.

The $\ell_1$ norm of a vector $\xb = [{\rm x}_1 \ {\rm x}_2 \ \dots \ {\rm x}_n]$ is $\|\xb\|_1 = \sum_{i} |{\rm x}_i|$ and the $\ell_2$ norm is  $\|\xb\|_2 = \left(\sum_{i} |{\rm x}_i|^2\right)^{1/2}$. We denote the nuclear norm (the sum of singular values) of a matrix $\bL$ by $\|\bL\|_*$ and the $1,2$ norm (the sum of column $\ell_2$ norms) of a matrix $\bC$ by $\|\bC\|_{1,2}$.  We denote the operator norm (the largest singular value) of a matrix $\bL$ by $\|\bL\|$. Superscript asterisks denote complex conjugate transpose.  

For positive integers $n$, we let $[n]$ denote the set of positive integers no greater than $n$; that is, $[n]=\{1,2,\dots,n\}$.  

\section{Main Results}\label{sec:main}

\subsection{Problem Statement}

Our specific problem of interest here may be formalized as follows.  We suppose $\Mb\in\RR^{n_1\times n_2}$ admits a decomposition of the form  $\Mb = \Lb + \Cb$, where $\Lb$ is a low-rank matrix having rank at most $r$, and $\Cb$ is a matrix having some $k\leq n_2$ nonzero columns that we will interpret as ``outliers'' from $\Lb$, in the sense that they do not lie (entirely) within the span of the columns of $\Lb$.  Formally, let $\cL$ denote the linear subspace of $\RR^{n_1}$ spanned by the columns of $\Lb$ (and having dimension at most $r$), denote its orthogonal complement in $\RR^{n_1}$ by $\cL^{\perp}$, and let $\Pb_{\cL}$ and $\Pb_{\cL^{\perp}}$ denote the orthogonal projection operators onto $\cL$ and $\cL^{\perp}$, respectively.  We assume that the nonzero columns of $\Cb$ are indexed by a set $\cI_{\Cb}$ of cardinality $k$, and that $i \in \cI_{\Cb}$ if and only if $\|\Pb_{\cL^{\perp}}\Cb_{:,i}\|_2 > 0$.  Aside from this assumption, the elements of the nonzero columns of $\Cb$ may be \emph{arbitrary}. 

Notice that \emph{without loss of generality}, we may assume that the columns of $\Lb$ are zero at the locations corresponding to the nonzero columns of $\Cb$ (since those columns of $\Lb$ can essentially be aggregated into the nonzero columns of $\Cb$, and the resulting column will still be an outlier according to our criteria above).  We adopt that model here, and assume $\Lb$ has a total of $n_{\Lb}$ nonzero columns\footnote{As we will see, the conditions under which our column subsampling in Step 1 succeeds will depend on the number of \emph{nonzero} columns in the low-rank component, since any all-zero columns are essentially non-informative for learning the low-rank subspace.  Thus, we make the distinction between $n_2$ and $n_{\Lb}$ explicit throughout.}, including all $k$ of the indices in $\cI_{\Cb}$ where $\Cb$ has a nonzero column, but also potentially others, to allow for the case where some $n_{\Lb} - k$ columns of $\Mb$ itself to be zero. Clearly $n_{\Lb} \leq n_2 - k$. 

Given this setup, our problem of interest here may be stated concisely -- our aim is to identify the set $\cI_{\Cb}$ containing the locations of the outlier columns.

\subsection{Assumptions}

It is well-known in the matrix completion and robust PCA literature that separation of low-rank and sparse matrices from observations of their sum may not be a well-posed task -- for example, matrices having only a single nonzero element are simultaneously low rank, sparse, column-sparse, row-sparse, etc.  To overcome these types of identifiability issues, it is common to assume that the linear subspace spanned by the rows and/or columns of the low-rank matrix be ``incoherent'' with the canonical basis (see, e.g., \cite{Candes:09, Chandrasekaran:11:Rank, Candes:11:PCA, Xu:12, Chen:11}).

In a similar vein, since our aim is to identify column outliers from an otherwise low-rank matrix we seek conditions that make the factors distinguishable so that any of the directions of the column space of $\Lb$ that we seek to identify are not defined by a single vector (stated another way, we would like the vectors whose columns comprise $\Lb$ to be ``spread out'' in the subspace spanned by columns of $\Lb$). To this end, we assume an incoherence condition on the row space of the low-rank component $\Lb$. We formalize this notion via the following definition from \cite{Xu:12}.
\begin{definition}[Column Incoherence Property]
Let $\Lb\in\RR^{n_1\times n_2}$ be a rank $r$ matrix with at most $n_{\Lb}\leq n_2$ nonzero columns, and compact singular value decomposition (SVD) $\Lb=\Ub\bSigma\Vb^*$, where $\Ub$ is $n_1\times r$, $\bSigma$ is $r\times r$, and $\Vb$ is $n_2\times r$.  The matrix $\Lb$ is said to satisfy the \emph{column incoherence property} with parameter $\mu_{\Lb}$ if
\begin{equation*}
\max_i \|\Vb^*\eb_i\|_2^2 \leq \mu_{\Lb} \frac{r}{n_{\Lb}},
\end{equation*} 
where $\{\eb_i\}$ are basis vectors of the canonical basis for $\RR^{n_2}$.
\end{definition}
Note that $\mu_{\Lb} \in[1,n_{\Lb}/r]$; the lower limit is achieved when all elements of $\Vb^*$ have the same amplitude, and the upper limit when any one element of $\Vb^*$ is equal to $1$ (i.e., when the row space of $\Lb$ is aligned with the canonical basis).  For our purposes, an undesirable case occurs when $\Vb^*$ is such that $\max_i \|\Vb^*\eb_i\|_2^2 =1$, since this implies that (at least) one of the directions in the span of the columns of $\Lb$ is described by only a single vector (and thus distinguishing that vector from a column outlier becomes ambiguous).

With this, we may state our assumptions concisely, as follows: we assume that the components $\Lb$ and $\Cb$ of the matrix $\Mb=\Lb + \Cb$ satisfy the following \emph{structural conditions}:
\begin{itemize}
\item[(\textbf{c1})] $\rk(\Lb) = r$,
\item[(\textbf{c2})] $\Lb$ has $n_{\Lb}$ nonzero columns,
\item[(\textbf{c3})] $\Lb$ satisfies the \emph{column incoherence property} with parameter $\mu_{\Lb}$, and
\item[(\textbf{c4})] $|\cI_{\Cb}|=k$, where $\cI_{\Cb} = \{i: \|\Pb_{\cL^{\perp}}\Cb_{:,i}\|_2 > 0, \Lb_{:,i} = \mathbf{0}\}$.
\end{itemize}

\subsection{Recovery Guarantees and Implications}

Our main results identify conditions under which the procedures outlined in Algorithm~\ref{alg:main} and Algorithm~\ref{alg:simple} succeed.  Our particular focus is on the case where the measurement matrices are random, and satisfy the following property.
\begin{definition}[Distributional Johnson-Lindenstrauss (JL) Property]
An $m\times n$  matrix $\bPhi$ is said to satisfy the \emph{distributional JL property} if for any fixed $\vb\in\RR^{n}$ and any $\epsilon\in(0,1)$,
\begin{equation}\label{eqn:distJL}
\Prob\left(\ \left| \ \|\bPhi\vb\|_2^2 - \|\vb\|_2^2 \ \right| \geq \epsilon \|\vb\|_2^2 \ \right) \leq 2e^{-mf(\epsilon)},
\end{equation} 
where $f(\epsilon)>0$ is a constant depending only on $\epsilon$ that is specific to the distribution of $\bPhi$.
\end{definition}
Random matrices satisfying the distributional JL property are those that preserve the length of any fixed vector to within a multiplicative factor of $(1\pm \epsilon)$ with probability at least $1-2e^{-mf(\epsilon)}$.  By a simple union bounding argument, such matrices can be shown to approximately preserve the lengths of a finite collection of vectors, all vectors in a linear subspace, all vectors in a union of subspaces, etc., provided the number of rows is sufficiently large.  As noted in \cite{Gilbert:12}, for many randomly constructed and appropriately normalized $\bPhi$, (e.g., such that entries of $\bPhi$ are  i.i.d. zero-mean Gaussian, or are drawn as an ensemble from any subgaussian distribution), $f(\epsilon)$ is quadratic\footnote{It was shown in \cite{Achlioptas:01}, for example, that $f(\epsilon) = \epsilon^2/4 - \epsilon^3/6$ for matrices whose elements are appropriately normalized Gaussian or symmetric Bernoulli random variables.} in $\epsilon$ as $\epsilon\rightarrow 0$.  This general framework also allows us to directly utilize other specially constructed \emph{fast} or \emph{sparse} JL transforms \cite{Ailon:06, Dasgupta:10}.

With this, we are in position to formulate our first main result. We state it here as a theorem; its proof appears in Section~\ref{sec:proof}.
\begin{thmi}[Accurate Recovery via ACOS]\label{thm:main}
Suppose $\Mb = \Lb + \Cb$, where the components $\Lb$ and $\Cb$ satisfy the structural conditions (\textbf{c1})-(\textbf{c4}) with
\begin{equation}\label{eqn:kmin}
k \leq \frac{1}{40(1+121\ r\mu_{\Lb})} \ n_2.
\end{equation}
For any $\delta\in(0,1)$, if the column subsampling parameter $\gamma$ satisfies
\begin{equation}\label{eqn:gammamin}
\gamma \geq \max\left\{\frac{1}{20}, \ \frac{200 \log(\frac{5}{\delta})}{n_{\Lb}}, \ \frac{24\log(\frac{10}{\delta})}{n_2},\ \frac{10 r \mu_{\Lb} \log(\frac{5r}{\delta})}{n_{\Lb}}   \right\},
\end{equation}
the measurement matrices are each drawn from any distribution satisfying \eqref{eqn:distJL} with
\begin{equation}\label{eqn:mmin}
m \geq \frac{5(r+1) + \log(k) + \log(2/\delta)}{f(1/4)}
\end{equation}
and
\begin{equation}\label{eqn:pmin}
p \geq \frac{11k + 2k\log(n_2/k) + \log(2/\delta)}{f(1/4)},
\end{equation}
the elements of $\bphi$ are i.i.d. realizations of any continuous random variable, and for any upper bound $k_{\rm ub}$ of $k$ the regularization parameter is set to $\lambda = \frac{3}{7\sqrt{ k_{\rm ub} }}$, then the following hold simultaneously with probability at least $1-3\delta$:
\begin{itemize}
\item the ACOS procedure in Algorithm~\ref{alg:main} correctly identifies the salient columns of $\bC$ (i.e., $\widehat{\cal I}_{\bC} = {\cal I}_{\bC}$), and
\item the total number of measurements collected is no greater than $\left(\frac{3}{2}\right)\gamma m n_2 + p$.
\end{itemize}
\end{thmi}

It is interesting to compare this result with that of \cite{Xu:12}, which established that the Outlier Pursuit procedure \eqref{eqn:OP} succeeds in recovering the true low-rank subspace and locations of the outlier columns provided $\Mb$ satisfy conditions analogous to (\textbf{c1})-(\textbf{c4}) with $k \leq n_2/(1+(121/9)\ r\mu_{\Lb})$. The sufficient condition \eqref{eqn:kmin} on the number of recoverable outliers that we identify for the ACOS procedure differs from the condition identified in that work by only constant factors. Further, the number of identifiable outliers could be as large as a fixed fraction of $n_2$ when both the rank $r$ and coherence parameter $\mu_{\Lb}$ are small.

It is also interesting to note the sample complexity improvements that are achievable using the ACOS procedure.  Namely, it follows directly from our analysis that for appropriate choice of the parameters $\gamma,m,$ and $p$ the ACOS algorithm correctly identifies the salient columns of $\Cb$ with high probability from relatively few observations, comprising only a fraction of the measurements required by other comparable (non-compressive) procedures \cite{Xu:12} that produce the same correct salient support estimate but operate directly on the full ($n_1 \times n_2$) matrix $\bM$.  Specifically, our analysis shows that the ACOS approach succeeds with high probability with an effective sampling rate of $\frac{\# {\rm obs}}{n_1 n_2} = {\cal O}\left( \ \max\left\{\frac{(r+ \log k) (n_2/n_{\Lb})\mu_{\Lb} r\log r}{n_1n_2}, \frac{(r+ \log k)}{n_1} \right\}   + \frac{k\log(n_2/k)}{n_1n_2}\ \right)$, which may be small when $r$ and $k$ are each small relative to the problem dimensions (and $n_{\Lb} \sim n_2$, so that $\Lb$ does not have a large number of zero columns outside of $\cI_{\Cb}$).

Another point of comparison for our result comes from the related work \cite{Chen:11}, which addresses a different (and in a sense, more difficult) task of identifying both the column space and the set of outlier columns of a matrix $\Mb = \Lb + \Cb$ from observations that take the form of samples of the elements of $\Mb$.  There, to deal with the fact that observations take the form of point samples of the matrix (rather than more general linear measurements as here), the authors of \cite{Chen:11} assume that $\Lb$ also satisfy a row incoherence property in addition to a column incoherence property, and show that in this setting that the column space of $\Lb$ and set of nonzero columns of $\Cb$ may be recovered from only $\cO\left(n_2 r^2 \mu^2 \log(n_2)\right)$ observations via a convex optimization, where $\mu\in[1,n_1/r]$ is the row incoherence parameter.  Normalizing this sample complexity by $n_1n_2$ facilitates comparison with our result above; we see that the sufficient conditions for the sample complexity of our approach are smaller than for the approach of \cite{Chen:11} by a factor of at least $1/r$, and, our approach does not require the row incoherence assumption.  We provide some additional, experimental, comparisons between our ACOS method and the RMC method in Section~\ref{sec:exp}.

We may also obtain performance guarantees for Algorithm~\ref{alg:simple} (in effect, using a simplified version of the analysis used to establish Theorem~\ref{thm:main}).  This yields the following corollary. 
\begin{cori}[Accurate Recovery via SACOS]\label{cor:main}
	Suppose $\Mb = \Lb + \Cb$, where the components $\Lb$ and $\Cb$ satisfy the structural conditions (\textbf{c1})-(\textbf{c4}) with $k$ as in \eqref{eqn:kmin}. Let the measurement matrix $\bPhi$ be drawn from a distribution satisfying \eqref{eqn:distJL}, and assume \eqref{eqn:gammamin} and \eqref{eqn:mmin} hold.  If for any upper bound $k_{\rm ub}$ of $k$ the regularization parameter is set to $\lambda = \frac{3}{7\sqrt{ k_{\rm ub} }}$, then the following hold simultaneously with probability at least $1-2\delta$:
\begin{itemize}
	\item the ACOS procedure in Algorithm~\ref{alg:simple} correctly identifies the salient columns of $\bC$ (i.e., $\widehat{\cal I}_{\bC} = {\cal I}_{\bC}$), and
	\item the total number of measurements collected is no greater than $m n_2 $.
\end{itemize}
\end{cori}
We leave the proof (which is straightforward, using the lemmata in the following section) to the interested reader.

\section{Proof of Theorem~\ref{thm:main}}\label{sec:proof}

First, we note that in both of the steps of Algorithm~\ref{alg:main} the prescribed observations are functions of $\Mb$ only through $\bPhi\Mb$; stated another way, $\Mb$ never appears in the algorithm in isolation from the measurement matrix $\bPhi$.  Motivated by this, we introduce
\begin{equation}\label{eqn:Mtilde}
\widetilde{\Mb} \triangleq \bPhi \Mb = \bPhi \Lb + \bPhi\Cb = \widetilde{\Lb} + \widetilde{\Cb},
\end{equation}
to effectively subsume the action of $\bPhi$ into $\widetilde{\Mb}$.  Now, our proof is a straightforward consequence of assembling three intermediate probabilistic results via a union bounding argument.  The first intermediate result establishes that for $\Mb = \Lb + \Cb$ with components $\Lb$ and $\Cb$ satisfying the structural conditions \textbf{(${\bc}$1})-\textbf{(${\bc}$4}), the components $\widetilde{\Lb}$ and $\widetilde{\Cb}$ of $\widetilde{\Mb}$ as defined in \eqref{eqn:Mtilde} satisfy analogous structural conditions provided that $m$, the number of rows of $\bPhi$, be sufficiently large.  We state this result here as a lemma; its proof appears in Appendix~\ref{a:lem1}.
\begin{lemmai}\label{lem:Mtilde}
Suppose $\Mb = \Lb + \Cb$, where $\Lb$ and $\Cb$ satisfy the structural conditions (\textbf{c1})-(\textbf{c4}).  Fix any $\delta\in(0,1)$, suppose $\bPhi$ is an $m\times n_1$ matrix drawn from a distribution satisfying the distributional JL property \eqref{eqn:distJL} with $m$ satisfying \eqref{eqn:mmin} and let $\widetilde{\Mb}=\widetilde{\Lb} + \widetilde{\Cb}$ be as defined in \eqref{eqn:Mtilde}. Then, the components $\widetilde{\Lb}$ and $\widetilde{\Cb}$ satisfy the following conditions simultaneously with probability at least $1-\delta$:
\begin{itemize}
\item[\textbf{($\widetilde{\bc}$1})] $\rk(\widetilde{\Lb}) = r$,
\item[\textbf{($\widetilde{\bc}$2})] $\widetilde{\Lb}$ has $n_{\Lb}$ nonzero columns,
\item[\textbf{($\widetilde{\bc}$3})] $\widetilde{\Lb}$ satisfies the column incoherence property with parameter $\mu_{\Lb}$, and
\item[\textbf{($\widetilde{\bc}$4})] ${\cI}_{\widetilde{\Cb}} \triangleq \{i: \|\Pb_{{\widetilde{\cL}}^{\perp}} \widetilde{\Cb}_{:,i}\|_2 > 0, \widetilde{\Lb}_{:,i} = \mathbf{0}\} = {\cI}_{\Cb}$, where $\widetilde{\cL}$ is the linear subspace of $\RR^{m}$ spanned by the columns of $\widetilde{\Lb}$, and $\Pb_{{\widetilde{\cL}}^{\perp}}$ denotes the orthogonal projection onto the orthogonal complement of $\widetilde{\cL}$ in $\RR^{m}$.
\end{itemize} 
\end{lemmai}

The second intermediate result guarantees two outcomes -- first, that Step 1 of Algorithm~\ref{alg:main} succeeds in identifying the correct column space of $\widetilde{\cL}$ (i.e., that $\widehat{\cL}_{(1)} = \widetilde{\cL}$) with high probability provided the components $\widetilde{\Lb}$ and $\widetilde{\Cb}$ of $\widetilde{\Mb}$ as specified in \eqref{eqn:Mtilde} satisfy the structural conditions \textbf{($\widetilde{\bc}$1})-\textbf{($\widetilde{\bc}$4}) and the column sampling probability parameter $\gamma$ be sufficiently large, and second, that the number of columns of the randomly generated sampling matrix $\Sbb$ be close to $\gamma n_2$. We also provide this result as a lemma; its proof appears in Appendix~\ref{a:lem2}.
\begin{lemmai}\label{lem:Step1}
Let $\widetilde{\Mb}=\widetilde{\Lb} + \widetilde{\Cb}$ be an $m\times n_2$ matrix, where the components $\widetilde{\Lb}$ and $\widetilde{\Cb}$ satisfy the conditions \textbf{($\widetilde{\bc}$1})-\textbf{($\widetilde{\bc}$4}) with $k$ satisfying \eqref{eqn:kmin}. Fix $\delta\in(0,1)$ and suppose the column sampling parameter $\gamma$ satisfies \eqref{eqn:gammamin}. When $\lambda = \frac{3}{7\sqrt{ k_{\rm ub} }}$ for any $k_{\rm ub}\geq |\cI_{\widetilde{\Cb}}|$, the following hold simultaneously with probability at least $1-\delta$:  $\Sbb$ has $|\cS|\leq (3/2) \gamma n_2$ columns, and the subspace $\widehat{\cL}_{(1)}$ resulting from Step 1 of Algorithm~\ref{alg:main} satisfies $\widehat{\cL}_{(1)} = \widetilde{\cL}$.
\end{lemmai}

Our third intermediate result shows that the support set of the vector $\widehat{\cbb}$ produced in Step 2 of Algorithm~\ref{alg:main} is the same as the set of salient columns of $\widetilde{\Cb}$, provided that $\widehat{\cL}_{(1)} = \widetilde{\cL}$ and that $p$, the number of rows of $\Ab$, is sufficiently large.  We state this result here as a lemma; its proof appears in Appendix~\ref{a:lem3}
\begin{lemmai}\label{lem:Step2}
$\widetilde{\Mb}=\widetilde{\Lb} + \widetilde{\Cb}$ be an $m\times n_2$ matrix, where the components $\widetilde{\Lb}$ and $\widetilde{\Cb}$ satisfy the conditions \textbf{($\widetilde{\bc}$1})-\textbf{($\widetilde{\bc}$4}) for any $k\leq n_2$, and suppose $\widehat{\cL}_{(1)} = \widetilde{\cL}$, the subspace spanned by the columns of $\widetilde{\Lb}$. Let $\bPhi\Mb = \widetilde{\Mb}$ in Step 2 of Algorithm~\ref{alg:main}.  Fix $\delta\in(0,1)$, suppose $\Ab$ is a $p\times n_2$ matrix drawn from a distribution satisfying the distributional JL property \eqref{eqn:distJL} with $p$ satisfying \eqref{eqn:pmin}, and suppose the elements of $\bphi$ are i.i.d. realizations of any continuous random variable. Then with probability at least $1-\delta$ the support $\cI_{\widehat{\cbb}}\triangleq \{i: \widehat{\rm c}_i \neq 0\}$ of the vector $\widehat{\cbb}$ produced by Step 2 of Algorithm~\ref{alg:main} satisfies $\cI_{\widehat{\cbb}} = \cI_{\widetilde{\Cb}}$.
\end{lemmai}

Our overall result follows from assembling these intermediate results via union bound.   In the event that the conclusion of Lemma~\ref{lem:Mtilde} holds, then so do the requisite conditions of Lemma~\ref{lem:Step1}.  Thus, with probability at least $1-2\delta$ the conclusions of Lemmata~\ref{lem:Mtilde} and \ref{lem:Step1} both hold.  This implies that the requisite conditions of Lemma~\ref{lem:Step2} hold also with probability at least $1-2\delta$, and so it follows that the conclusions of all three Lemmata hold with probability at least $1-3\delta$.

\section{Experimental Evaluation}\label{sec:exp}

In this section we provide a comprehensive experimental evaluation of the performance of our approaches for both synthetically generated and real data, the latter motivated by a stylized application of saliency map estimation in an image processing task.  We compare our methods with the Outlier Pursuit (OP) approach of \cite{Xu:12} and the Robust Matrix Completion (RMC) approach of \cite{Chen:11}, each of which employs a convex optimization to identify both the subspace in which the columns of the low rank matrix lie, and the locations of the nonzero columns in the outlier matrix.  We implement the RMC method using an accelerated approximate alternating direction method of multipliers (ADMM) method inspired by \cite{Goldstein:12} (as well as \cite{Xu:12,Beck:09}). We implement the OP methods (as well as the intermediate execution of the OP-like optimization in Step 1 of our approach) using the procedure in \cite{Chen:11}.  We implement the $\ell_1$-regularized estimation in Step 2 of our procedure by casting it as a LASSO problem and using an accelerated proximal gradient method \cite{Beck:09}.

\begin{figure}[!t]
\footnotesize
\centering
\subfloat[2.1\%]{
	\includegraphics[width=0.28\linewidth]{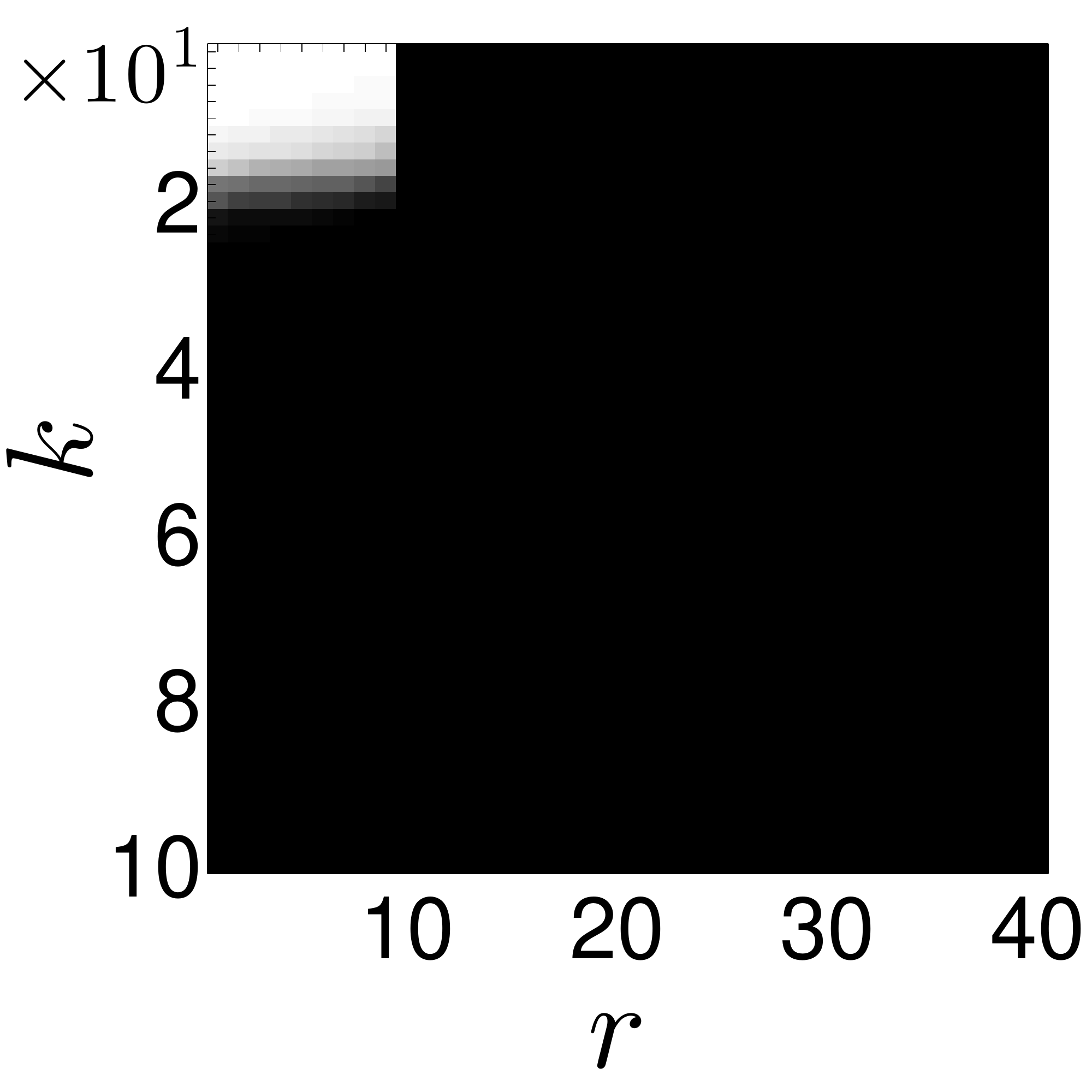}
}
\subfloat[2.2\%]{
	\includegraphics[width=0.28\linewidth]{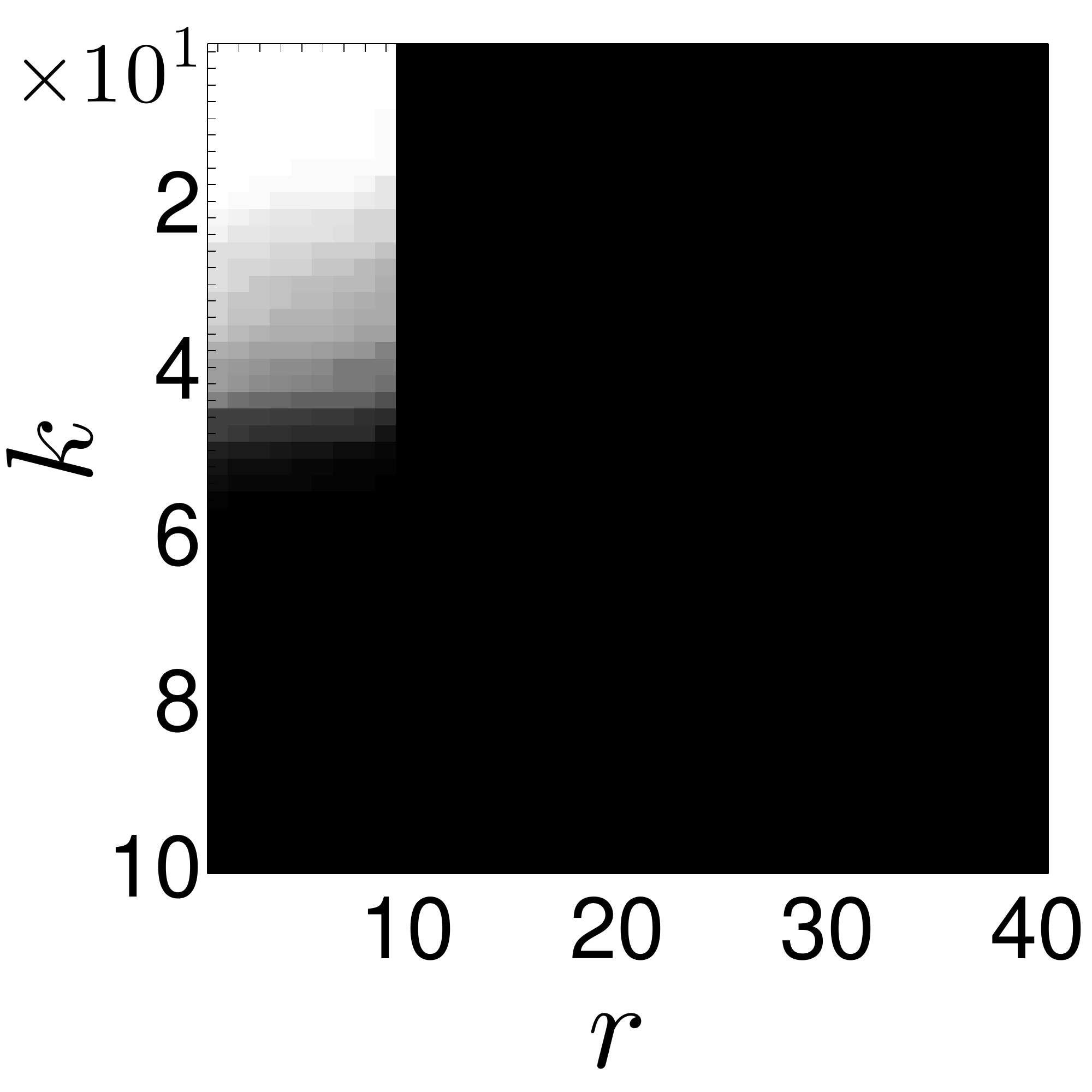}
}
\subfloat[2.3\%]{
	\includegraphics[width=0.28\linewidth]{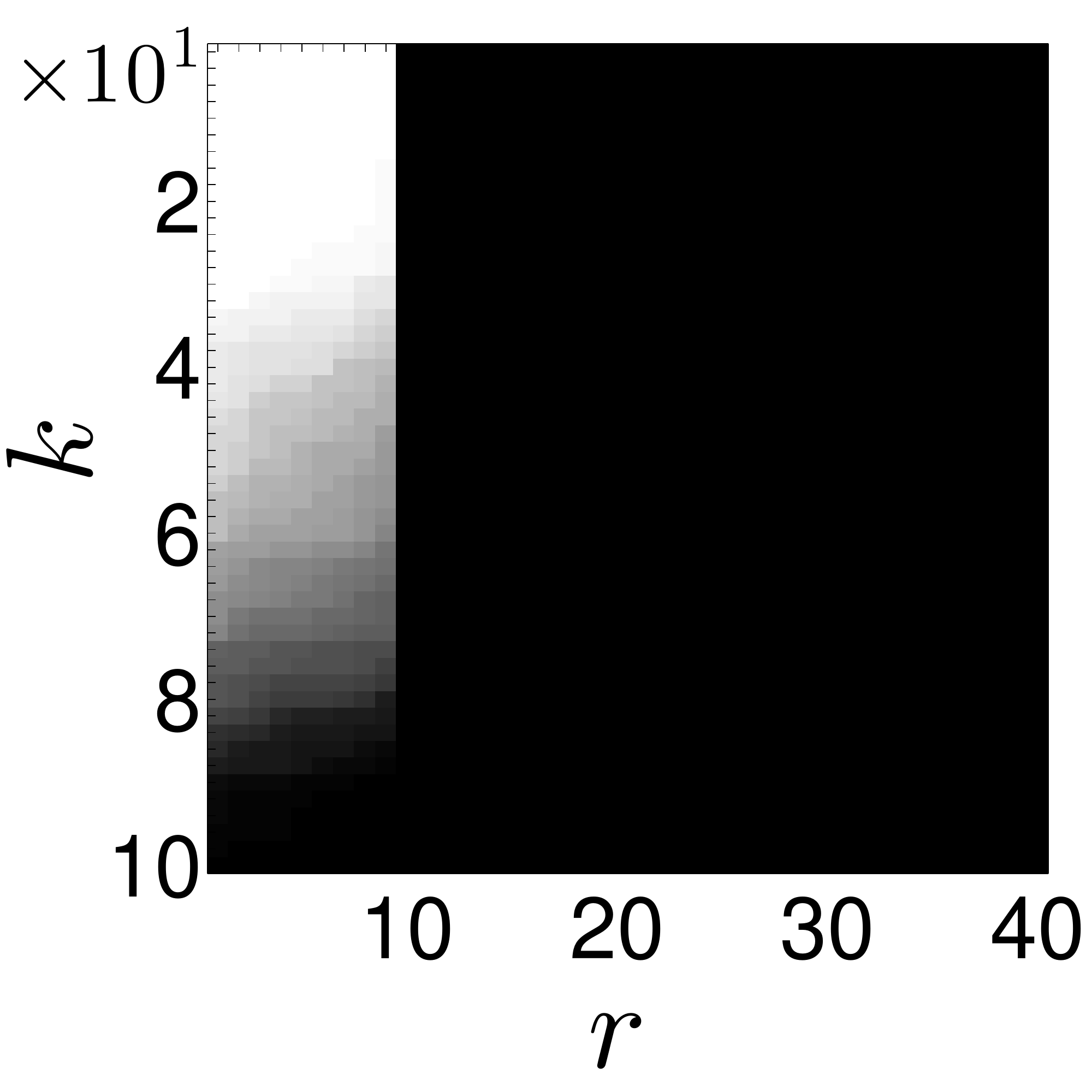}
} \vspace{-0.3em}\\
\subfloat[4.1\%]{
	\includegraphics[width=0.28\linewidth]{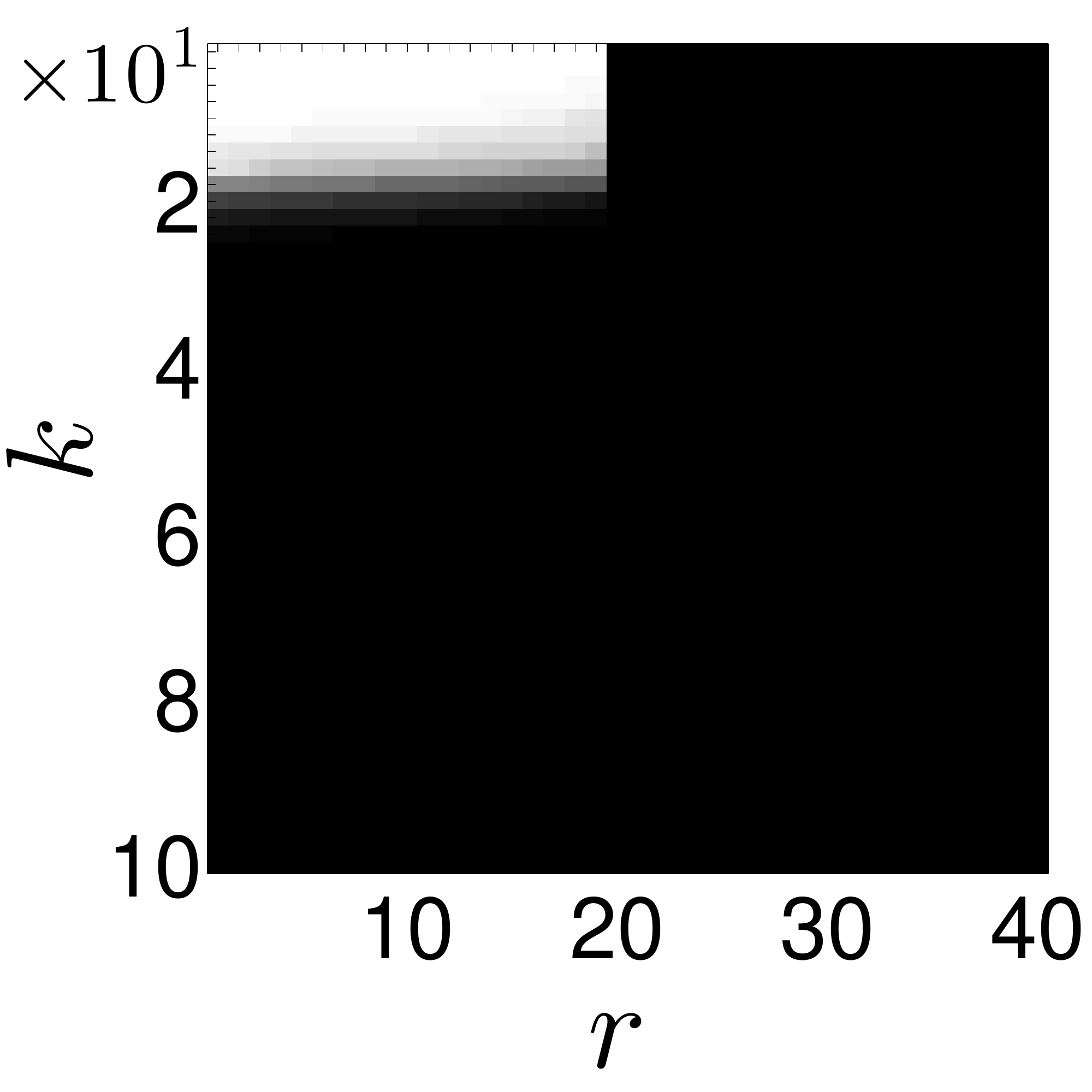}
}
\subfloat[4.2\%]{
	\includegraphics[width=0.28\linewidth]{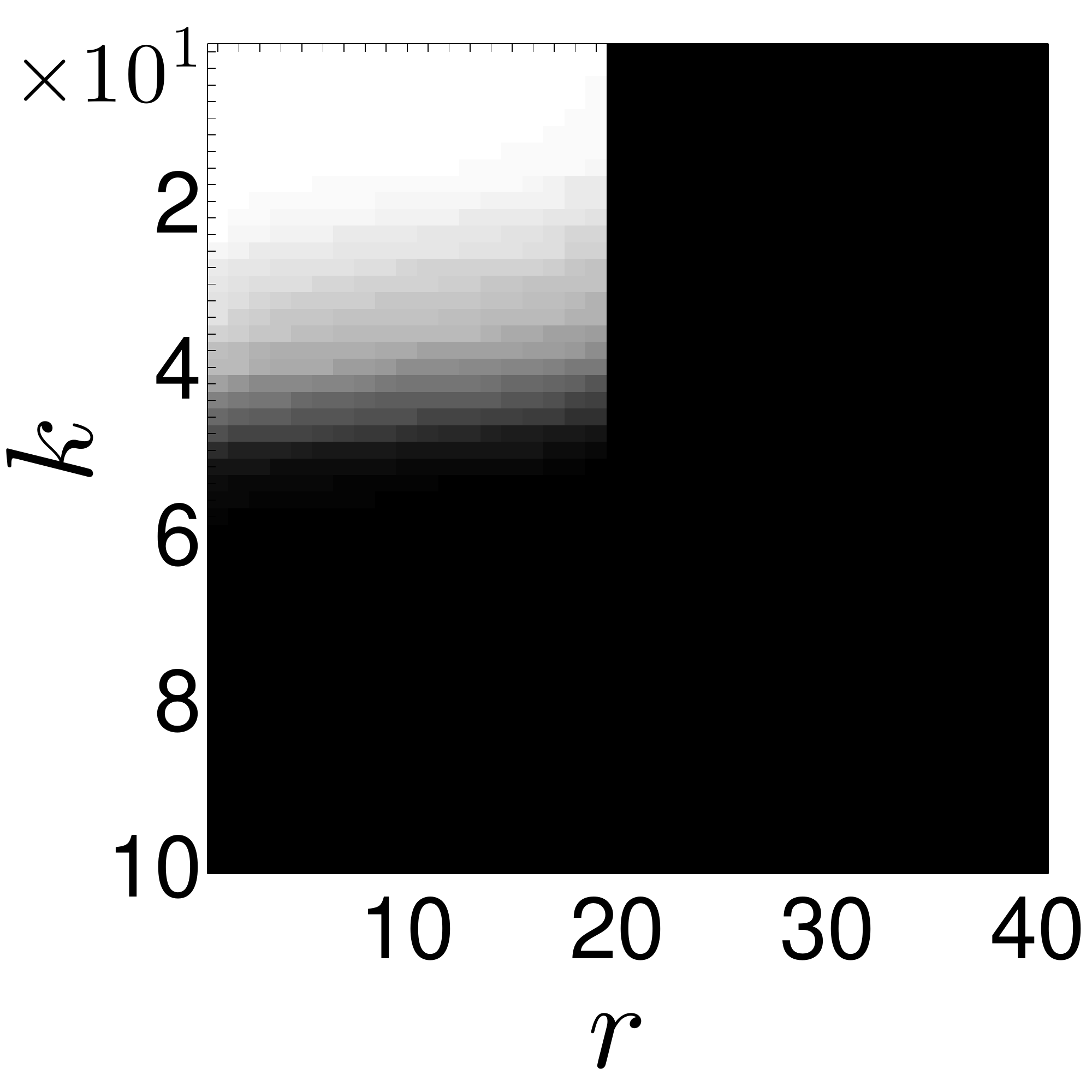}
}\subfloat[4.3\%]{
	\includegraphics[width=0.28\linewidth]{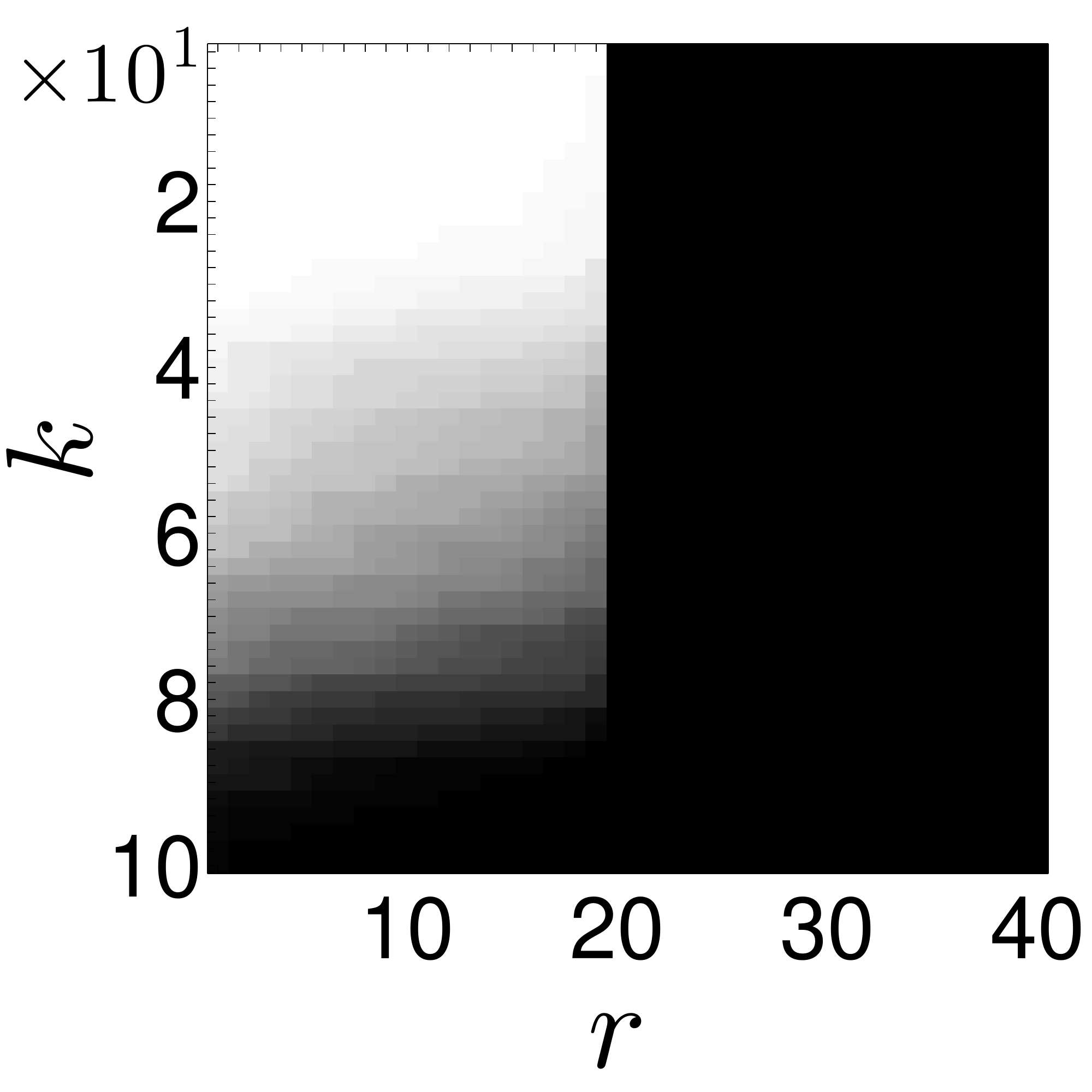}
}\vspace{-0.3em}\\
\subfloat[6.1\%]{
	\includegraphics[width=0.28\linewidth]{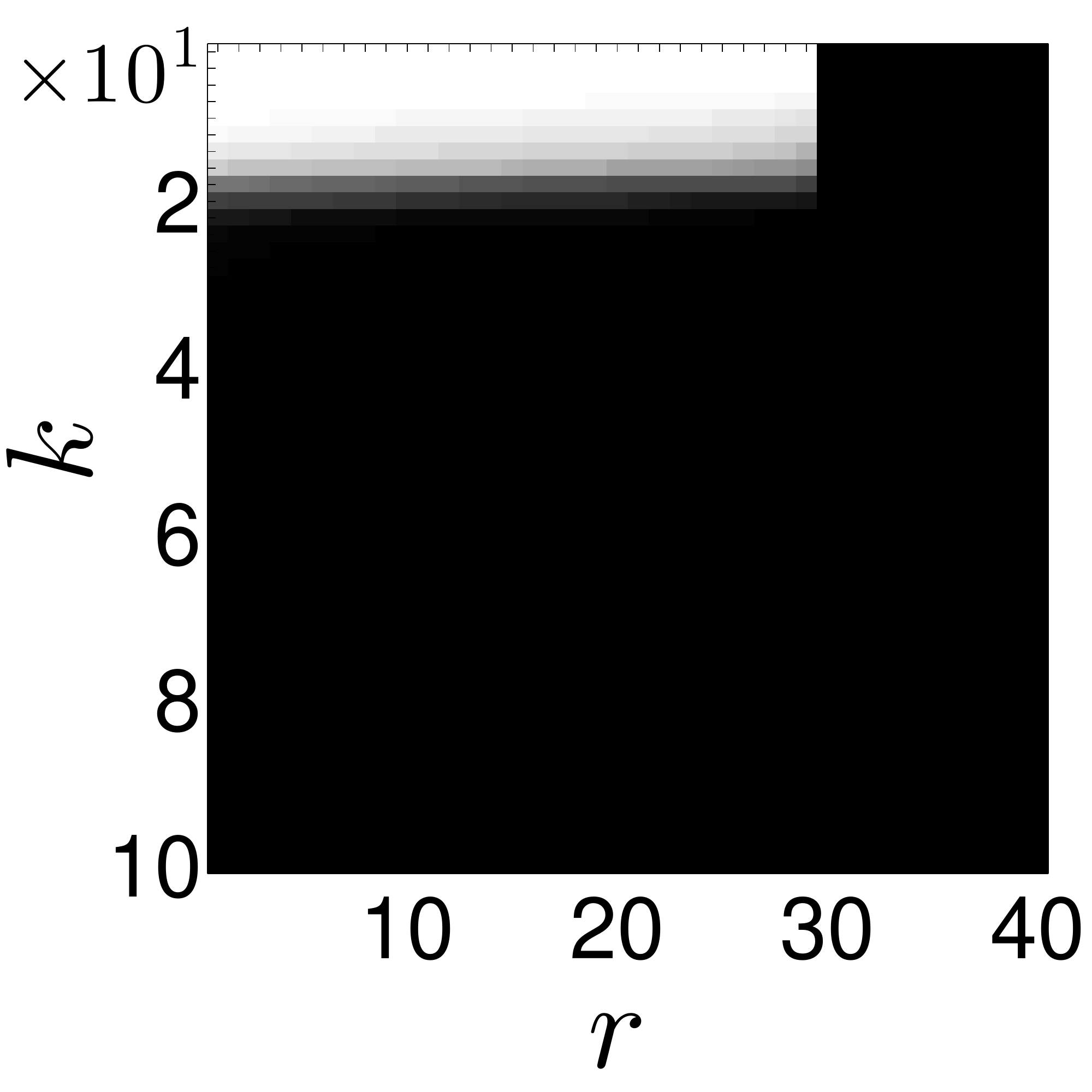}
}
\subfloat[6.2\%]{
	\includegraphics[width=0.28\linewidth]{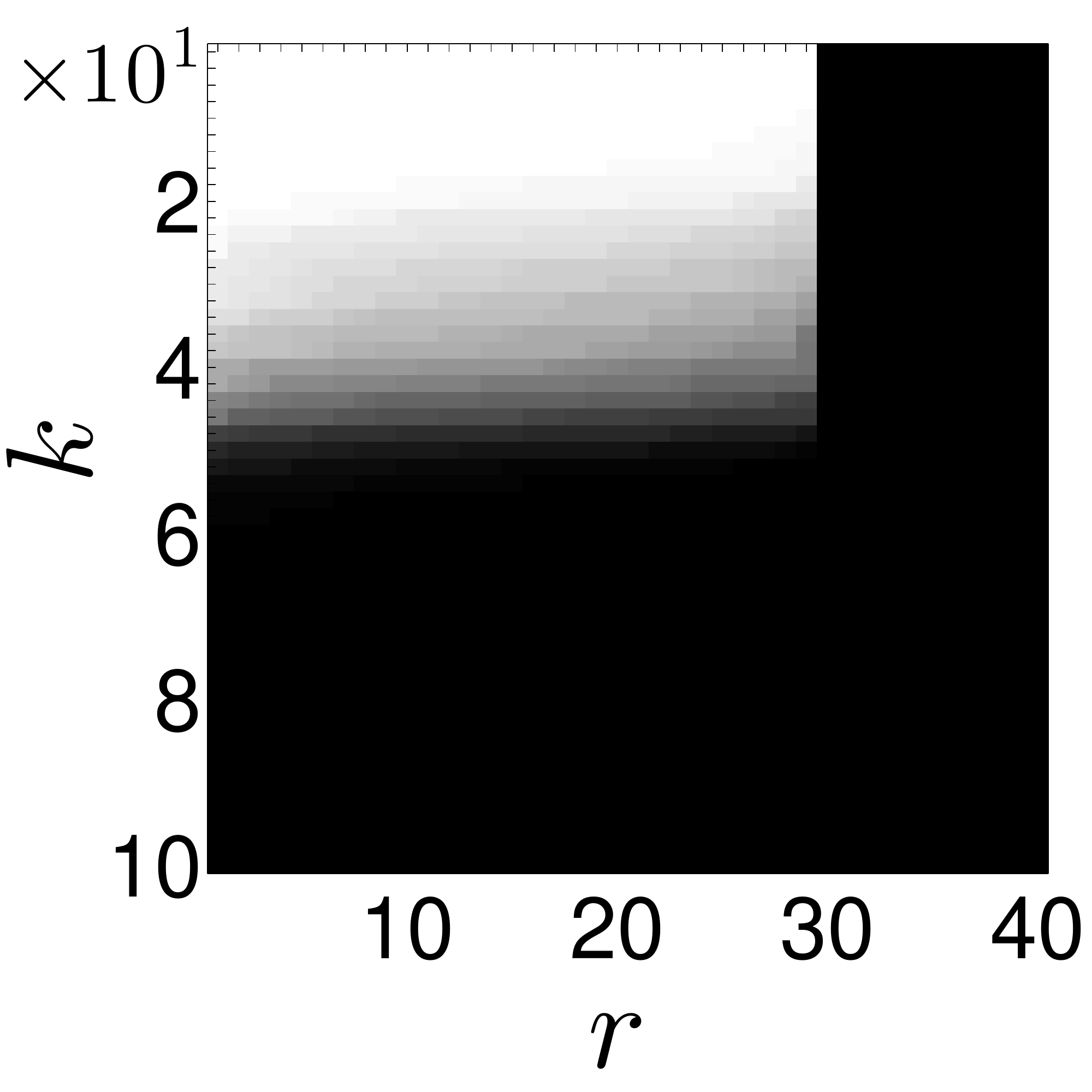}
}
\subfloat[6.3\%]{
	\includegraphics[width=0.28\linewidth]{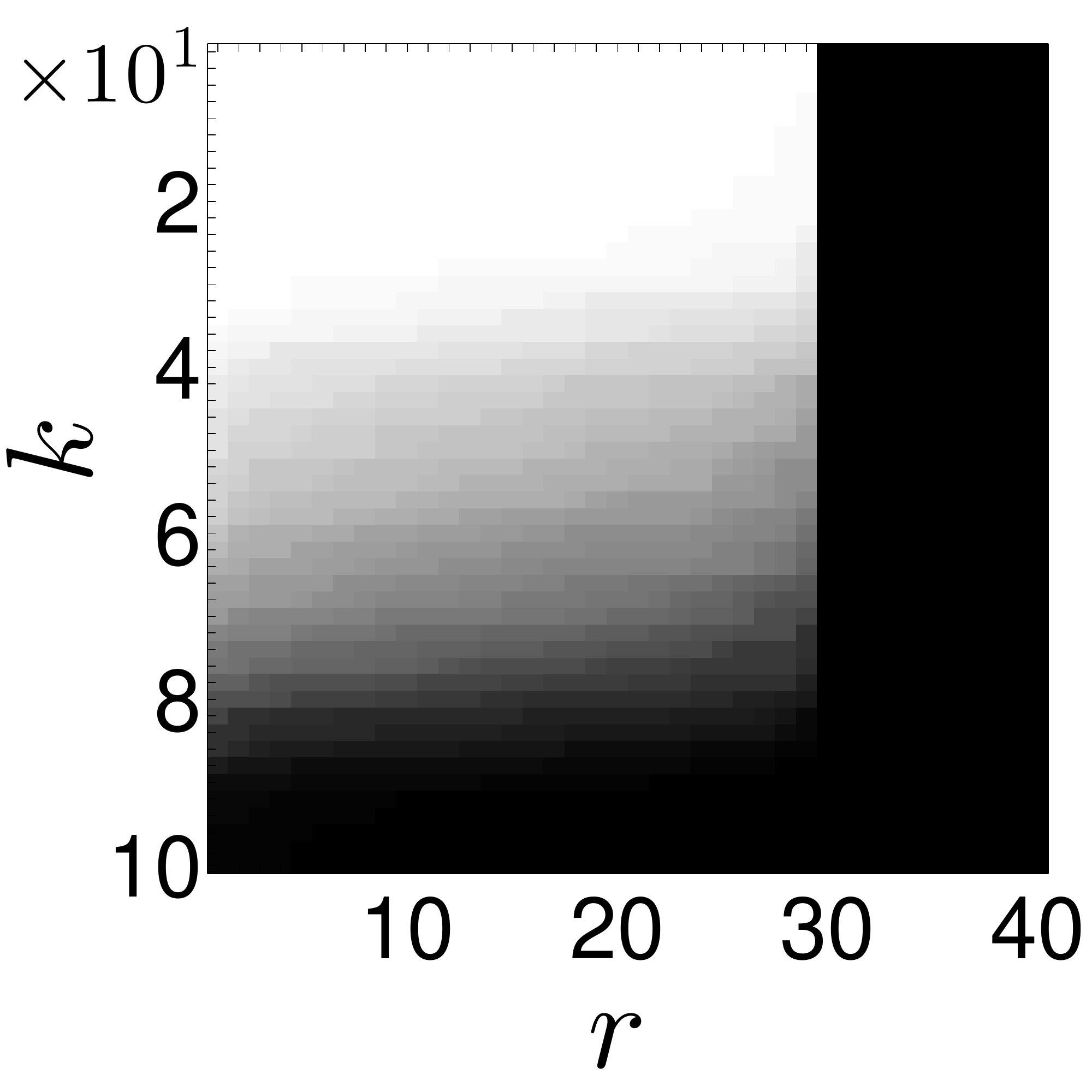}
}\\
%\vspace{-0.1in}
\caption{Outlier recovery phase transitions plots for ACOS (white regions correspond to successful recovery). Each row of the figure corresponds to a different level of compression of rows of $\Mb$, where $m = 0.1n_1, 0.2n_1$ and $0.3n_1$, respectively, from top to bottom.  Each column corresponds to a different level of compression of rows of $\Mb$ in Step 2 of Algorithm~\ref{alg:main}, with $p = 0.1n_2, 0.2n_2$ and $0.3n_2$, respectively, from left to right.  The fraction of observations obtained (as a percentage, relative to the full dimension) is provided as a caption below each figure.  As expected, increasing $m$ (top to bottom) facilitates accurate estimation for increasing rank $r$ of $\Lb$, while increasing $p$ (left to right) allows for recovery of increasing numbers $k$ of outlier columns.}
\label{fig:sim1}
\vspace{-0.1in}
\end{figure}

\subsection{Synthetic Data}\label{sec:syn}

We experiment on synthetically generated $n_1 \times n_2$ matrices $\Mb$, with $n_1 = 100$ and $n_2=1000$, formed as follows.  For a specified rank $r$ and number of outliers $k$, we let the number of nonzero columns of $\Lb$ be $n_{\Lb} = n_2-k$, generate two random matrices $\Ub \in \RR^{n_1 \times r}$ and $\Vb \in \RR^{n_{\Lb} \times r}$ with i.i.d. $\cN \left( 0,1 \right)$ entries, and we take $\Lb = [\Ub\Vb^{T} \ \bZero_{n_1\times k}]$.  We generate the outlier matrix $\Cb$ as $\Cb=[\bZero_{n_1\times n_{\Lb}} \ \Wb]$ where $\Wb\in\RR^{n_1\times k}$ has i.i.d. $\cN(0,r)$ entries (which are also independent of entries of $\Ub$ and $\Vb$).  Then, we set $\Mb = \Lb + \Cb$. Notice that the outlier vector elements have been scaled, so that all columns of $\Mb$ have the same squared $\ell_2$ norm, in expectation. In all experiments we generate $\bphi$, $\bPhi$, and $\Ab$ with i.i.d. zero-mean Gaussian entries.

Our first experiment investigates the ``phase transition'' behavior of our ACOS approach; our experimental setting is as follows.  First, we set the average sampling rate by fixing the column downsampling fraction $\gamma=0.2$, and choosing a row sampling parameter $m\in\{0.1n_1, 0.2n_1, 0.3n_1\}$ and column sampling parameter $p\in\{0.1n_2, 0.2n_2, 0.3n_2\}$.  Then, for each $(r,k)$ pair with $r\in\{1, 2, 3, \dots,40\}$ and $k\in\{2, 4, 6, \dots,100\}$ we generate a synthetic matrix $\Mb$ as above, and for each of $3$ different values of the regularization parameter $\lambda \in \{0.3, 0.4, 0.5\}$ we perform $100$ trials of Algorithm~\ref{alg:main} recording in each whether the recovery approach succeeded\footnote{We solve the optimization associated with Step 2 of our approach as a LASSO problem, with $10$ different choices of regularization parameter $\mu\in(0,1)$. We deem any trial a success if for at least one value of $\mu$, there exists a threshold $\tau>0$ such that $\min_{i\in\cI_{\Cb}} |\widehat{{\rm c}}_i(\mu)| > \tau > \max_{j\notin \cI_{\Cb}} |\widehat{{\rm c}}_j(\mu)|$ for the estimate $\widehat{\cbb}({\mu})$ produced in Step 2. An analogous threshold-based methodology was employed to assess the outlier detection performance of the Outlier Pursuit approach in \cite{Xu:12}.} in identifying the locations of the true outliers  for that value of $\lambda$, and associate to each $(r,k)$ pair the (empirical) average success rate.  Then, at each $(r,k)$ point examined we identify the point-wise maximum of the average success rates for the $3$ different values of $\lambda$; in this way, we assess whether recovery for that $(r,k)$ is achievable by our method for the specified sampling regime for \emph{some} choice of regularization parameters.  The results in Figure~\ref{fig:sim1} depict the outcome of this experiment for the $9$ different sampling regimes examined. For easy comparison, we provide the average sampling rate as fraction of observations obtained (relative to the full matrix dimension) in the caption in each figure.

The results of this experiment provide an interesting, and somewhat intuitive, illustration of the efficacy of our approach. Namely, we see that increasing the parameter $m$ of the matrix $\bPhi$ in Step 1 of our algorithm while keeping the other sampling parameters fixed (i.e., moving from top to bottom in any one column) facilitates accurate recovery for increasing ranks $r$ of the matrix $\Lb$.  Similarly, increasing the parameter $p$ of the matrix $\Ab$ in Step 2 of our algorithm while keeping the other sampling parameters fixed (i.e., moving from left to right in any one row) facilitates accurate recovery for an increasing number $k$ of outlier columns.  Overall, our approach can successfully recover the locations of the outliers for non-trivial regimes of $r$ and $k$ using very few measurements -- see, for instance, panel $(i)$, where $\sim30$ outlier columns can be accurately identified in the presence of a rank $\sim30$ background using an effective sampling rate of only $\sim6.3\%$.

\begin{figure}[t]
\centering
\begin{minipage}[b]{1\linewidth}
	\footnotesize
	\centering
	\subfloat[10\%]{
		\includegraphics[width=0.28\linewidth]{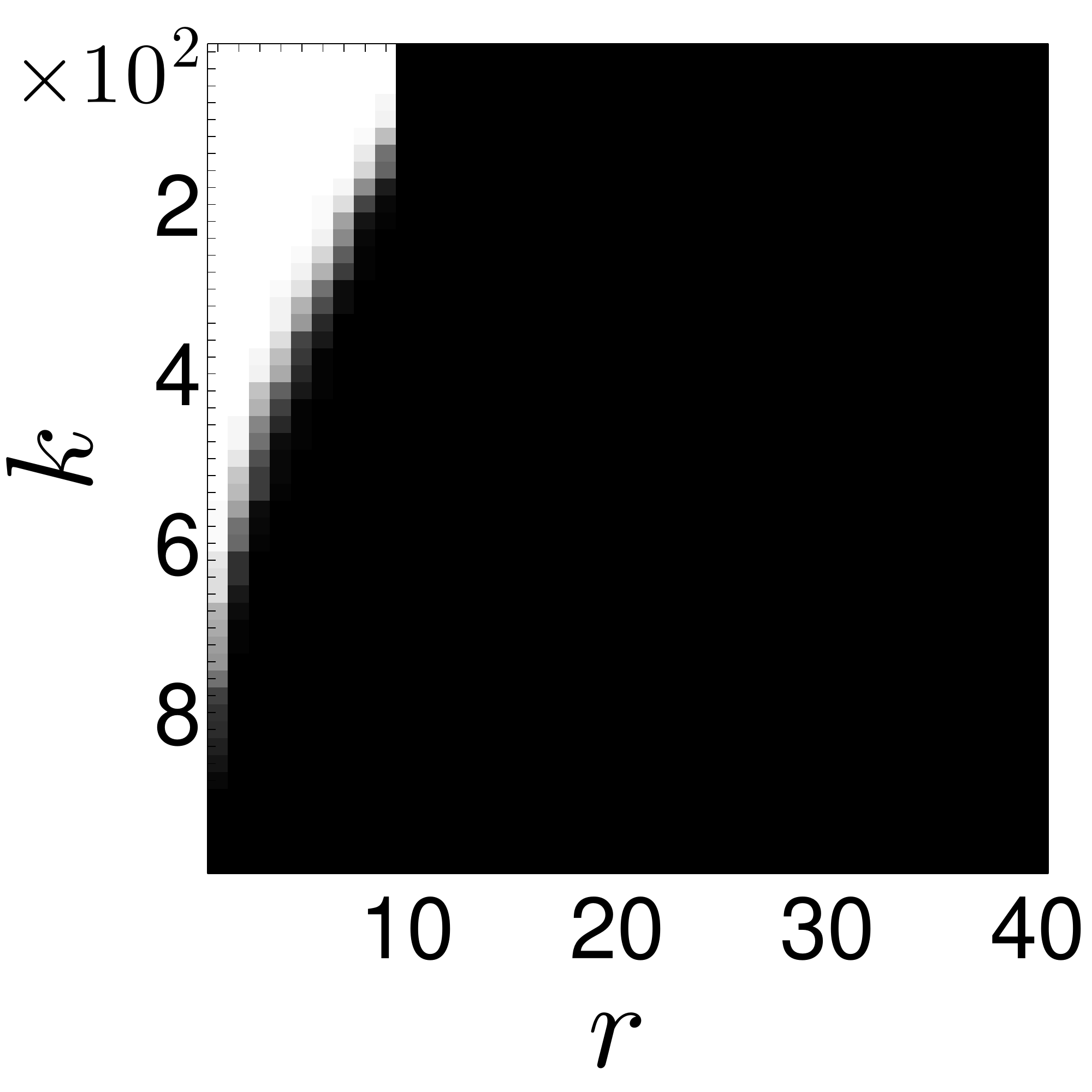}
	}
	\subfloat[20\%]{
		\includegraphics[width=0.28\linewidth]{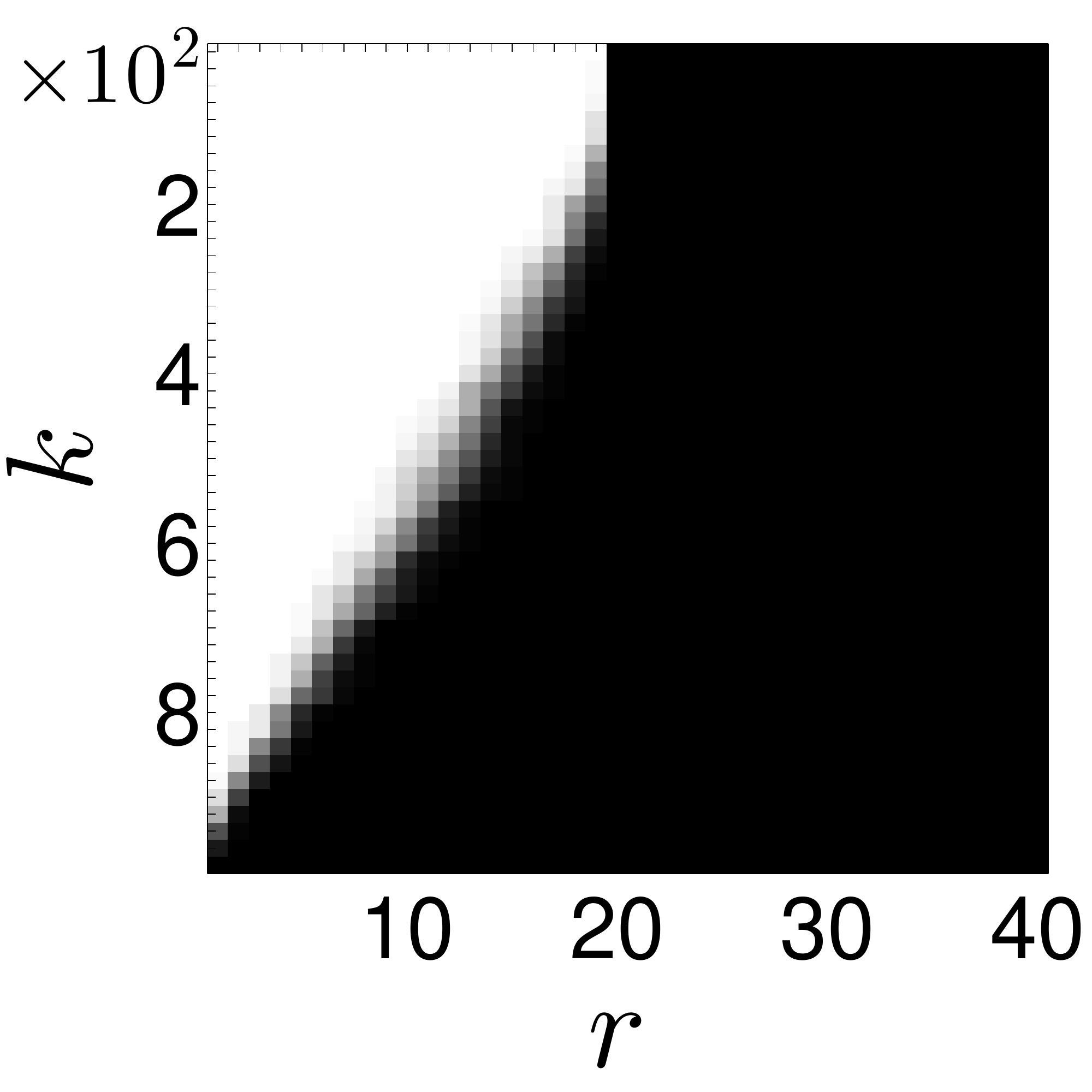}
	}
	\subfloat[30\%]{
		\includegraphics[width=0.28\linewidth]{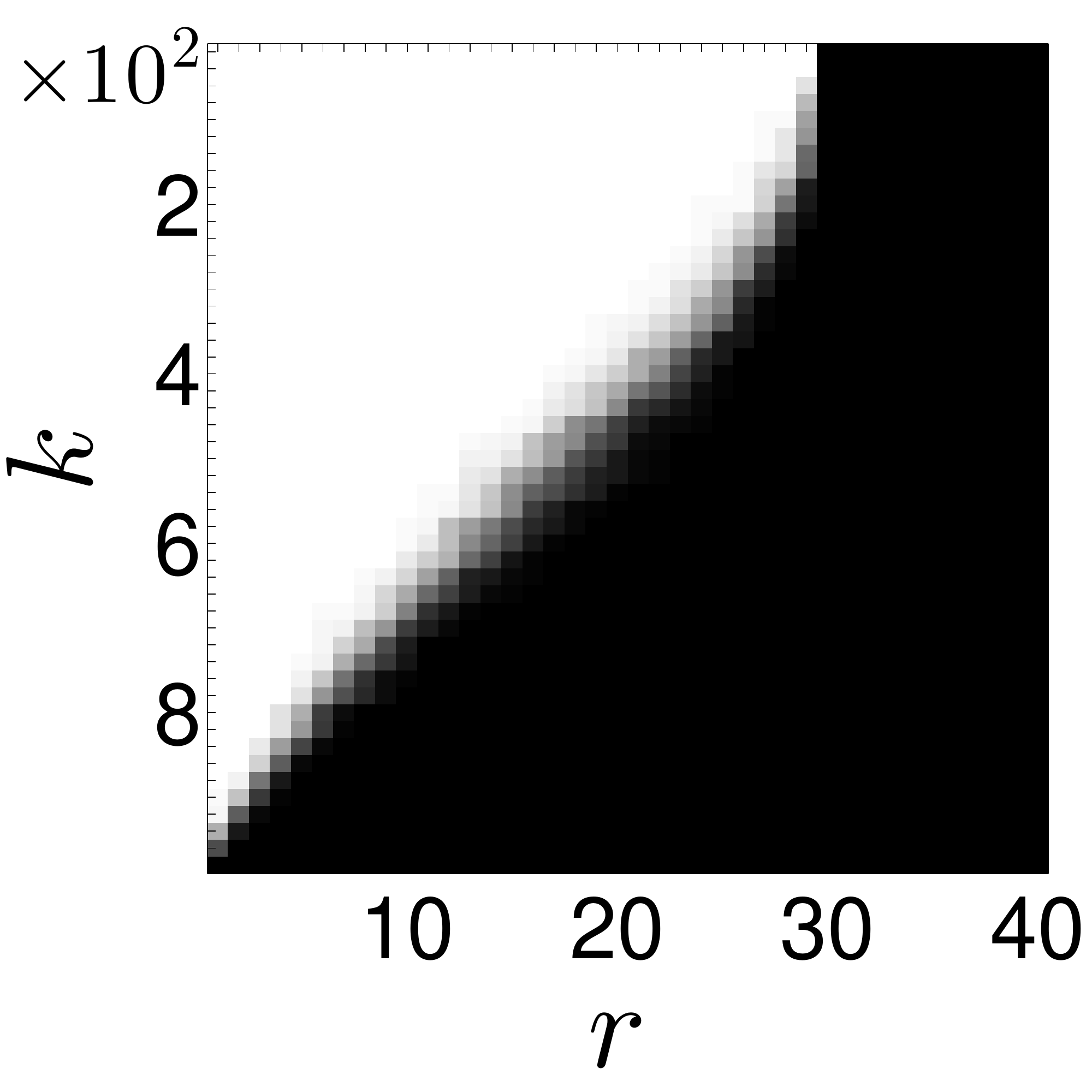}
	}
	%\vspace{-0.08in}
	\caption{Outlier recovery phase transitions plots for SACOS (white regions correspond to successful recovery). The row sampling parameters are $m=0.1n_1,~0.2n_1$, and $0.3n_1$ respectively, from left to right.  Increasing $m$ in SACOS enables accurate estimation for larger rank and increasing numbers of outlier columns. The sampling rate is provided below each plot.}
	\label{fig:sim_simple}
\end{minipage}
	
\begin{minipage}[b]{1\linewidth}
	\footnotesize
	\centering
	\subfloat[5\% (RMC)]{
		\includegraphics[width=0.28\linewidth]{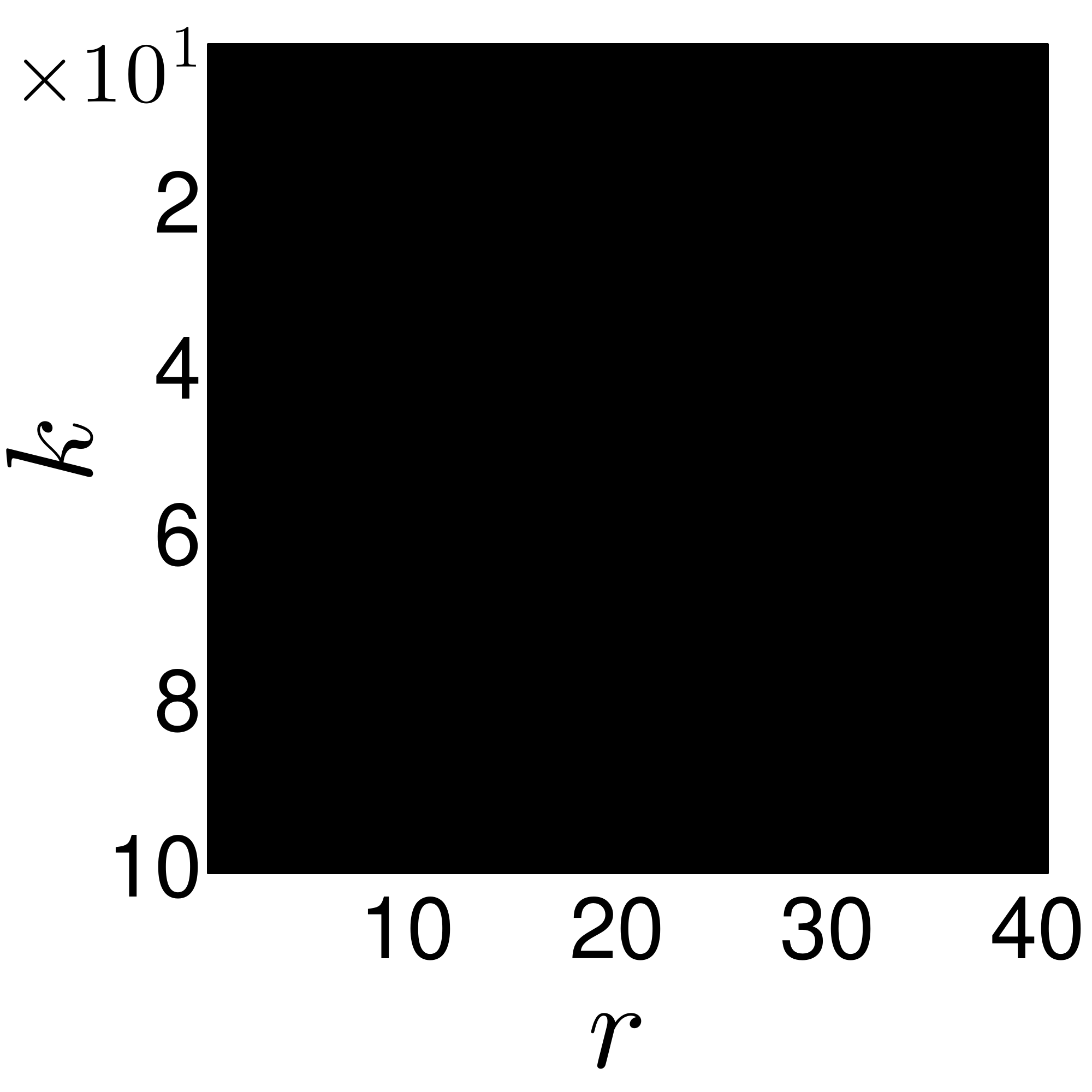}
	}
	\subfloat[10\% (RMC)]{
		\includegraphics[width=0.28\linewidth]{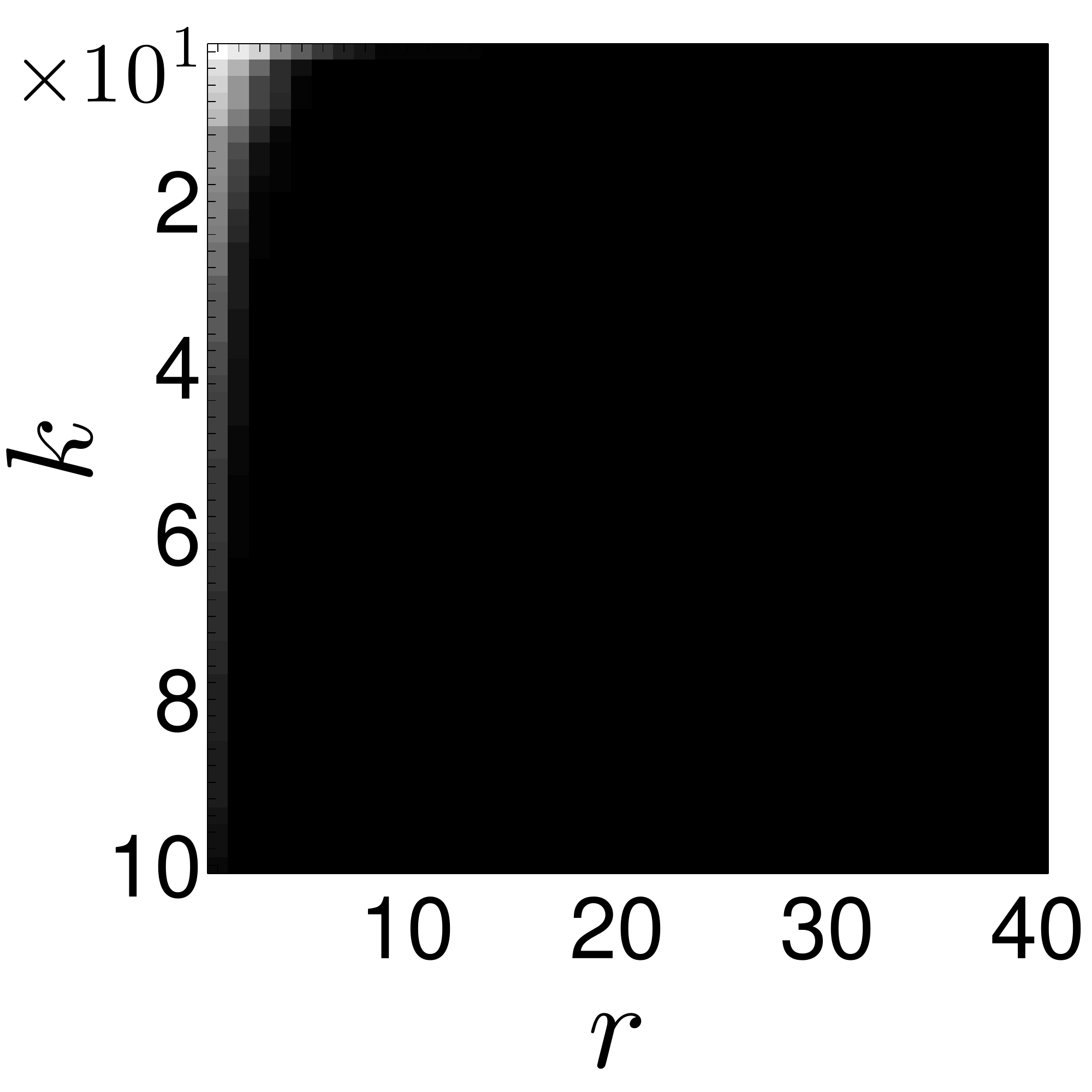}
	}
	\subfloat[20\% (RMC)]{
		\includegraphics[width=0.28\linewidth]{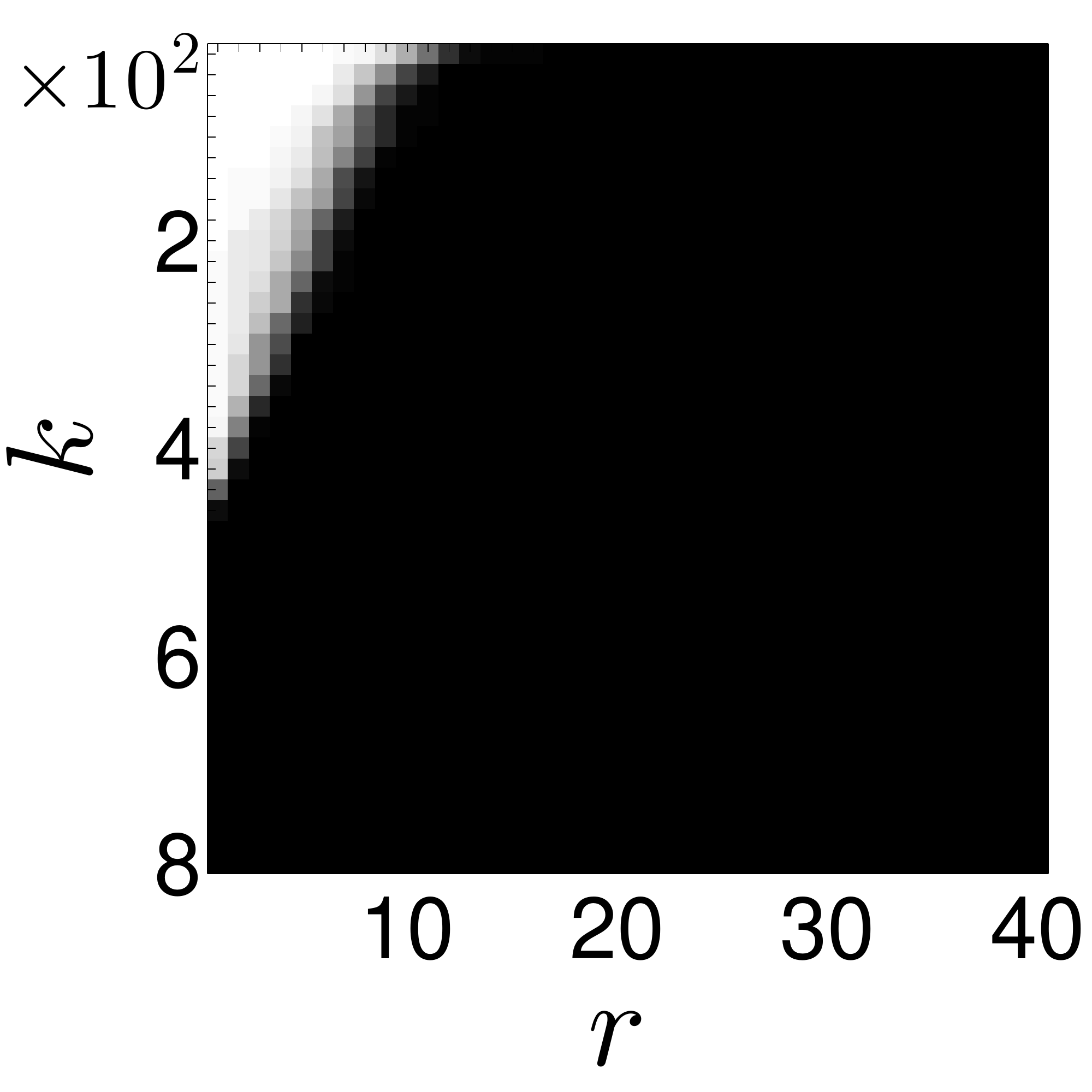}
	}
	%\vspace{-0.1in}
	\caption{Outlier recovery phase transitions plots for RMC.  The average sampling rates are 5\%, 10\% and 20\%, from left to right.  Note that the vertical ($k$) scale in panels (a) and (b) matches that of Figure~\ref{fig:sim1}, while the scale on panel (c) matches that of Figure~\ref{fig:sim_simple}.  Further, comparing panels (a) and (b) here  with Figure~\ref{fig:sim1} shows that ACOS outperforms RMC at low sampling rates, while comparing panels (b) and (c) here with panels (a) and (b) of Figure~\ref{fig:sim_simple} shows that SACOS yields correct outlier identification for a larger portion of the parameter space than RMC for the same average sampling rates.}
\label{fig:rmc}
\end{minipage}
\vspace{-0.1in}
\end{figure}

We adopt a similar methodology to evaluate the Simplified ACOS approach, except that we set $k\in \{20,40,60,\ldots,980\}$ (and the parameter $p$ is no longer applicable, since there is no additional compression in Step 2 for this method). The results are shown in Figure~\ref{fig:sim_simple}. As noted above the SACOS approach has a higher average sampling rate than ACOS for the same $m$, but the results show this facilitates recovery of much larger numbers $k$ of outlier columns (notice the difference in the vertical scales in Figures~\ref{fig:sim1} and \ref{fig:sim_simple}). Overall, we may view ACOS and SACOS as complementary; when the number $k$ of outlier columns is relatively small and low sampling ratio $\frac{\# {\rm obs}}{n_1 n_2}$ is a primary focus, ACOS may be preferred, while if the number $k$ of outlier columns is relatively large, SACOS is more favorable (at the cost of increased sample complexity). 

We also compute phase transition curves for RMC using a similar methodology to that described above.  The results are provided in Figure~\ref{fig:rmc} .  We observe\footnote{Our evaluation of RMC here agrees qualitatively with results in \cite{Chen:11}, where sampling rates around $10\%$ yielded successful recovery for small $r$.} that RMC approach is viable for identifying the outliers from subsampled data provided the sampling rate exceeds about $10\%$, but even then only for small values of the rank $r$.  As alluded in the discussion in previous sections, the relative difference in performance is likely due in large part to the difference in the observation models between the two approaches -- the RMC approach is inherently operating in the presence of ``missing data'' (a difficult scenario!) while our approach permits us to observe linear combinations of any row or column of the entire matrix (i.e., we are allowed to ``see'' each entry of the matrix, albeit not necessarily individually, throughout our approach).

\begin{figure*}[t]
\footnotesize
\begin{center}
\begin{tabular}{ccccccccccc}
\hspace{-0.14in} 
\subfloat{
	\includegraphics[width=0.08\linewidth]{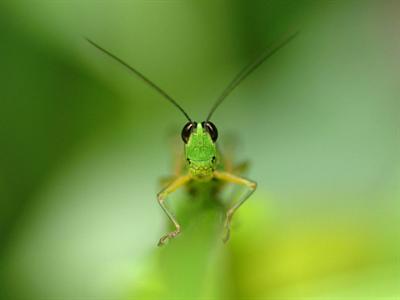}
}\hspace{-0.112in}   & \hspace{-0.112in}   
\subfloat{
	\includegraphics[width=0.08\linewidth]{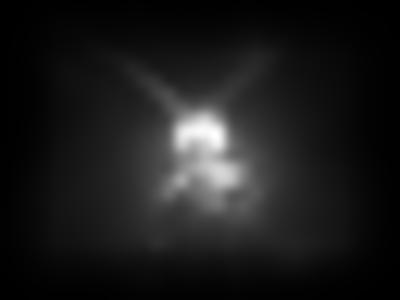}
}\hspace{-0.112in}   & \hspace{-0.112in}   
\subfloat{
	\includegraphics[width=0.08\linewidth]{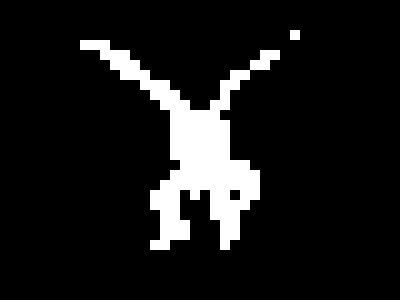}
}\hspace{-0.112in}   & \hspace{-0.112in}   
\subfloat{
	\includegraphics[width=0.08\linewidth]{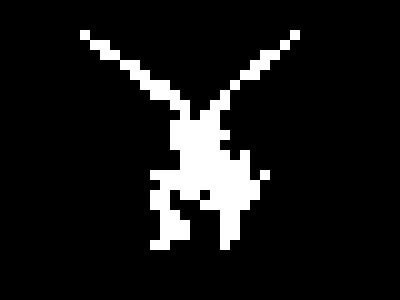}
}\hspace{-0.112in}   & \hspace{-0.112in}   
\subfloat{
	\includegraphics[width=0.08\linewidth]{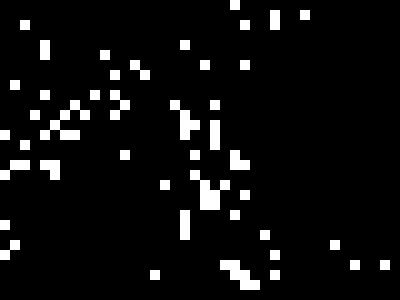}
}\hspace{-0.112in}   & \hspace{-0.112in}   
\subfloat{
	\includegraphics[width=0.08\linewidth]{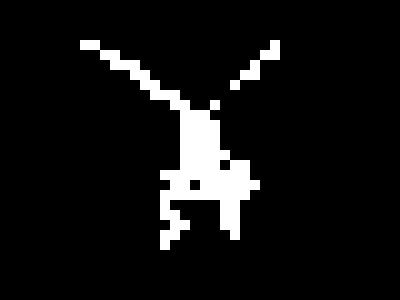}
}\hspace{-0.112in}   & \hspace{-0.112in}   
\subfloat{
	\includegraphics[width=0.08\linewidth]{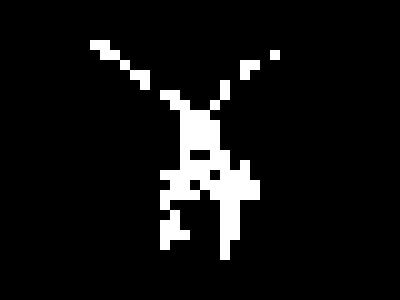}
}\hspace{-0.112in}   & \hspace{-0.112in}   
\subfloat{
	\includegraphics[width=0.08\linewidth]{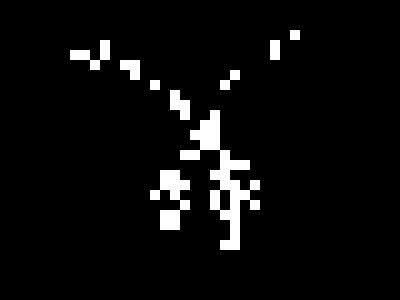}
}\hspace{-0.112in}   & \hspace{-0.112in}   
\subfloat{
	\includegraphics[width=0.08\linewidth]{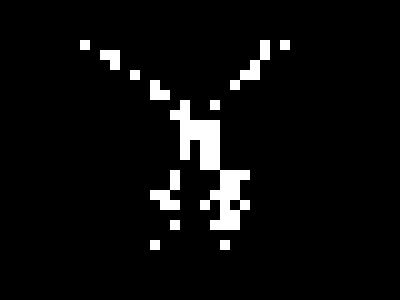}
}\hspace{-0.112in}   & \hspace{-0.112in}   
\subfloat{
	\includegraphics[width=0.08\linewidth]{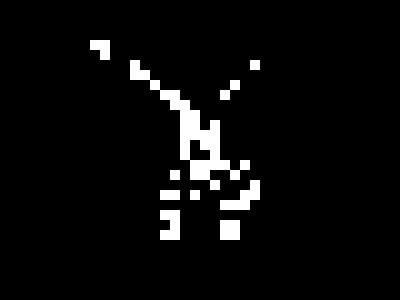}
}\hspace{-0.112in}   & \hspace{-0.112in}   
\subfloat{
	\includegraphics[width=0.08\linewidth]{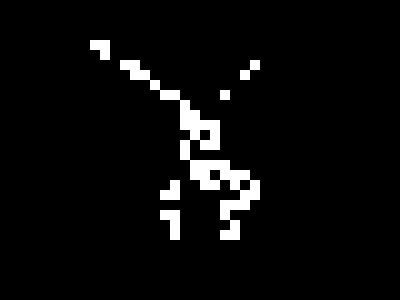}
}\hspace{-0.112in}  \vspace{-0.09in}\\
\hspace{-0.14in} 
\subfloat{
	\includegraphics[width=0.08\linewidth]{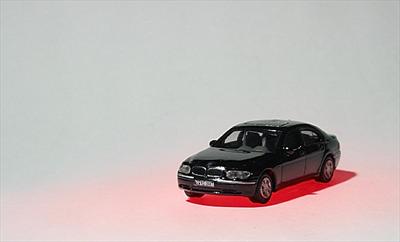}
}\hspace{-0.112in}   & \hspace{-0.112in}   
\subfloat{
	\includegraphics[width=0.08\linewidth]{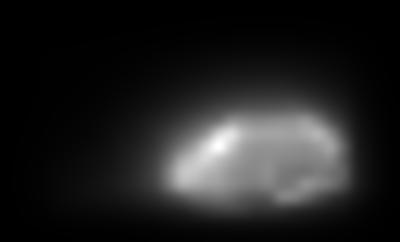}
}\hspace{-0.112in}   & \hspace{-0.112in}   
\subfloat{
	\includegraphics[width=0.08\linewidth]{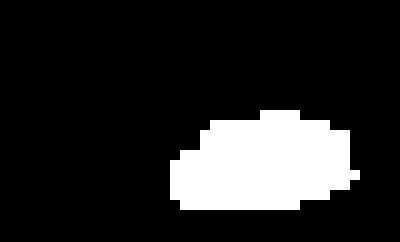}
}\hspace{-0.112in}   & \hspace{-0.112in}   
\subfloat{
	\includegraphics[width=0.08\linewidth]{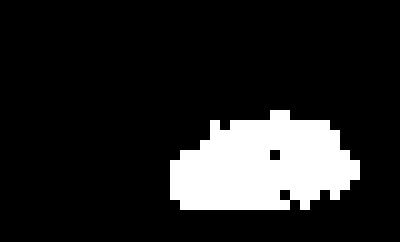}
}\hspace{-0.112in}   & \hspace{-0.112in}   
\subfloat{
	\includegraphics[width=0.08\linewidth]{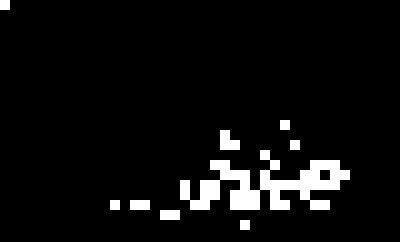}
}\hspace{-0.112in}   & \hspace{-0.112in}   
\subfloat{
	\includegraphics[width=0.08\linewidth]{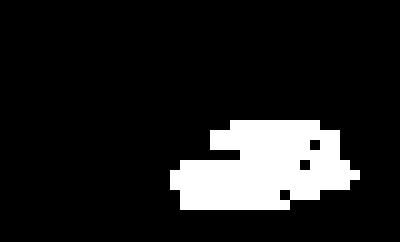}
}\hspace{-0.112in}   & \hspace{-0.112in}   
\subfloat{
	\includegraphics[width=0.08\linewidth]{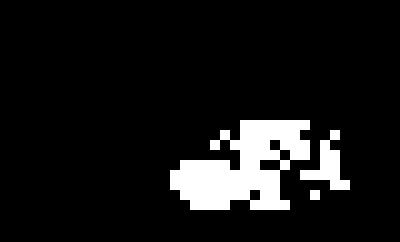}
}\hspace{-0.112in}   & \hspace{-0.112in}   
\subfloat{
	\includegraphics[width=0.08\linewidth]{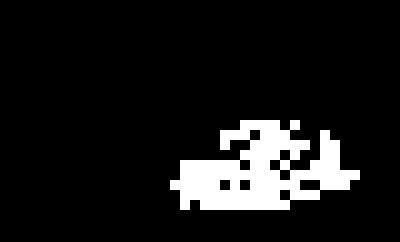}
}\hspace{-0.112in}   & \hspace{-0.112in}   
\subfloat{
	\includegraphics[width=0.08\linewidth]{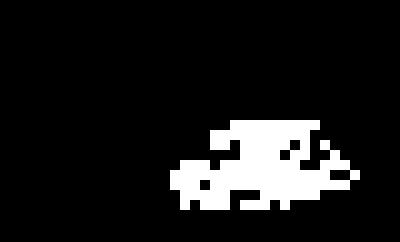}
}\hspace{-0.112in}   & \hspace{-0.112in}   
\subfloat{
	\includegraphics[width=0.08\linewidth]{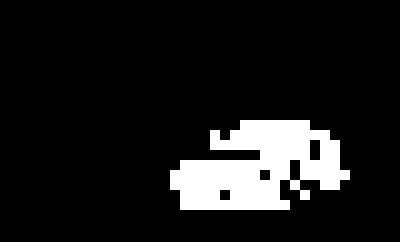}
}\hspace{-0.112in}  & \hspace{-0.112in}   
\subfloat{
	\includegraphics[width=0.08\linewidth]{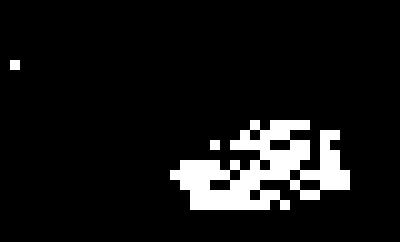}
}\hspace{-0.112in} \vspace{-0.09in}\\
\hspace{-0.14in} 
\subfloat{
	\includegraphics[width=0.08\linewidth]{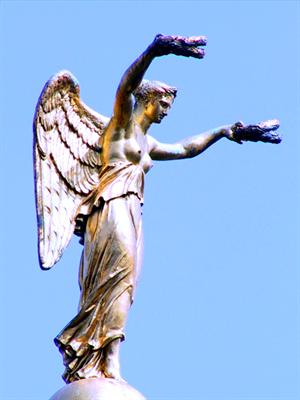}
}\hspace{-0.112in}   & \hspace{-0.112in}   
\subfloat{
	\includegraphics[width=0.08\linewidth]{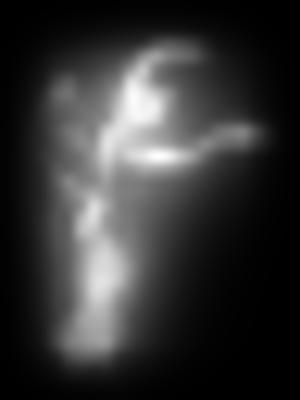}
}\hspace{-0.112in}   & \hspace{-0.112in} 
\subfloat{
	\includegraphics[width=0.08\linewidth]{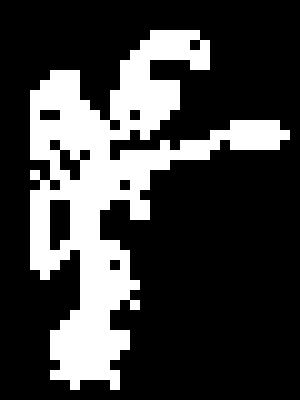}
}\hspace{-0.112in}   & \hspace{-0.112in}   
\subfloat{
	\includegraphics[width=0.08\linewidth]{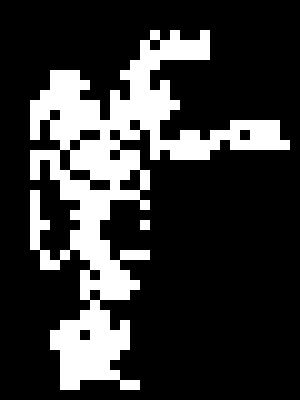}
}\hspace{-0.112in}   & \hspace{-0.112in}   
\subfloat{
	\includegraphics[width=0.08\linewidth]{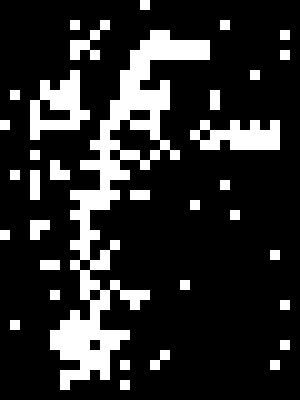}
}\hspace{-0.112in}   & \hspace{-0.112in}   
\subfloat{
	\includegraphics[width=0.08\linewidth]{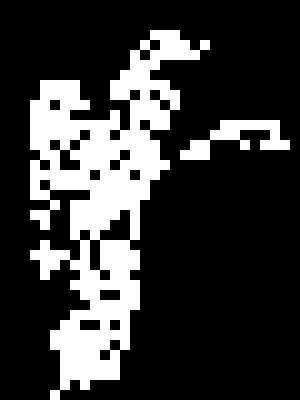}
}\hspace{-0.112in}   & \hspace{-0.112in}   
\subfloat{
	\includegraphics[width=0.08\linewidth]{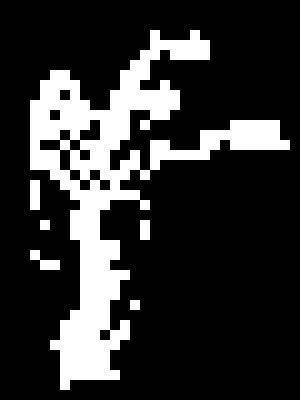}
}\hspace{-0.112in}   & \hspace{-0.112in}   
\subfloat{
	\includegraphics[width=0.08\linewidth]{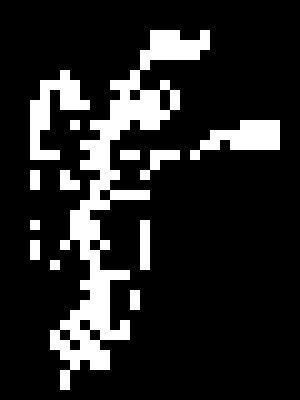}
}\hspace{-0.112in}   & \hspace{-0.112in} 
\subfloat{
	\includegraphics[width=0.08\linewidth]{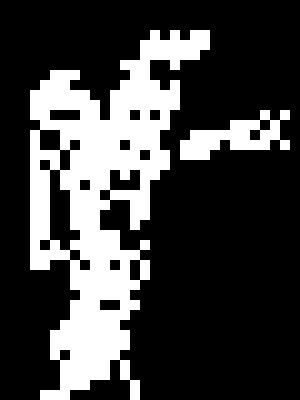}
}\hspace{-0.112in}   & \hspace{-0.112in}   
\subfloat{
	\includegraphics[width=0.08\linewidth]{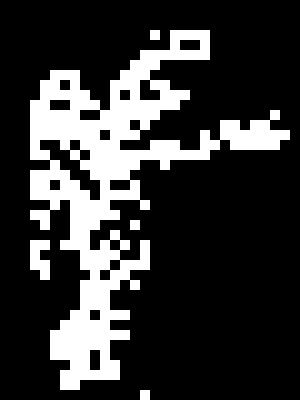}
}\hspace{-0.112in}  & \hspace{-0.112in}   
\subfloat{
	\includegraphics[width=0.08\linewidth]{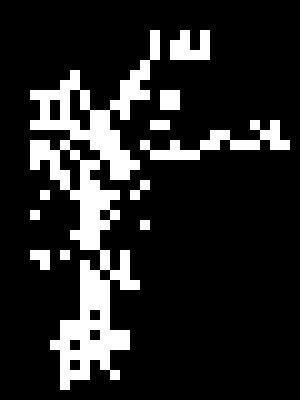}
}\hspace{-0.112in}  \vspace{-0.09in}\\
\hspace{-0.14in} 
\subfloat{
	\includegraphics[width=0.08\linewidth]{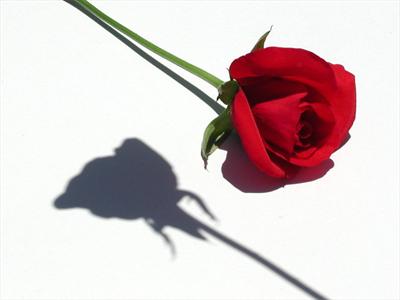}
}\hspace{-0.112in}   & \hspace{-0.112in}   
\subfloat{
	\includegraphics[width=0.08\linewidth]{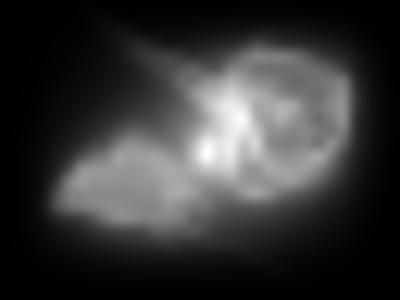}
}\hspace{-0.112in}   & \hspace{-0.112in}   
\subfloat{
	\includegraphics[width=0.08\linewidth]{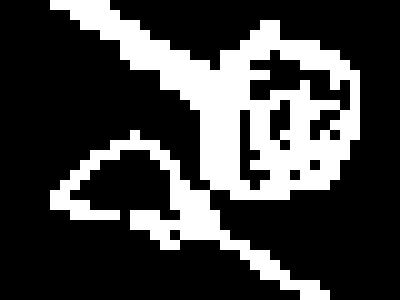}
}\hspace{-0.112in}   & \hspace{-0.112in}   
\subfloat{
	\includegraphics[width=0.08\linewidth]{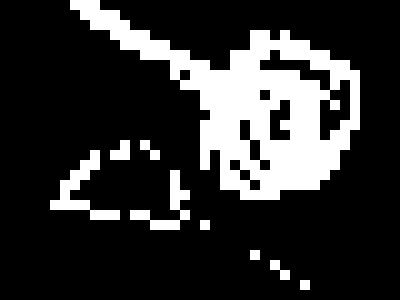}
}\hspace{-0.112in}   & \hspace{-0.112in}   
\subfloat{
	\includegraphics[width=0.08\linewidth]{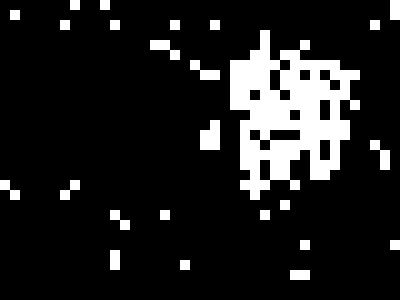}
}\hspace{-0.112in}   & \hspace{-0.112in}   
\subfloat{
	\includegraphics[width=0.08\linewidth]{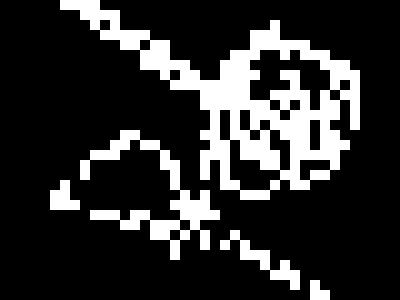}
}\hspace{-0.112in}   & \hspace{-0.112in}   
\subfloat{
	\includegraphics[width=0.08\linewidth]{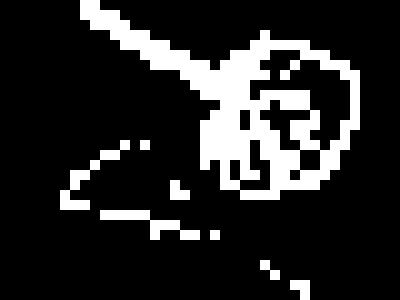}
}\hspace{-0.112in}   & \hspace{-0.112in}   
\subfloat{
	\includegraphics[width=0.08\linewidth]{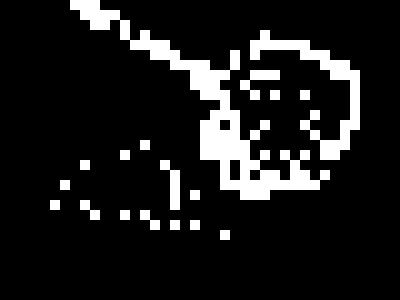}
}\hspace{-0.112in}   & \hspace{-0.112in}   
\subfloat{
	\includegraphics[width=0.08\linewidth]{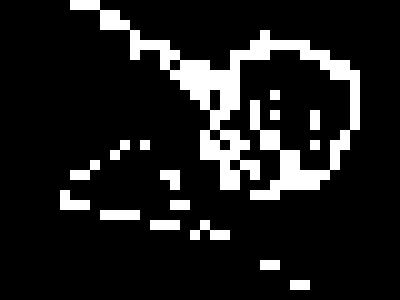}
}\hspace{-0.112in}   & \hspace{-0.112in}   
\subfloat{
	\includegraphics[width=0.08\linewidth]{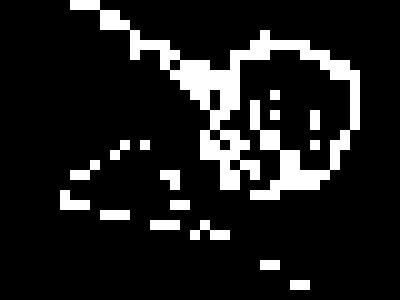}
}\hspace{-0.112in}  & \hspace{-0.112in}   
\subfloat{
	\includegraphics[width=0.08\linewidth]{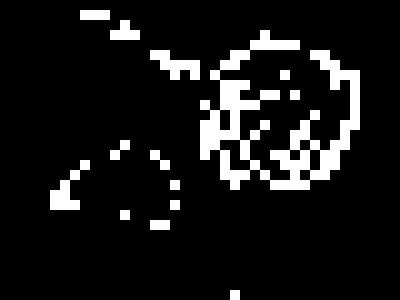}
}\hspace{-0.112in}  \vspace{-0.09in}\\
\hspace{-0.14in} 
\subfloat{
	\includegraphics[width=0.08\linewidth]{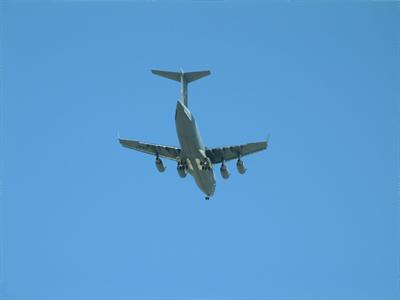}
}\hspace{-0.112in}   & \hspace{-0.112in}   
\subfloat{
	\includegraphics[width=0.08\linewidth]{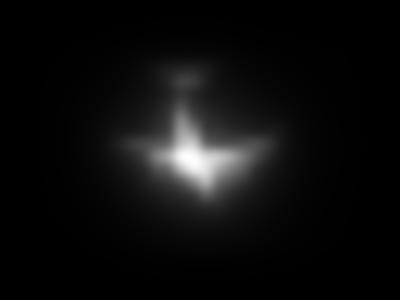}
}\hspace{-0.112in}   & \hspace{-0.112in}   
\subfloat{
	\includegraphics[width=0.08\linewidth]{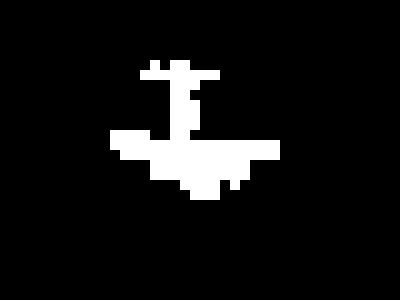}
}\hspace{-0.112in}   & \hspace{-0.112in}   
\subfloat{
	\includegraphics[width=0.08\linewidth]{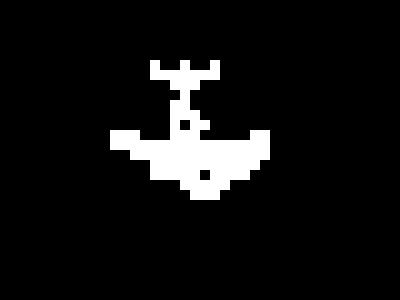}
}\hspace{-0.112in}   & \hspace{-0.112in}   
\subfloat{
	\includegraphics[width=0.08\linewidth]{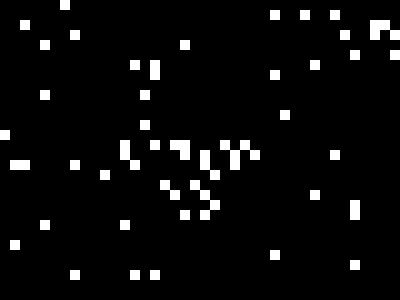}
}\hspace{-0.112in}   & \hspace{-0.112in}   
\subfloat{
	\includegraphics[width=0.08\linewidth]{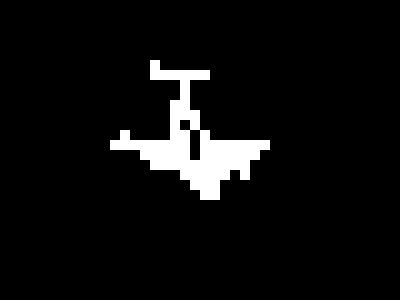}
}\hspace{-0.112in}   & \hspace{-0.112in}   
\subfloat{
	\includegraphics[width=0.08\linewidth]{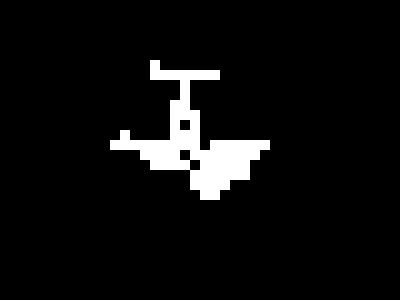}
}\hspace{-0.112in}   & \hspace{-0.112in}   
\subfloat{
	\includegraphics[width=0.08\linewidth]{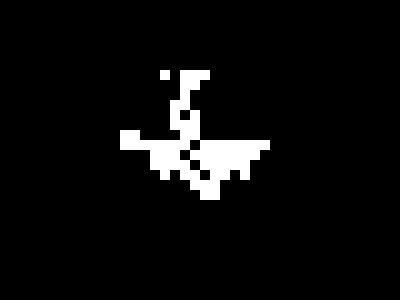}
}\hspace{-0.112in}   & \hspace{-0.112in}   
\subfloat{
	\includegraphics[width=0.08\linewidth]{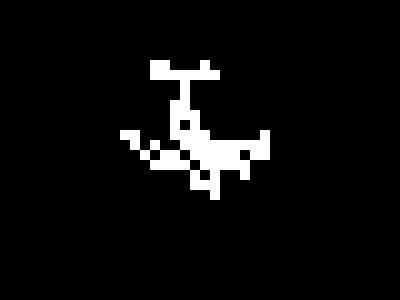}
}\hspace{-0.112in}   & \hspace{-0.112in}   
\subfloat{
	\includegraphics[width=0.08\linewidth]{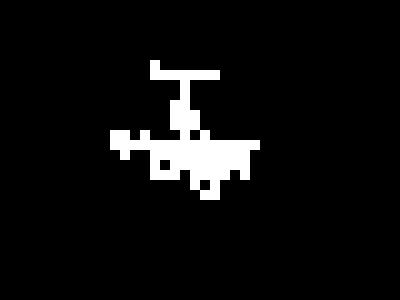}
}\hspace{-0.112in}  & \hspace{-0.112in}   
\subfloat{
	\includegraphics[width=0.08\linewidth]{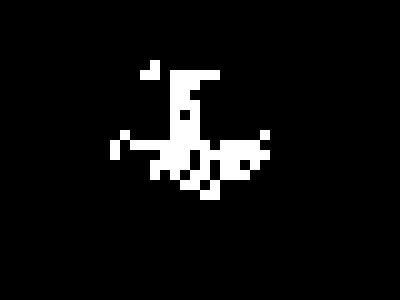}
}\hspace{-0.112in}  \vspace{-0.09in}\\
\hspace{-0.14in} 
\subfloat{
	\includegraphics[width=0.08\linewidth]{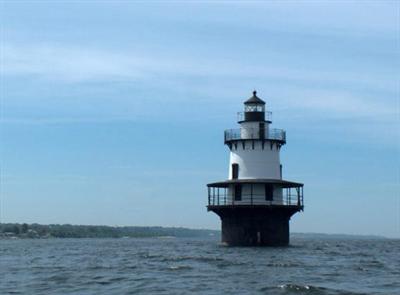}
}\hspace{-0.112in}   & \hspace{-0.112in}   
\subfloat{
	\includegraphics[width=0.08\linewidth]{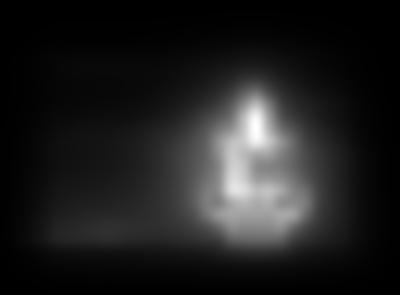}
}\hspace{-0.112in}   & \hspace{-0.112in}   
\subfloat{
	\includegraphics[width=0.08\linewidth]{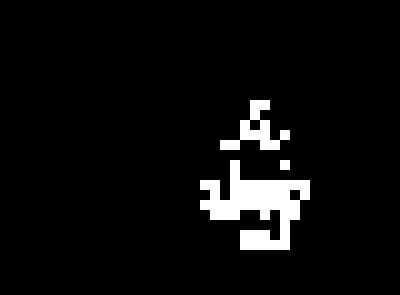}
}\hspace{-0.112in}   & \hspace{-0.112in}   
\subfloat{
	\includegraphics[width=0.08\linewidth]{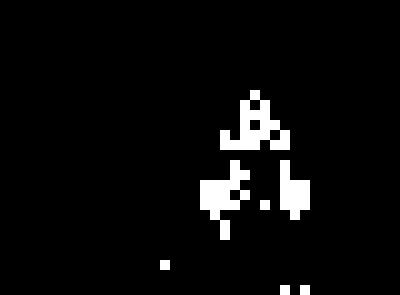}
}\hspace{-0.112in}   & \hspace{-0.112in}   
\subfloat{
	\includegraphics[width=0.08\linewidth]{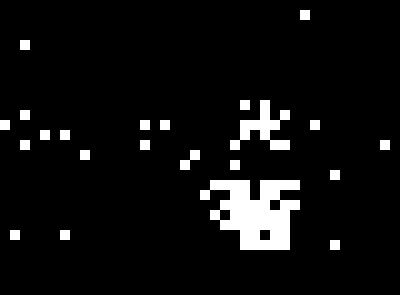}
}\hspace{-0.112in}   & \hspace{-0.112in}   
\subfloat{
	\includegraphics[width=0.08\linewidth]{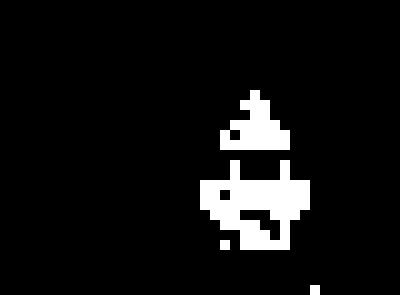}
}\hspace{-0.112in}   & \hspace{-0.112in}   
\subfloat{
	\includegraphics[width=0.08\linewidth]{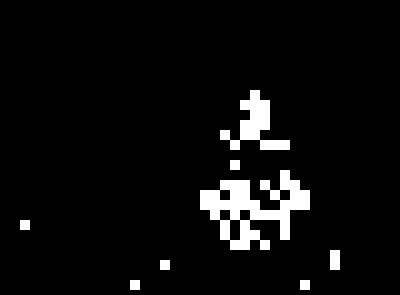}
}\hspace{-0.112in}   & \hspace{-0.112in}  
\subfloat{
	\includegraphics[width=0.08\linewidth]{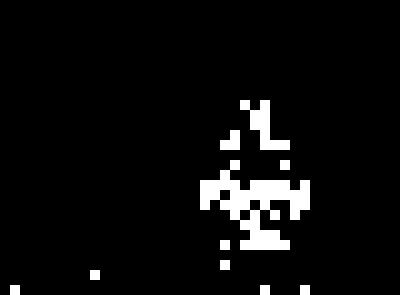}
}\hspace{-0.112in}   & \hspace{-0.112in}  
\subfloat{
	\includegraphics[width=0.08\linewidth]{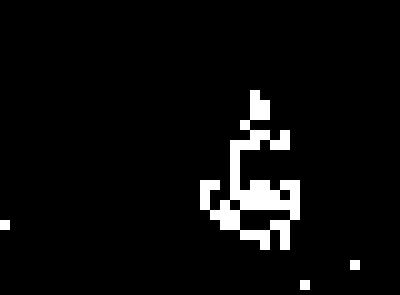}
}\hspace{-0.112in}   & \hspace{-0.112in}   
\subfloat{
	\includegraphics[width=0.08\linewidth]{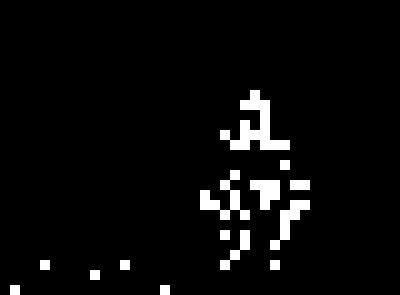}
}\hspace{-0.112in}  & \hspace{-0.112in}   
\subfloat{
	\includegraphics[width=0.08\linewidth]{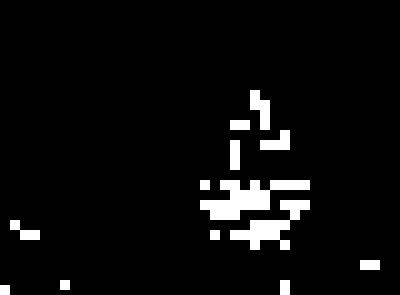}
}\hspace{-0.112in}  \vspace{-0.09in}\\
\hspace{-0.14in} 
\subfloat{
	\includegraphics[width=0.08\linewidth]{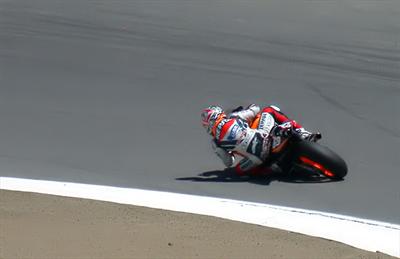}
}\hspace{-0.112in}   & \hspace{-0.112in}   
\subfloat{
	\includegraphics[width=0.08\linewidth]{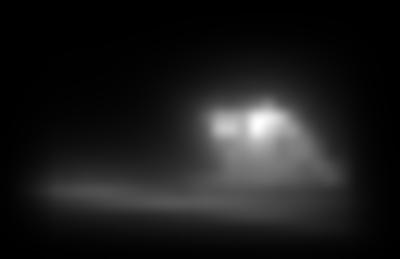}
}\hspace{-0.112in}   & \hspace{-0.112in}   
\subfloat{
	\includegraphics[width=0.08\linewidth]{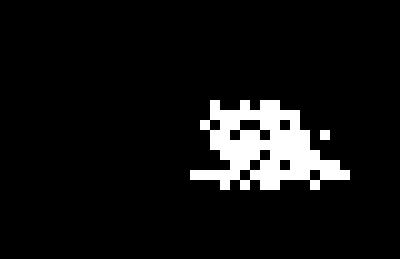}
}\hspace{-0.112in}   & \hspace{-0.112in}   
\subfloat{
	\includegraphics[width=0.08\linewidth]{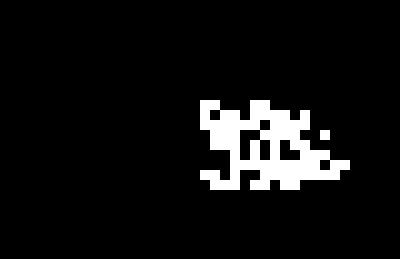}
}\hspace{-0.112in}   & \hspace{-0.112in}   
\subfloat{
	\includegraphics[width=0.08\linewidth]{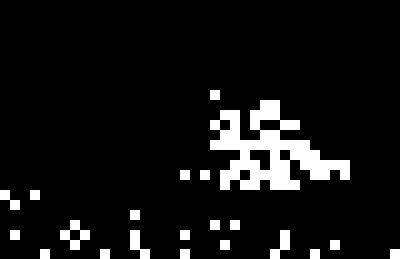}
}\hspace{-0.112in}   & \hspace{-0.112in}   
\subfloat{
	\includegraphics[width=0.08\linewidth]{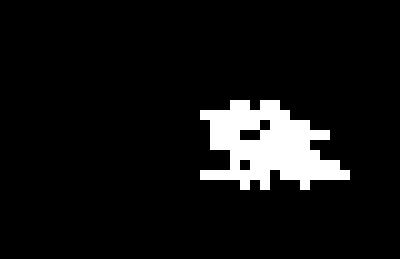}
}\hspace{-0.112in}   & \hspace{-0.112in}   
\subfloat{
	\includegraphics[width=0.08\linewidth]{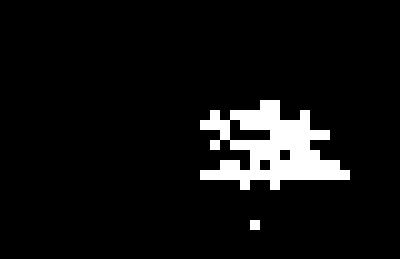}
}\hspace{-0.112in}   & \hspace{-0.112in}  
\subfloat{
	\includegraphics[width=0.08\linewidth]{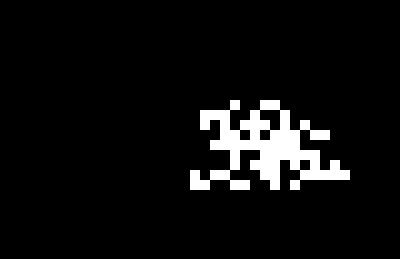}
}\hspace{-0.112in}   & \hspace{-0.112in}  
\subfloat{
	\includegraphics[width=0.08\linewidth]{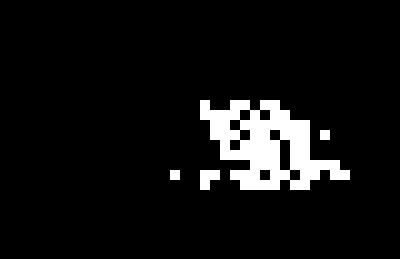}
}\hspace{-0.112in}   & \hspace{-0.112in}   
\subfloat{
	\includegraphics[width=0.08\linewidth]{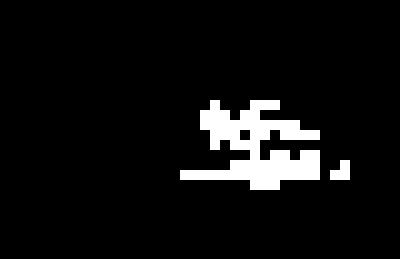}
}\hspace{-0.112in}  & \hspace{-0.112in}   
\subfloat{
	\includegraphics[width=0.08\linewidth]{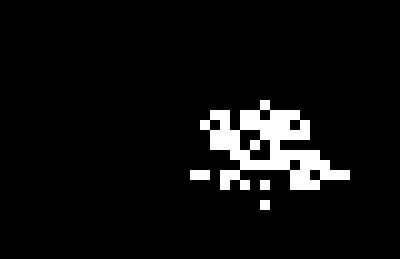}
}\hspace{-0.112in}  \vspace{-0.09in}\\
\hspace{-0.14in} 
\subfloat{
	\includegraphics[width=0.08\linewidth]{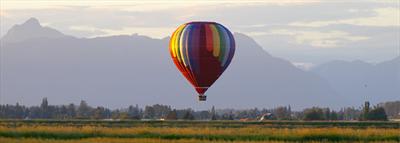}
}\hspace{-0.112in}   & \hspace{-0.112in}   
\subfloat{
	\includegraphics[width=0.08\linewidth]{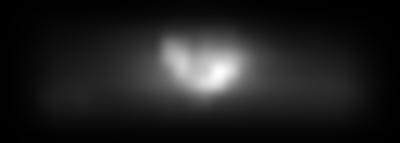}
}\hspace{-0.112in}   & \hspace{-0.112in}  
\subfloat{
	\includegraphics[width=0.08\linewidth]{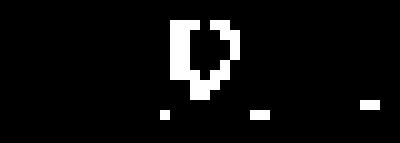}
}\hspace{-0.112in}   & \hspace{-0.112in}   
\subfloat{
	\includegraphics[width=0.08\linewidth]{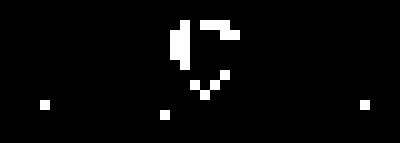}
}\hspace{-0.112in}   & \hspace{-0.112in}   
\subfloat{
	\includegraphics[width=0.08\linewidth]{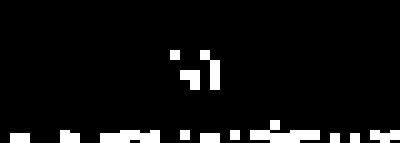}
}\hspace{-0.112in}   & \hspace{-0.112in}   
\subfloat{
	\includegraphics[width=0.08\linewidth]{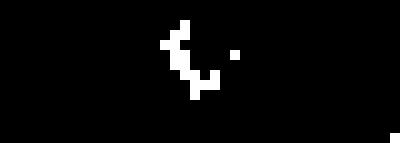}
}\hspace{-0.112in}   & \hspace{-0.112in}   
\subfloat{
	\includegraphics[width=0.08\linewidth]{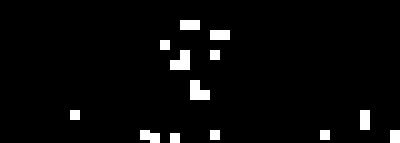}
}\hspace{-0.112in}   & \hspace{-0.112in}   
\subfloat{
	\includegraphics[width=0.08\linewidth]{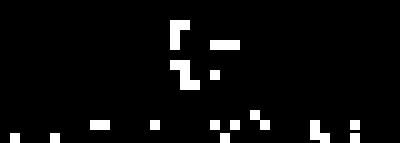}
}\hspace{-0.112in}   & \hspace{-0.112in}   
\subfloat{
	\includegraphics[width=0.08\linewidth]{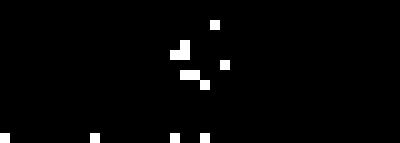}
}\hspace{-0.112in}   & \hspace{-0.112in}   
\subfloat{
	\includegraphics[width=0.08\linewidth]{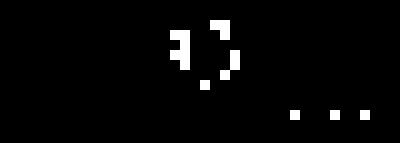}
}\hspace{-0.112in}  & \hspace{-0.112in}   
\subfloat{
	\includegraphics[width=0.08\linewidth]{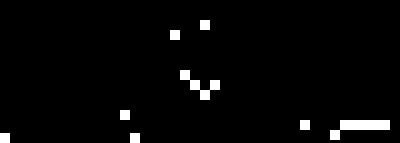}
}\hspace{-0.112in}  \vspace{-0.09in}\\
\hspace{-0.14in} 
\subfloat{
	\includegraphics[width=0.08\linewidth]{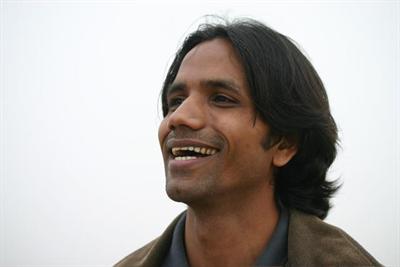}
}\hspace{-0.112in}   & \hspace{-0.112in}   
\subfloat{
	\includegraphics[width=0.08\linewidth]{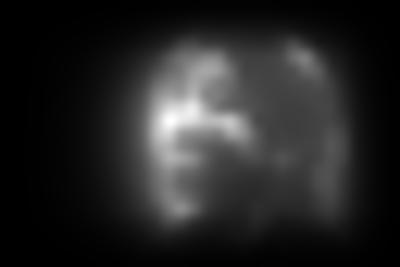}
}\hspace{-0.112in}   & \hspace{-0.112in}   
\subfloat{
	\includegraphics[width=0.08\linewidth]{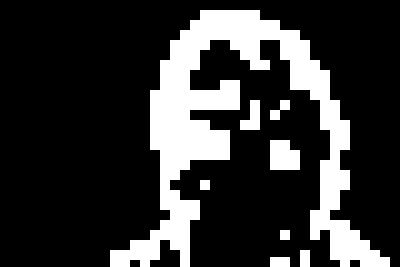}
}\hspace{-0.112in}   & \hspace{-0.112in}   
\subfloat{
	\includegraphics[width=0.08\linewidth]{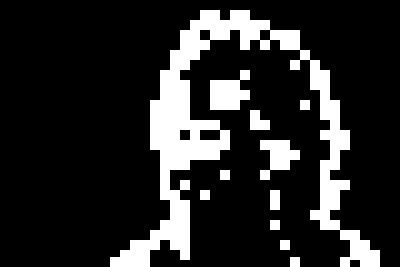}
}\hspace{-0.112in}   & \hspace{-0.112in}   
\subfloat{
	\includegraphics[width=0.08\linewidth]{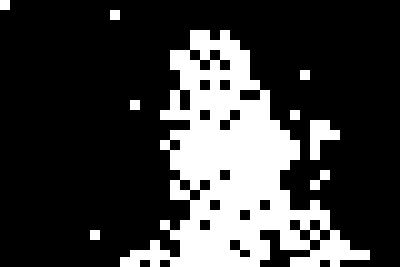}
}\hspace{-0.112in}   & \hspace{-0.112in}   
\subfloat{
	\includegraphics[width=0.08\linewidth]{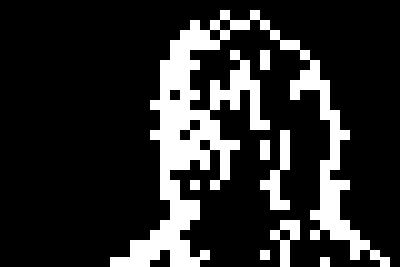}
}\hspace{-0.112in}   & \hspace{-0.112in}   
\subfloat{
	\includegraphics[width=0.08\linewidth]{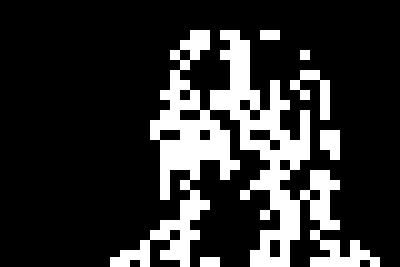}
}\hspace{-0.112in}   & \hspace{-0.112in}   
\subfloat{
	\includegraphics[width=0.08\linewidth]{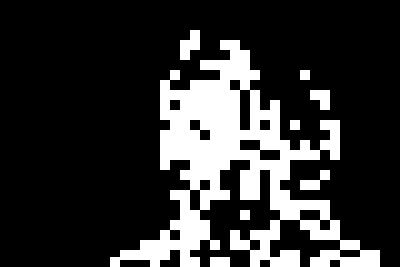}
}\hspace{-0.112in}   & \hspace{-0.112in}   
\subfloat{
	\includegraphics[width=0.08\linewidth]{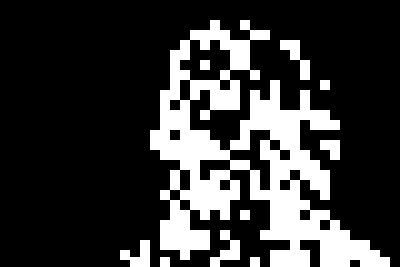}
}\hspace{-0.112in}   & \hspace{-0.112in}   
\subfloat{
	\includegraphics[width=0.08\linewidth]{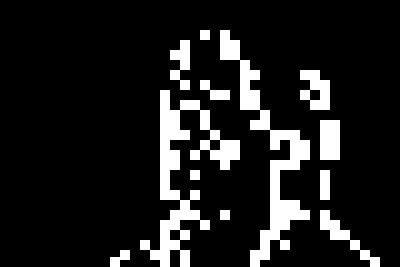}
}\hspace{-0.112in}  & \hspace{-0.112in}   
\subfloat{
	\includegraphics[width=0.08\linewidth]{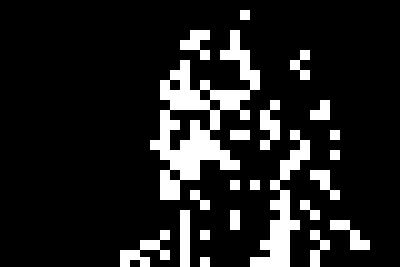}
}\hspace{-0.112in}  \vspace{-0.09in}\\
\hspace{-0.14in} 
\subfloat{
	\includegraphics[width=0.08\linewidth]{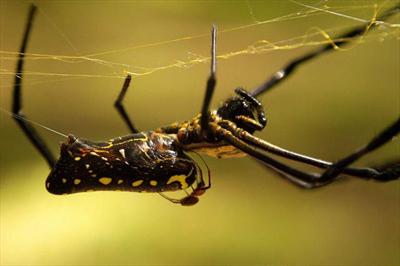}
}\hspace{-0.112in}   & \hspace{-0.112in}   
\subfloat{
	\includegraphics[width=0.08\linewidth]{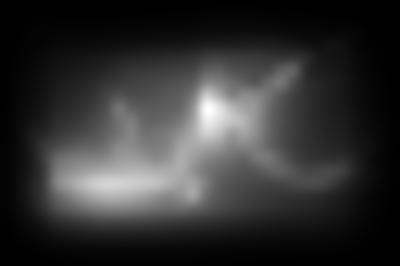}
}\hspace{-0.112in}   & \hspace{-0.112in}
\subfloat{
	\includegraphics[width=0.08\linewidth]{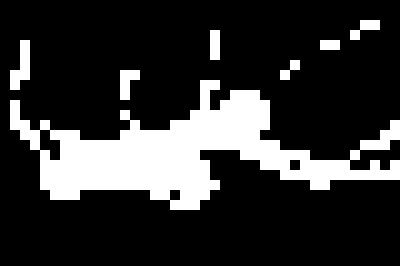}
}\hspace{-0.112in}   & \hspace{-0.112in}   
\subfloat{
	\includegraphics[width=0.08\linewidth]{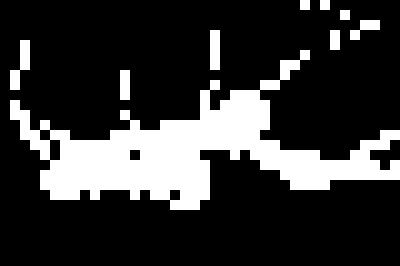}
}\hspace{-0.112in}   & \hspace{-0.112in}   
\subfloat{
	\includegraphics[width=0.08\linewidth]{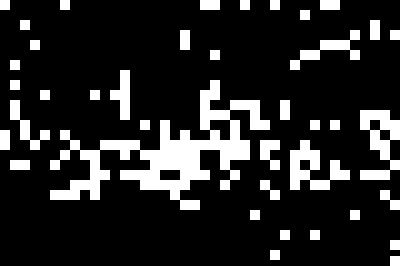}
}\hspace{-0.112in}   & \hspace{-0.112in}   
\subfloat{
	\includegraphics[width=0.08\linewidth]{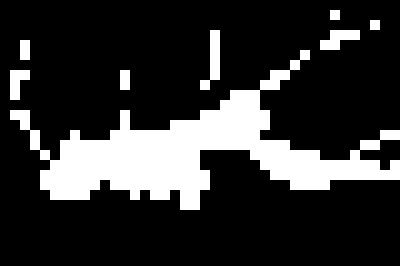}
}\hspace{-0.112in}   & \hspace{-0.112in}   
\subfloat{
	\includegraphics[width=0.08\linewidth]{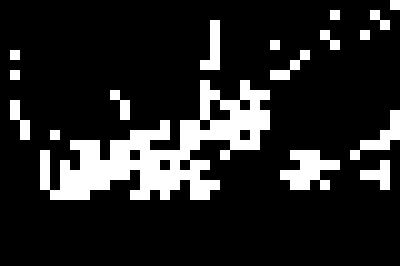}
}\hspace{-0.112in}   & \hspace{-0.112in}  
\subfloat{
	\includegraphics[width=0.08\linewidth]{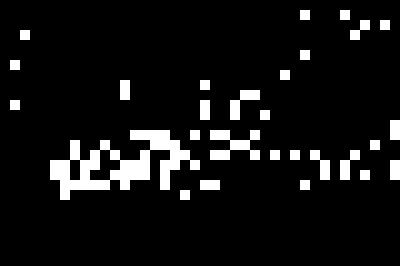}
}\hspace{-0.112in}   & \hspace{-0.112in}  
\subfloat{
	\includegraphics[width=0.08\linewidth]{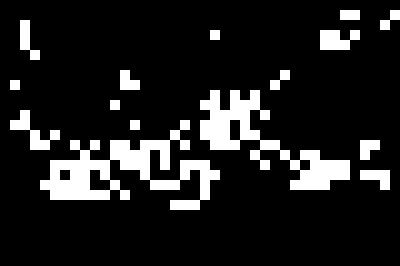}
}\hspace{-0.112in}   & \hspace{-0.112in}   
\subfloat{
	\includegraphics[width=0.08\linewidth]{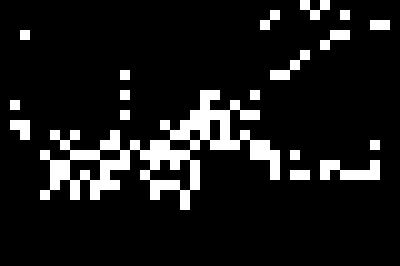}
}\hspace{-0.112in}   & \hspace{-0.112in}   
\subfloat{
	\includegraphics[width=0.08\linewidth]{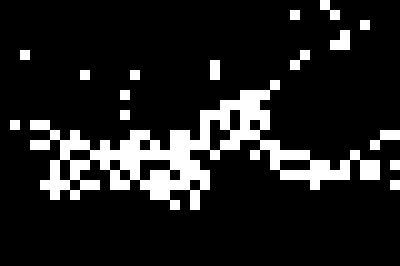}
}\hspace{-0.112in}   \vspace{-0.0in}\\
%\hspace{-0.14in} 
%(a) \hspace{-0.112in} & \hspace{-0.112in}   
%(b) \hspace{-0.112in} & \hspace{-0.112in}   
%(c) \hspace{-0.112in} & \hspace{-0.112in}   
%(d) \hspace{-0.112in} & \hspace{-0.112in}   
%%(e) \hspace{-0.112in} & \hspace{-0.112in}   
%(e) \hspace{-0.112in} & \hspace{-0.112in}   
%(f) \hspace{-0.112in}  \vspace{-0.0in}\\
\hspace{-0.14in} 
Method \hspace{-0.112in} & \hspace{-0.112in}   
GBVS \hspace{-0.112in} & \hspace{-0.112in}   
OP \hspace{-0.112in} & \hspace{-0.112in}   
RMC \hspace{-0.112in} & \hspace{-0.112in}   
RMC \hspace{-0.112in} & \hspace{-0.112in}   
%20\% \hspace{-0.112in} & \hspace{-0.112in}    
SACOS \hspace{-0.112in} & \hspace{-0.112in}    
SACOS \hspace{-0.112in} & \hspace{-0.112in}    
SACOS \hspace{-0.112in} & \hspace{-0.112in}   
ACOS \hspace{-0.112in} & \hspace{-0.112in}   
ACOS \hspace{-0.112in} & \hspace{-0.112in}   
ACOS \hspace{-0.112in} \vspace{-0.03in}\\
\hspace{-0.14in} 
Sampling \hspace{-0.112in} & \hspace{-0.112in}   
100\% \hspace{-0.112in} & \hspace{-0.112in}   
100\% \hspace{-0.112in} & \hspace{-0.112in}   
20\% \hspace{-0.112in} & \hspace{-0.112in}   
5\% \hspace{-0.112in} & \hspace{-0.112in}   
20\% \hspace{-0.112in} & \hspace{-0.112in}    
5\% \hspace{-0.112in} & \hspace{-0.112in}    
3\% \hspace{-0.112in} & \hspace{-0.112in}   
4.5\% \hspace{-0.112in} & \hspace{-0.112in}   
2.5\% \hspace{-0.112in} & \hspace{-0.112in}   
1.5\% \hspace{-0.112in} \vspace{-0.1in}
\end{tabular}
\end{center}
\caption{Detection results for the MSRA Salient Object Database for various methods.  Our ACOS approach produces results comparable to the ``full sampling'' OP method using an average sampling rate below $5\%$.  The performance of the RMC approach appears to degrade at low sampling rates.}
\label{fig:real}
\vspace{-1.0em}
\end{figure*}

\begin{table*}[!t]
\begin{center}
\caption{Timing analysis for detection experiments on 1000 images from MRSA Database.\hspace{23em}Each entry is the mean execution time in seconds with the standard deviation in parenthesis.}
\vspace{-0.04in}
\begin{tabular}{lcccccccccc}
\toprule
Method & GBVS & OP & RMC & RMC & SACOS  & SACOS & SACOS  & ACOS & ACOS  & ACOS  \\ 
Sampling & 100\% & 100\% & 20\% & 5\%  & 20\% & 5\%  & 3\% & 4.5\% & 2.5\% & 1.5\%\\
\midrule
Step 1 & 0.9926 & 2.9441& 2.6324 & 2.7254 & 0.0538 & 0.0107 & 0.0074 & 0.0533 & 0.0214 & 0.0105 \\
 & (0.2742) & (0.3854) & (0.3237) & (0.3660) & (0.0121) & (0.0034) & (0.0017) & (0.0118) & (0.0056) & (0.0025) \\
Step 2 & -- & -- & -- & -- & 0.0015 & 0.0011 & 0.0009 & 0.2010 & 0.2014 & 0.2065 \\
 & -- & -- & -- & -- & (0.0003) & (0.0003) & (0.0003) & (0.0674) & (0.0692) & (0.0689) \\
\bottomrule
\end{tabular}\label{table:timing}
\end{center}\vspace{-1.5em}
\end{table*}

\subsection{Real Data}

We also evaluate the performance of our proposed methods on real data in the context of a stylized image processing task that arises in many computer vision and automated surveillance -- that of identifying the ``saliency map'' of an image.  For this, we use images from the MSRA Salient Object Database \cite{Liu:07}, available online at \url{http://research.microsoft.com/en-us/um/people/jiansun/SalientObject/salient\_object.htm}. 

As discussed above, our approach here is based on representing each test image as a collection of (vectorized) non-overlapping image patches.  We transform each (color) test image to gray scale, decompose it into non-overlapping $10 \times 10$-pixel patches, vectorize each patch into a $100 \times 1$ column vector, and assemble the column vectors into a matrix.  Most of the images in the database are of the size $300 \times 400$ (or $400 \times 300$), which here yields matrices of size $100\times 1200$, corresponding to $1200$ patches.  Notice that we only used gray scale values of image as the input feature rather than any high-level images feature -- this facilitates the use of our approach, which is based on collecting linear measurements of the data (e.g., using a spatial light modulator, or an architecture like the \emph{single pixel camera} \cite{Duarte:08:SPC}).

Here, our experimental approach is (somewhat necessarily) a bit more heuristic than for the synthetic data experiments above, due in large part to the fact that the data here may not adhere exactly to the low-rank plus outlier model.  To compensate for this, we augment Step 1 of Algorithm~\ref{alg:main} and Algorithm~\ref{alg:simple} with an additional ``rank reduction'' step, where we further reduce the dimension of the subspace spanned by the columns of the learned $\widehat{\Lb}_{(1)}$ by truncating its SVDs to retain the smallest number of leading singular values whose sum is at least $0.95 \times \|\widehat{\Lb}_{(1)}\|_*$. Further, we generalize Step 2 of each procedure by declaring an image patch to be salient when its (residual) column norm is sufficiently large, rather than strictly nonzero.  We used visual heuristics to determine the ``best'' outputs for Step 2 of each method, selecting LASSO parameters (for ACOS) or thresholds (for SACOS) in order to qualitatively trade off false positives with misses.

We implement our ACOS and SACOS methods using three different sampling regimes for each, with the fixed column downsampling parameter $\gamma=0.2$ throughout. For ACOS, we examine settings where $m=0.2n_1$, $0.1n_1$ and $0.05n_1$ with $p=0.5n_2$, which result in average sampling rates of 4.5\%, 2.5\% and 1.5\%, respectively. For SACOS, we examine settings where $m=0.2n_1$, $0.05n_1$ and $0.03n_1$, resulting in average sampling rates of 20\%, 5\% and 3\%, respectively. As before, we generate the $\bPhi$ and $\Ab$ matrices to have i.i.d. zero-mean Gaussian entries. We compare our approaches with two ``benchmarks'' -- the Graph-based visual saliency (GBVS) method from the computer vision literature \cite{Harel:06} and the OP approach (both of which use the full data) -- as well as with the RMC approach at sampling rates of $20\%$ and $5\%$.

The results of this experiment are provided in Figure~\ref{fig:real}. We note first that the OP approach performs fairly well at identifying the visually salient regions in the image, essentially identifying the same salient regions as the GBVS procedure and providing evidence to validate the use of the low-rank plus outlier model for visual saliency (see also \cite{Shen:12}).  Next, comparing the results of the individual procedures, we see that the OP approach appears to uniformly give the best detection results, which is reasonable since it is using the full data as input.  The RMC approach performs well at the 20\% sampling rate, but its performance appears to degrade at the 5\% sampling rate.  The SACOS approach, on the other hand, still produces reasonably accurate results using only 3\% sampling. Moreover, ACOS provides acceptable results even with 1-2\% sampling rate.

We also compare implementation times of the algorithms on this saliency map estimation task.  Table~\ref{table:timing} provides the average execution times (and standard deviations) for each approach, evaluated over 1000 images in the MSRA database\footnote{Timing comparisons were done with {\tt MATLAB} R2013a on an iMac with a 3.4 GHz Intel Core i7 processor, 32 GB memory, and running OS X 10.8.5.}. Here, we only execute each procedure for one choice of regularization parameter, and we also include the additional ``rank reduction'' step discussed above for the ACOS and SACOS methods.  Overall, we see the ACOS approach is up to $4\times$ faster than the GBVS method and $15\times$ faster than the OP and RMC methods, while the SACOS approach could result overall in relative speedups of $100\times$ over GBVS and $300\times$ over the OP and RMC methods. Overall, our results suggest a significant improvement obtained via ACOS and SACOS for both detection consistency and timing, which may have a promising impact in a variety of salient signal detection tasks.

\begin{figure}[t]
\centering
\footnotesize
\subfloat{
	\includegraphics[width=0.28\linewidth]{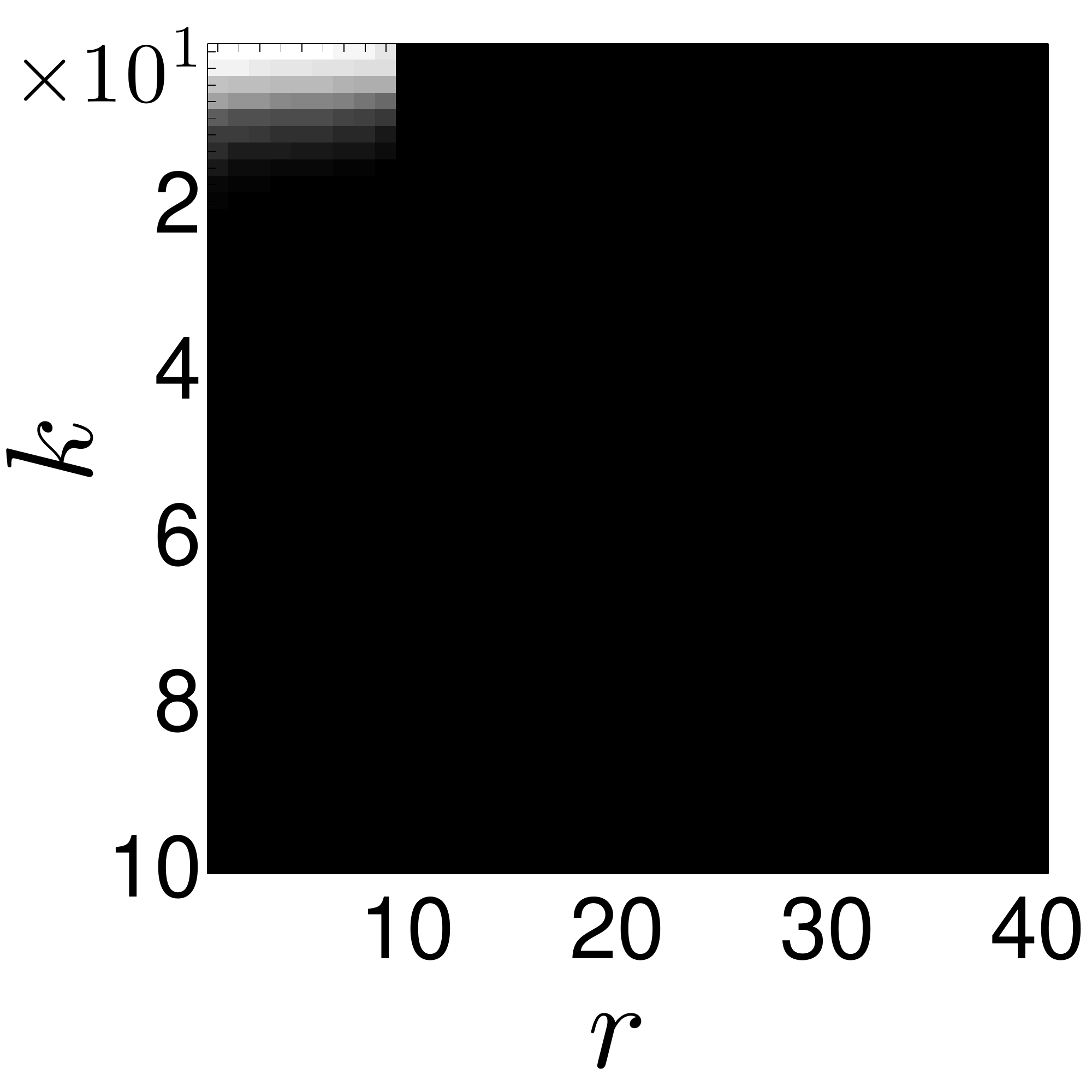}
	}
\subfloat{
	\includegraphics[width=0.28\linewidth]{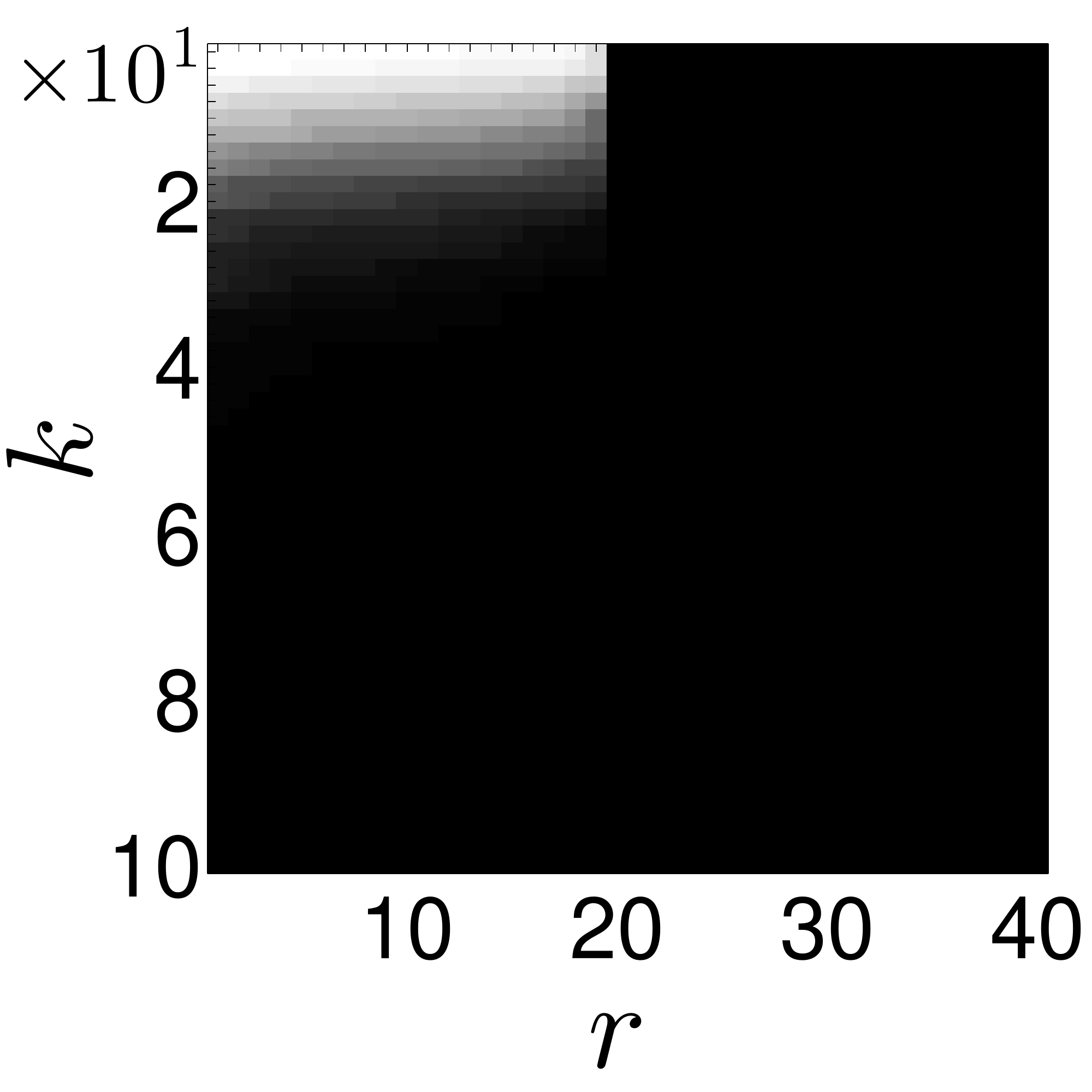}
}
\subfloat{
	\includegraphics[width=0.28\linewidth]{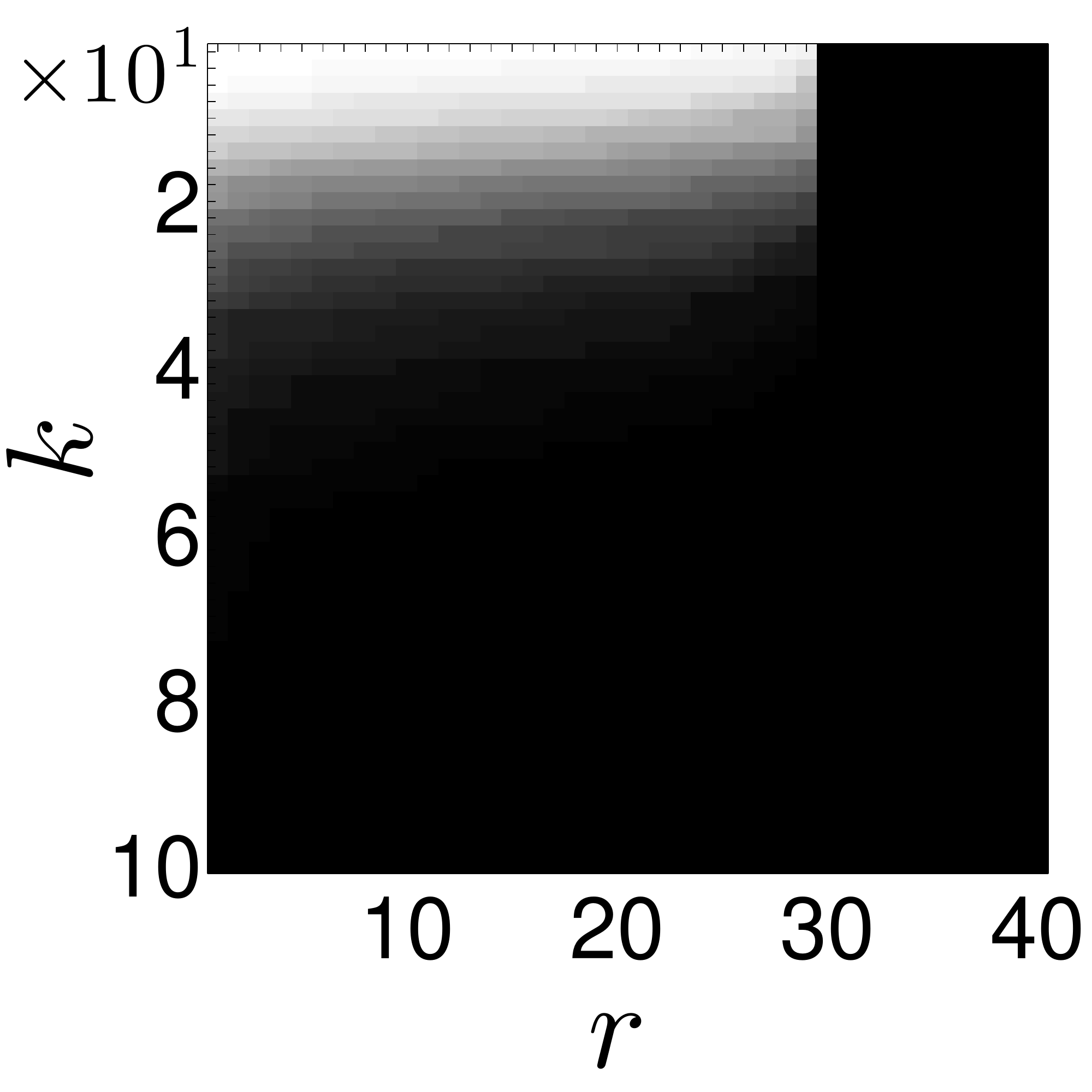}
}\vspace{-0.0em}\\
\subfloat{
	\includegraphics[width=0.28\linewidth]{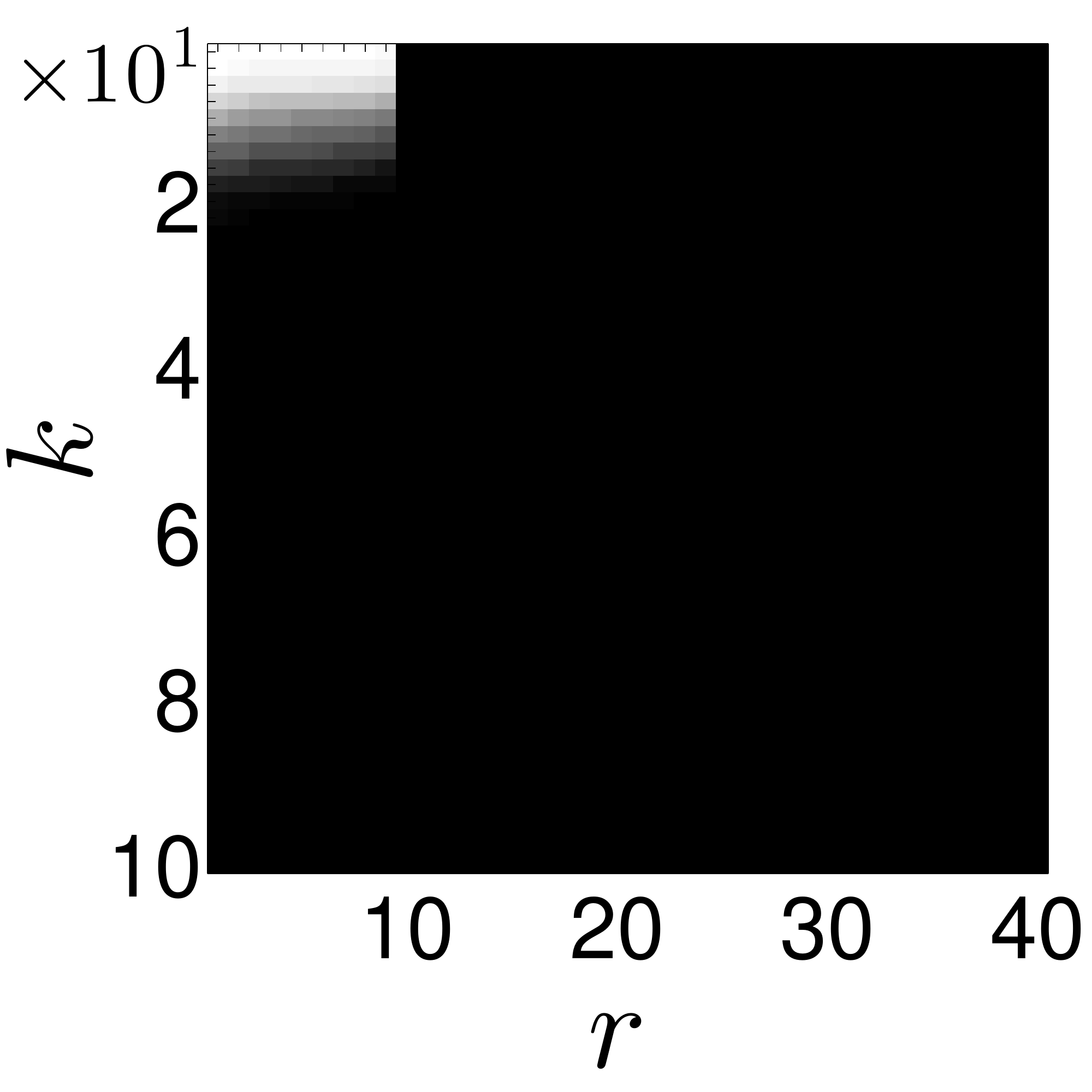}
}
\subfloat{
	\includegraphics[width=0.28\linewidth]{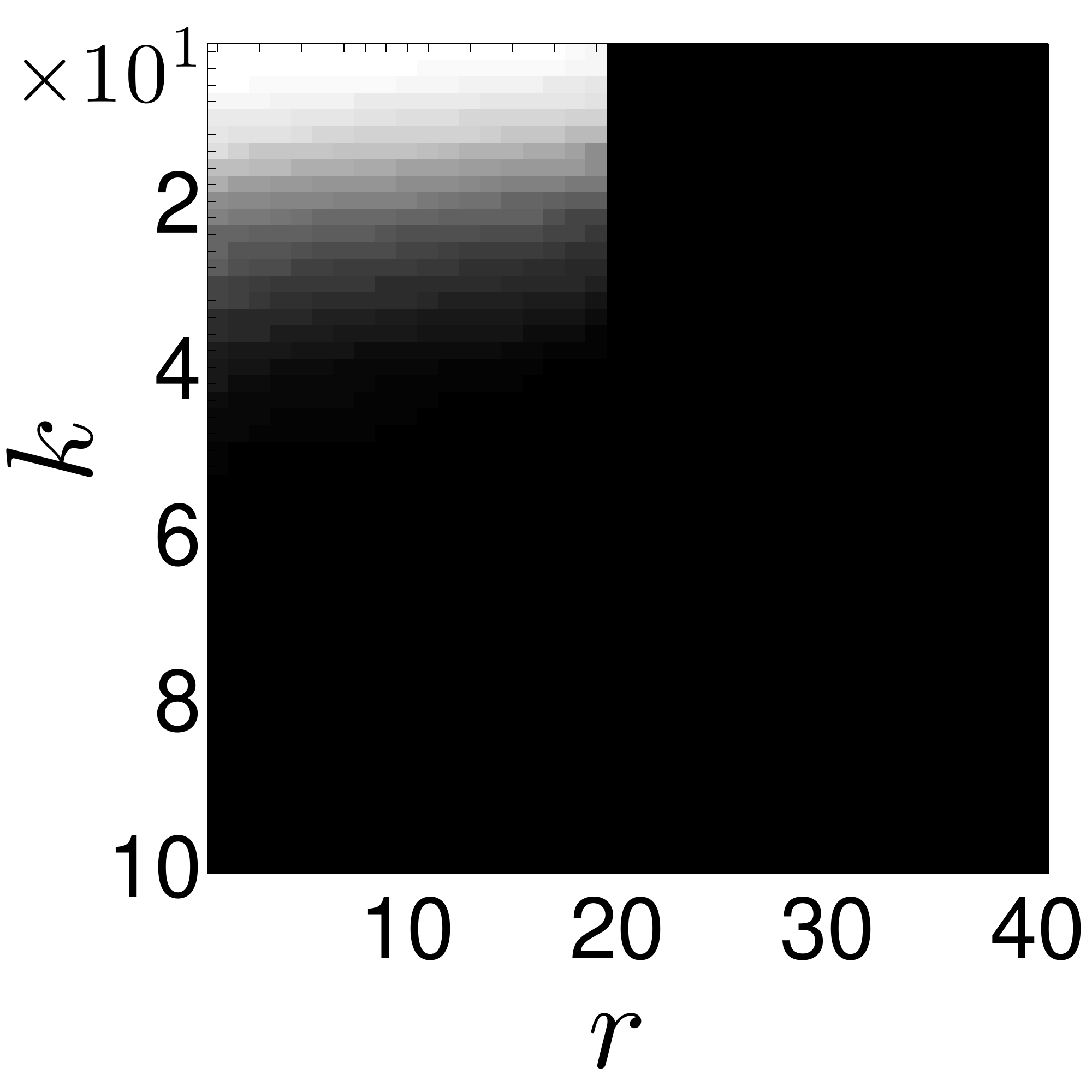}
}
\subfloat{
	\includegraphics[width=0.28\linewidth]{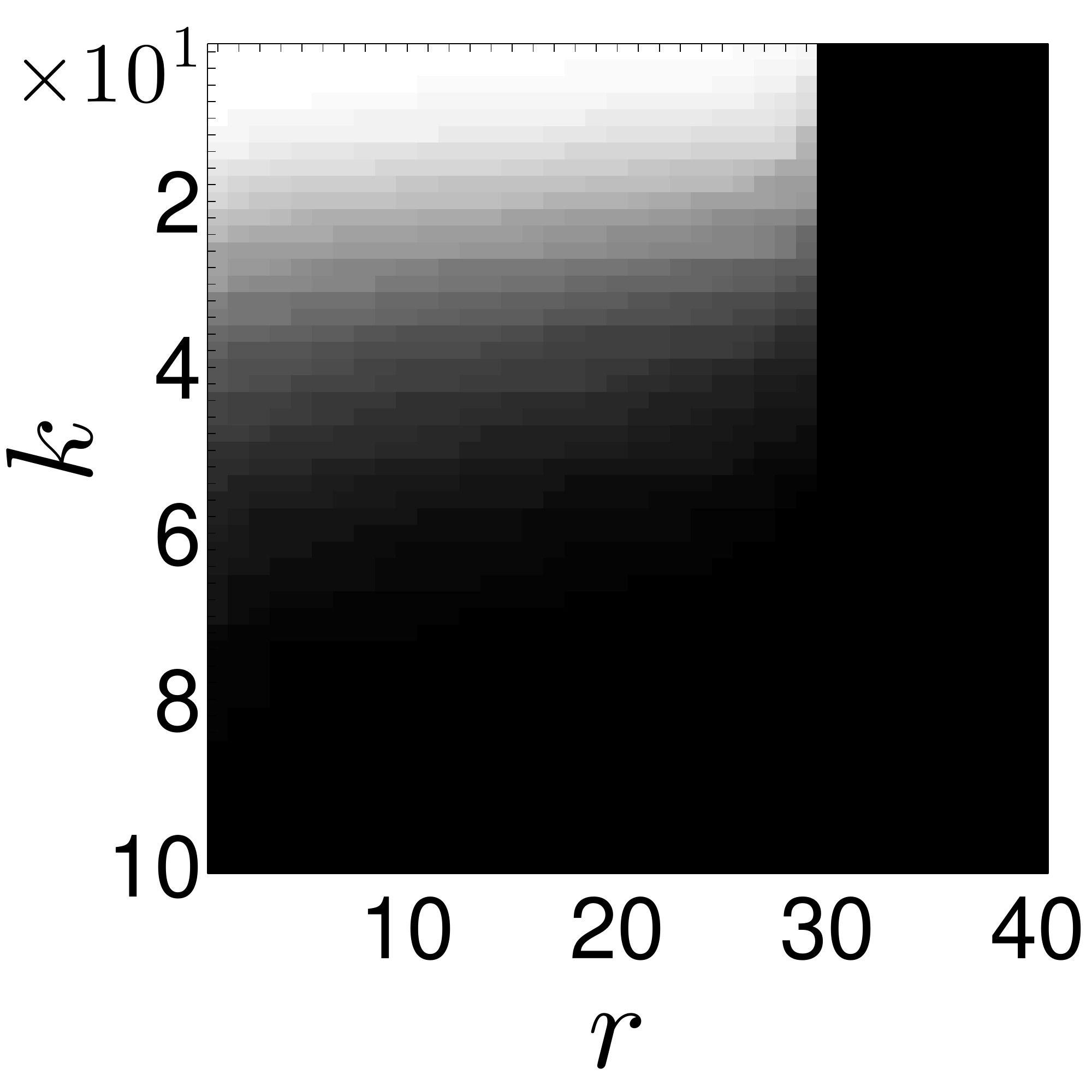}
}\vspace{-0.0em}\\
\subfloat{
	\includegraphics[width=0.28\linewidth]{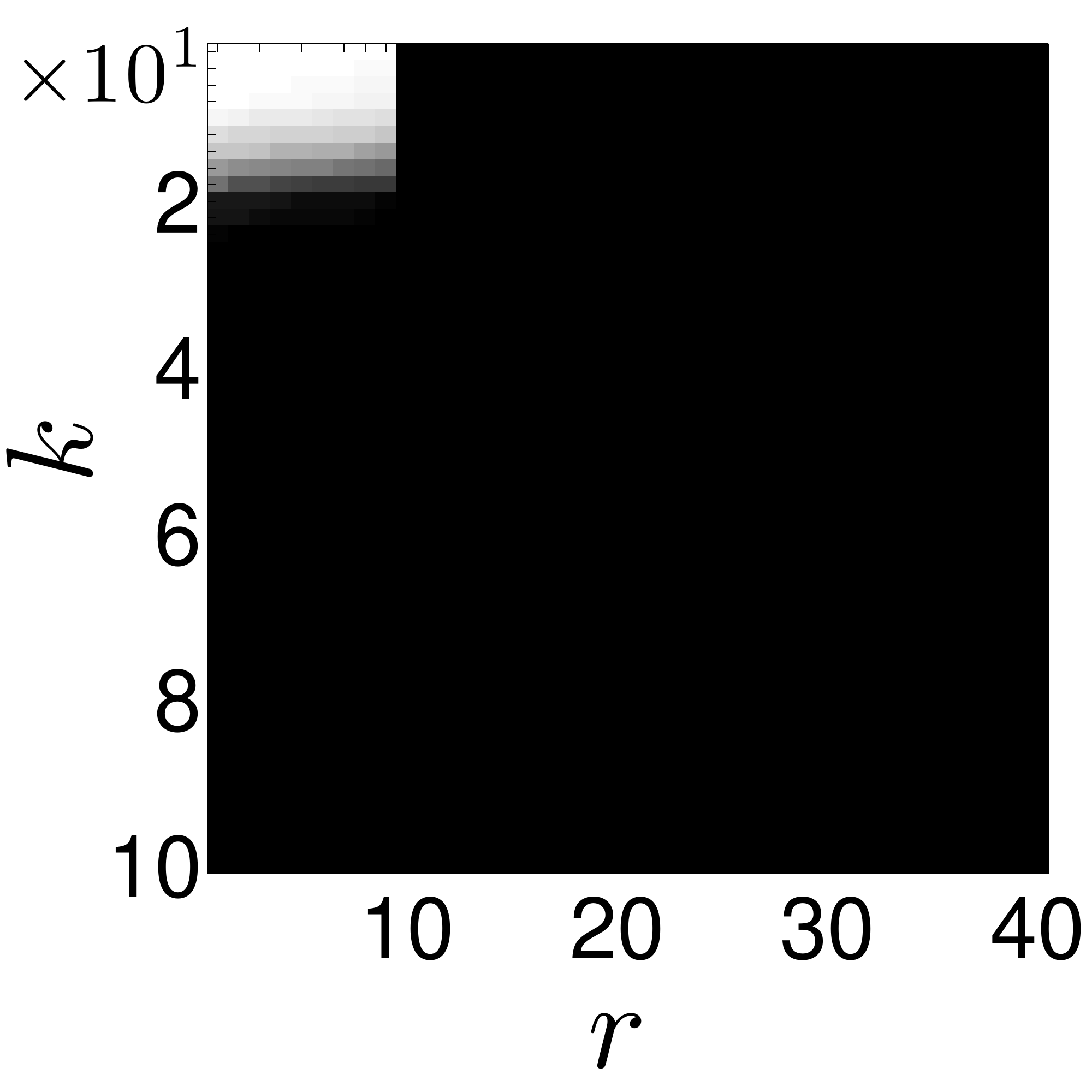}
}
\subfloat{
	\includegraphics[width=0.28\linewidth]{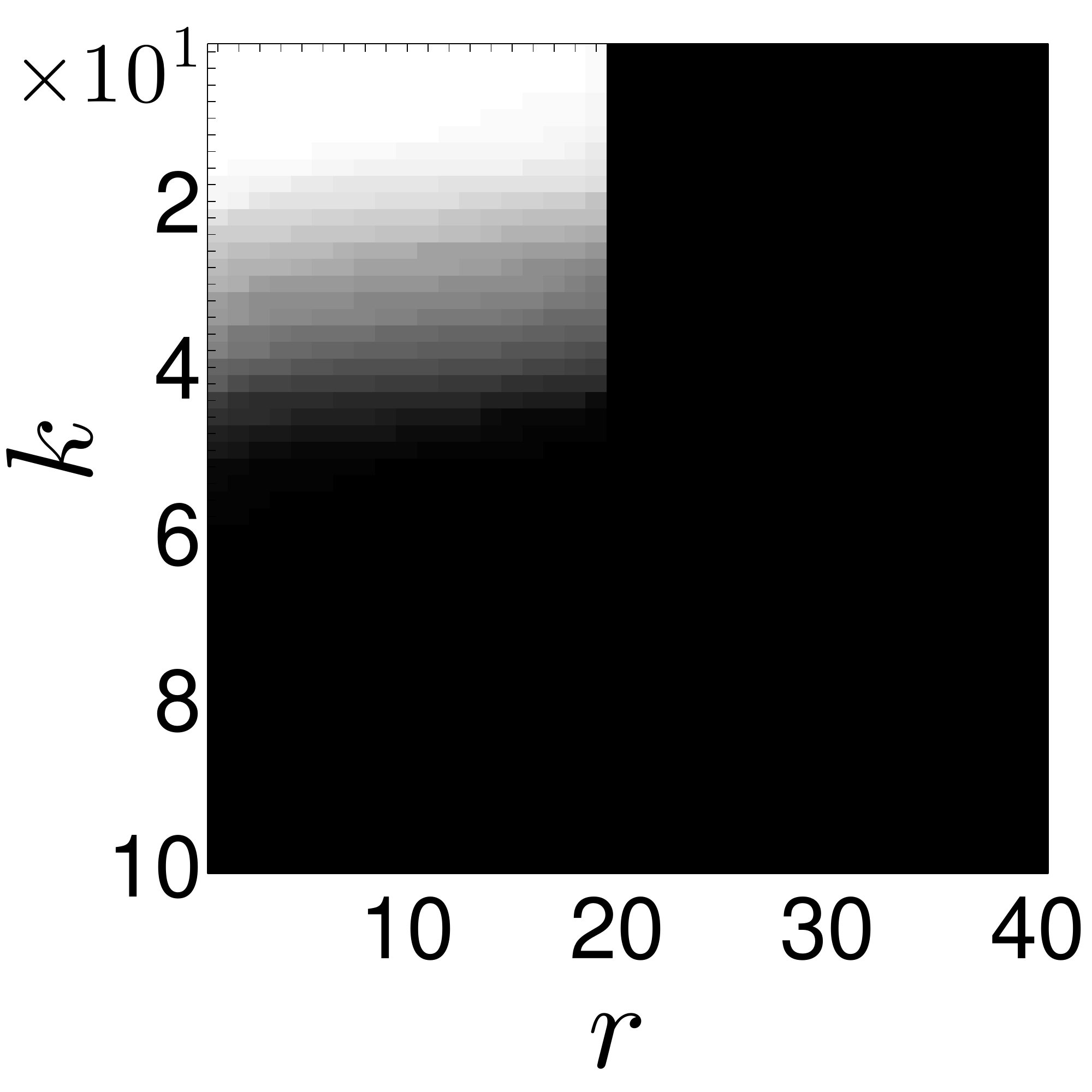}
}
\subfloat{
	\includegraphics[width=0.28\linewidth]{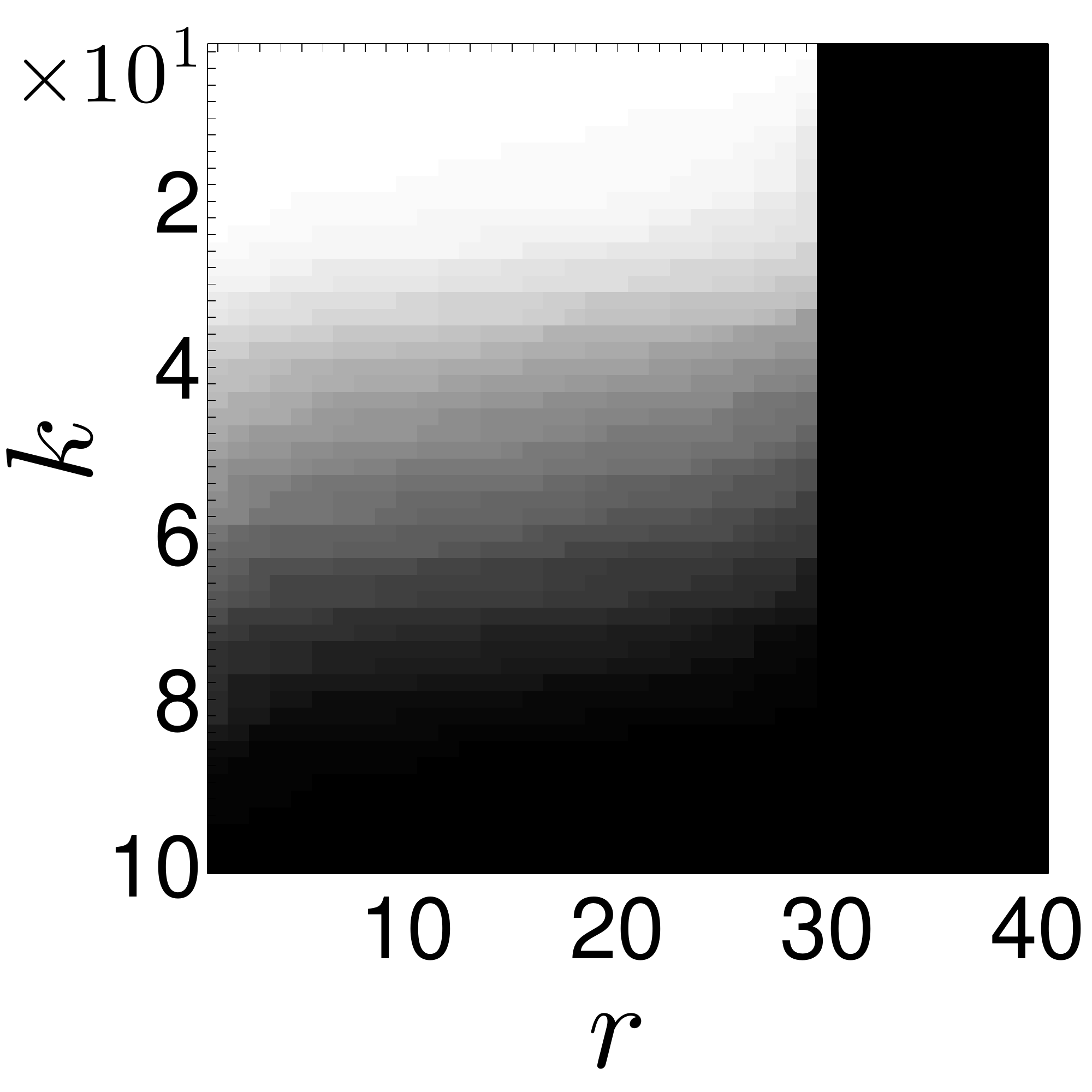}
}\\
\hspace{0.0in}
(a) 2.1\%  \hspace{0.6in} (b) 4.2\% \hspace{0.6in} (c) 6.3\%
\vspace{-0.05in}
\caption{Outlier recovery phase transitions plots for ACOS for noisy settings (white regions correspond to successful recovery). Rows correspond to $\sigma = 0.001$, 0.0005 and 0.0001 respectively, from top to bottom; columns correspond to the settings $m=0.1n_1,~p=0.1n_2$;  $m=0.2n_1,~p=0.2n_2$; and $m=0.3n_1,~p=0.3n_2$ respectively, from left to right. The fraction of observations obtained is provided below each figure. As in Figure~\ref{fig:sim1}, larger $m$ and $p$ promote accurate recovery for increasing rank $r$ and numbers $k$ of outlier columns.  Here, however, increasing noise variance degrades the estimation results, especially with respect to the number $k$ of outliers that can be accurately identified.}
\label{fig:noisy}
%\vspace{-0.1in}
\end{figure}

\section{Extensions}\label{extension}

%In this section we investigate extensions of our approach to noisy settings, and scenarios characterized by missing data.  

\subsection{Noisy Observations}
	
We demonstrate the outlier detection performance of our approaches under the scenario when $\Mb$ is contaminated by unknown random noise or modeling error. Formally, we consider the setting where $\Lb$ and $\Cb$ are as above, but
\begin{align}
\Mb =\Lb + \Cb + \Nb,
\end{align}
where $\Nb$ has i.i.d. $\cN(0,\sigma^2)$ entries. 

\begin{figure}[!t]
\centering
%\begin{minipage}[b]{1\linewidth}
	\footnotesize
	\centering
	\subfloat{
		\includegraphics[width=0.28\linewidth]{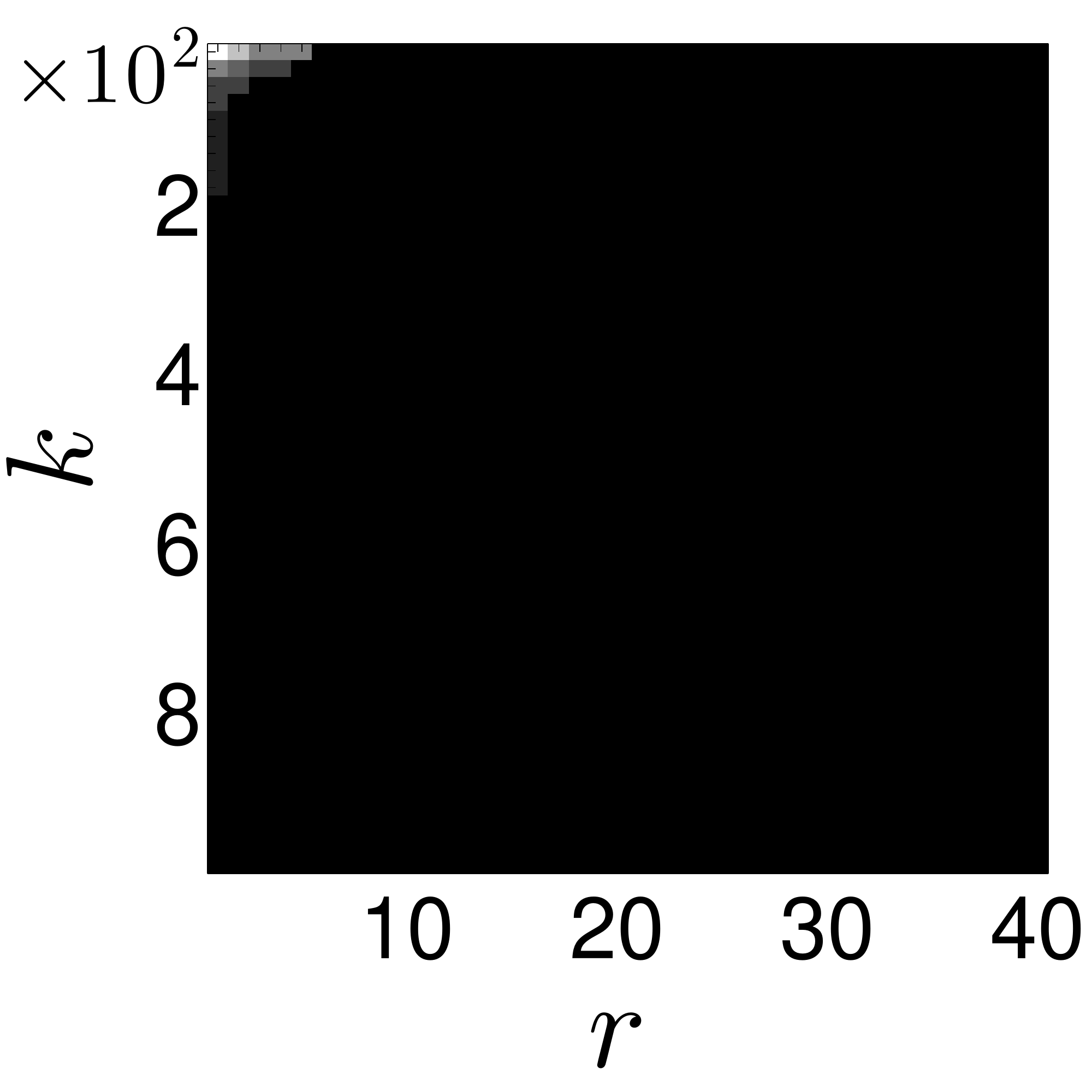}
	}\subfloat{
		\includegraphics[width=0.28\linewidth]{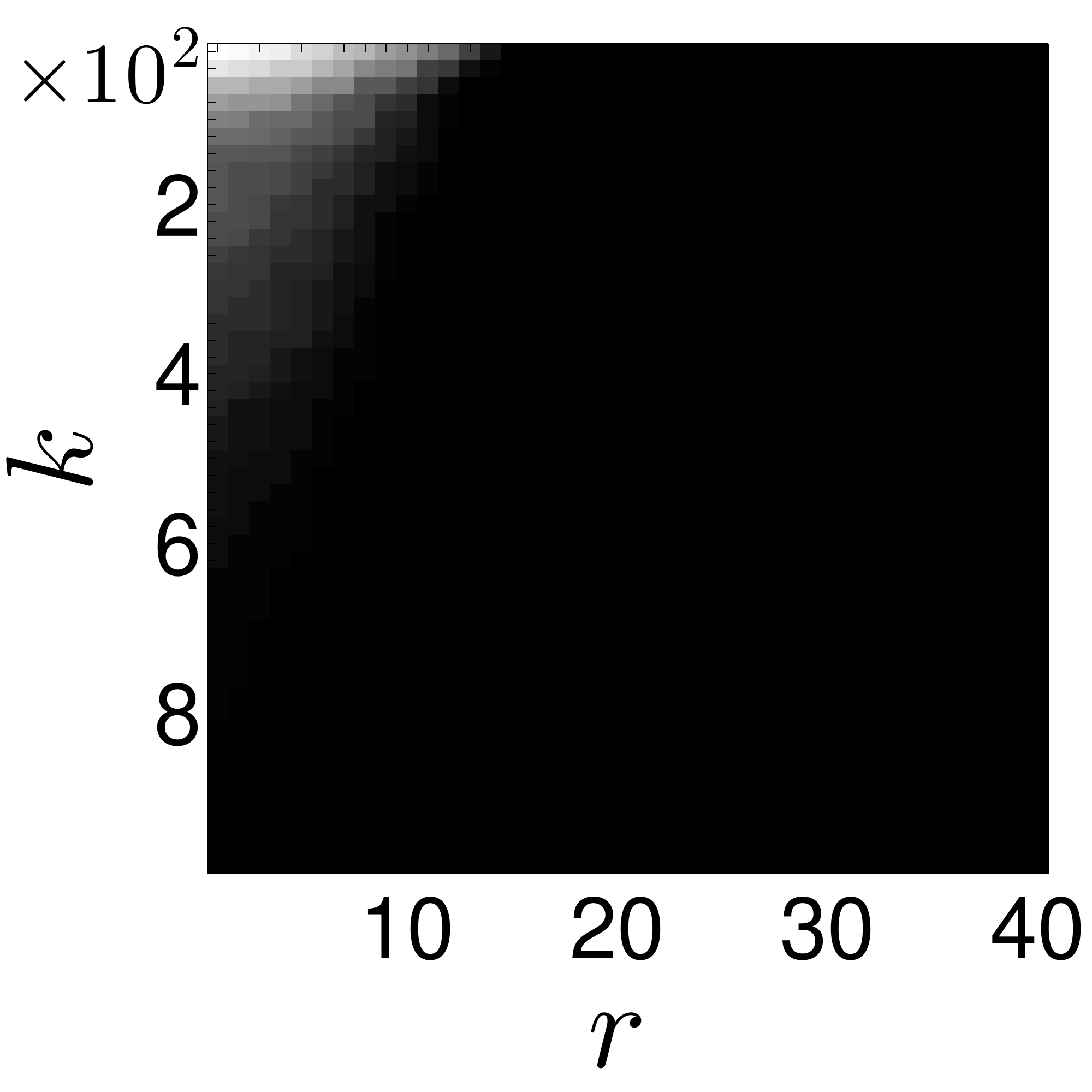}
	}\subfloat{
		\includegraphics[width=0.28\linewidth]{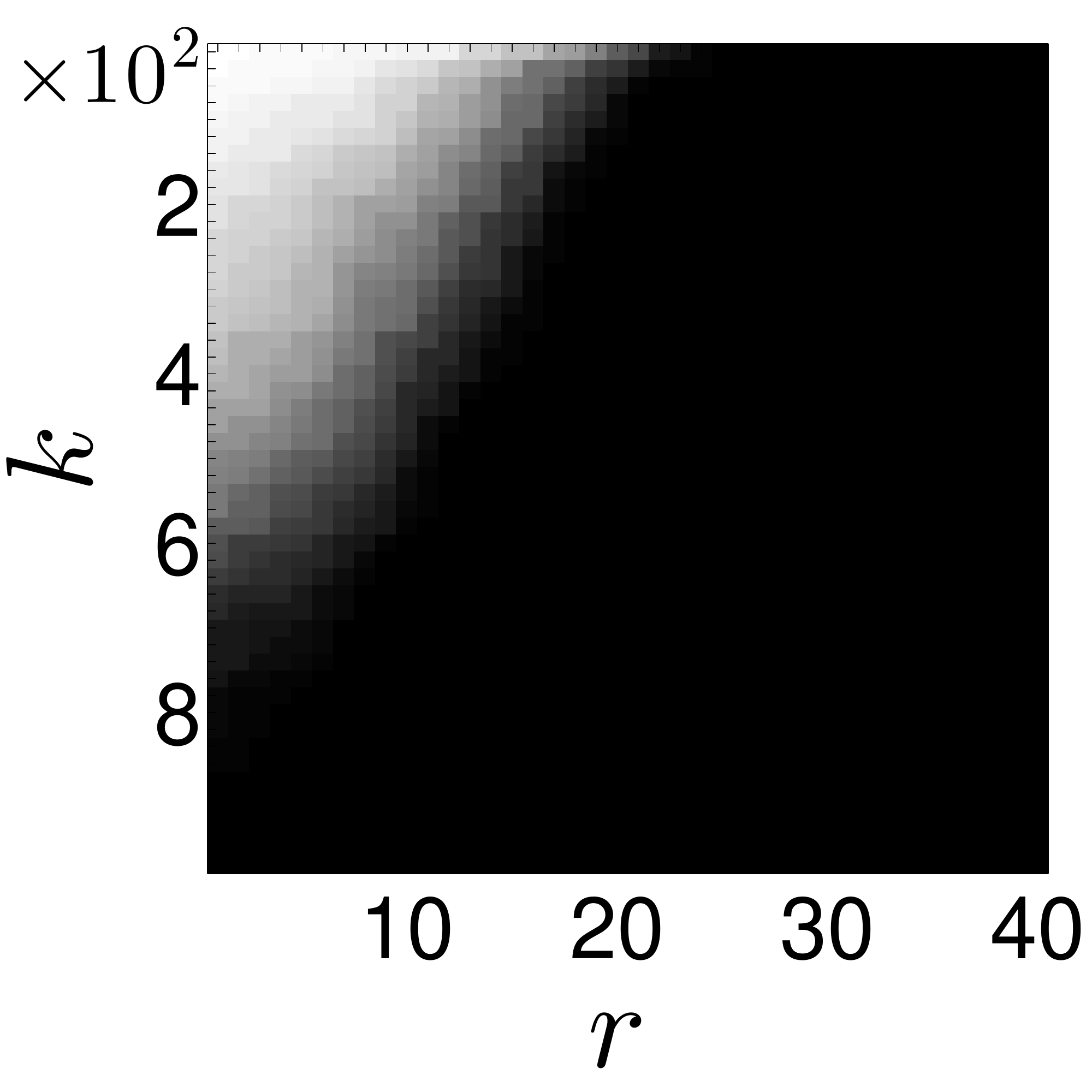}
	}\\
	\subfloat{
		\includegraphics[width=0.28\linewidth]{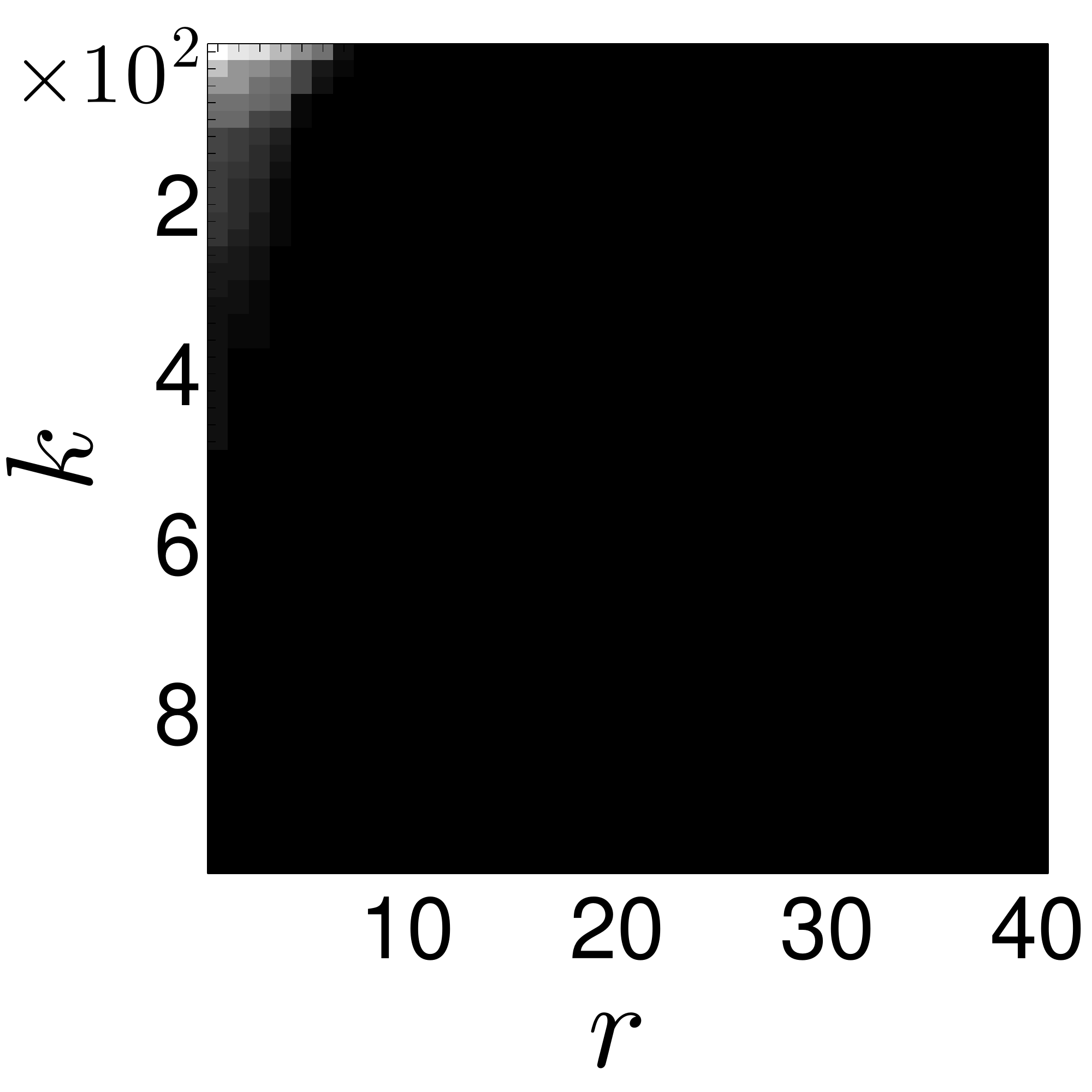}
	}\subfloat{
		\includegraphics[width=0.28\linewidth]{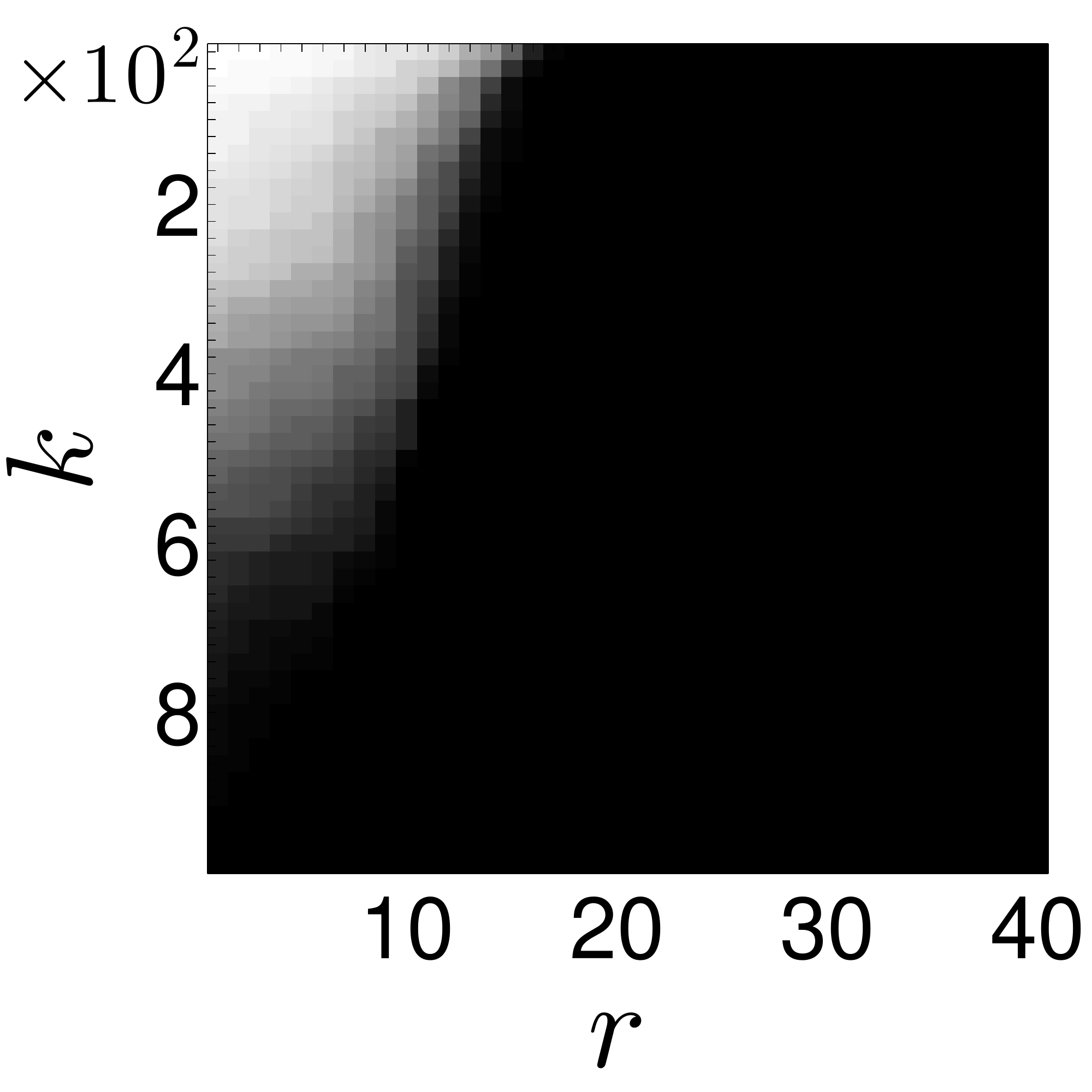}
	}\subfloat{
		\includegraphics[width=0.28\linewidth]{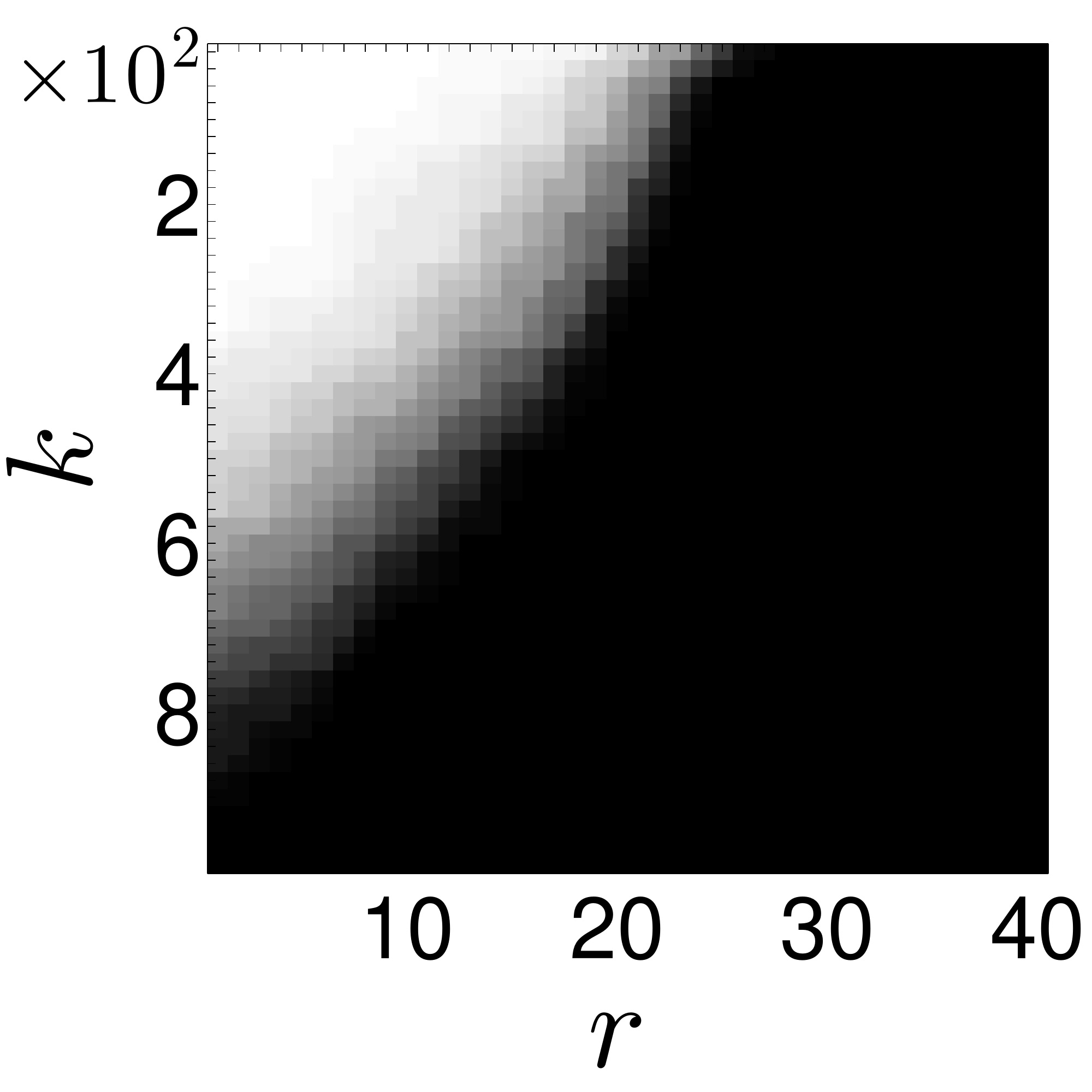}
	}\\
	\subfloat{
		\includegraphics[width=0.28\linewidth]{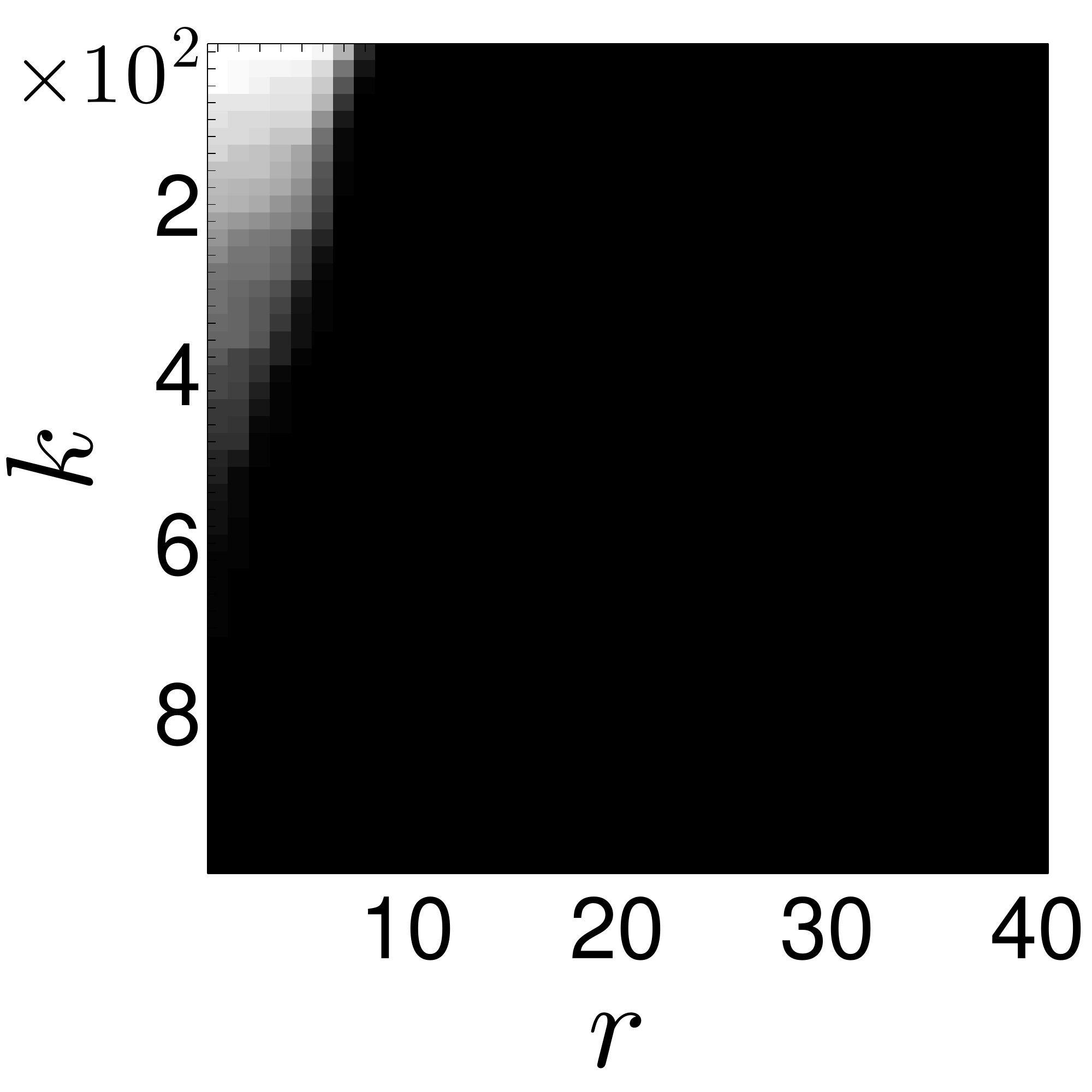}
	}\subfloat{
		\includegraphics[width=0.28\linewidth]{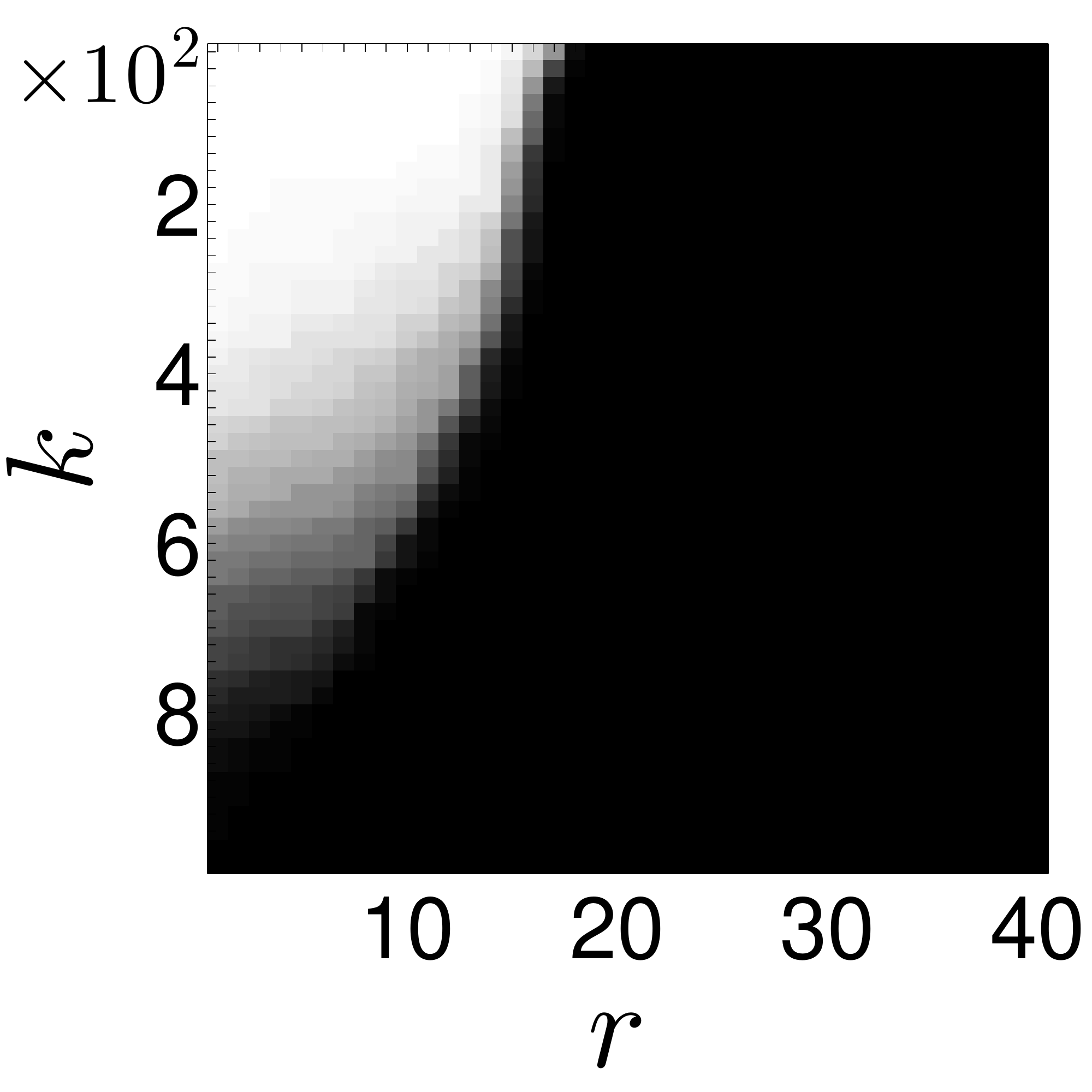}
	}\subfloat{
		\includegraphics[width=0.28\linewidth]{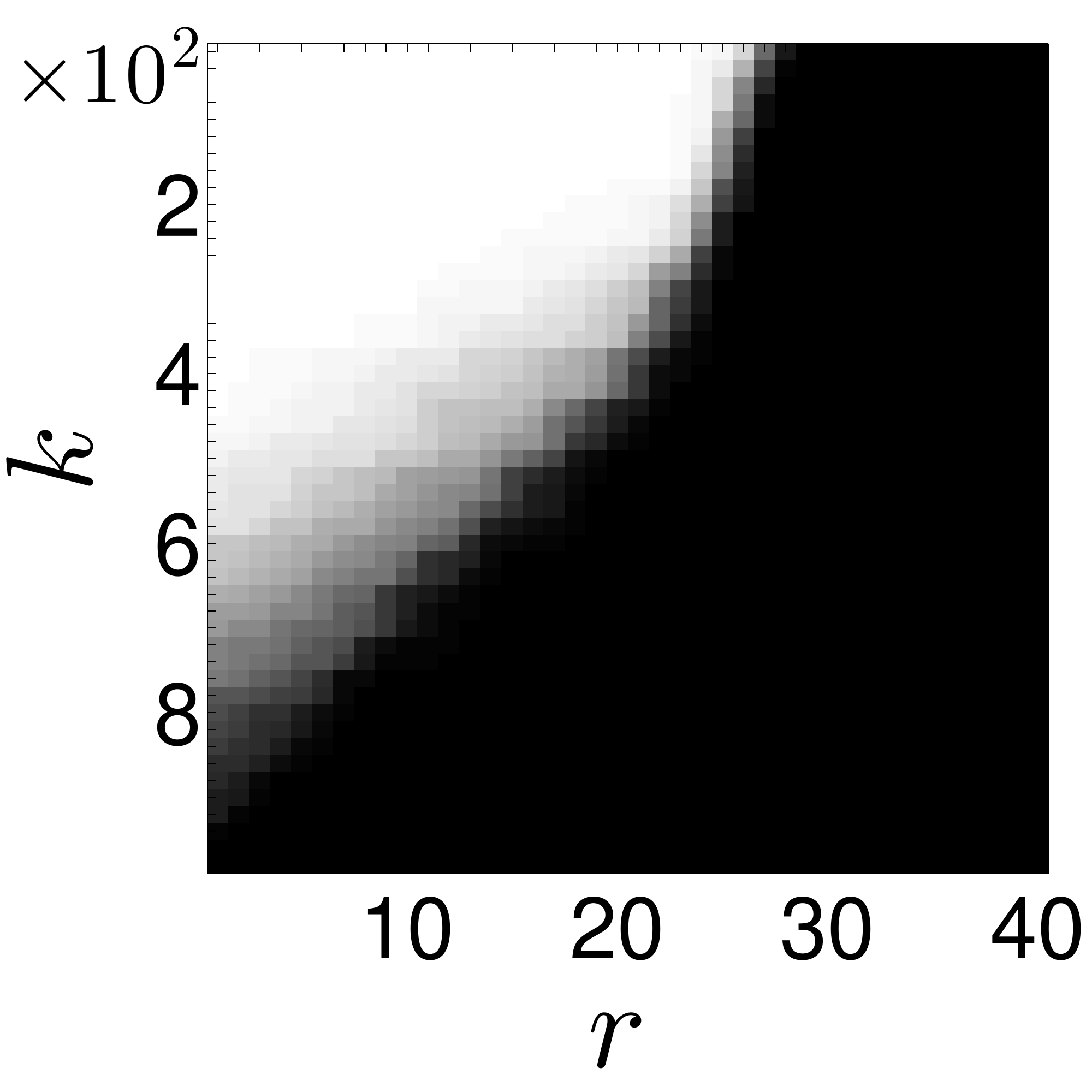}
	}\\
	\hspace{0.0in}
	(a) 10\%  \hspace{0.6in} (b) 20\% \hspace{0.6in} (c) 30\%
	\vspace{-0.05in}
	\caption{Outlier recovery phase transitions plots for SACOS for noisy settings (white regions correspond to successful recovery). Rows of the figure correspond to $\sigma=$ 0.03, 0.02 and 0.01 respectively, from top to bottom; columns correspond to $m=0.1n_1,~0.2n_1$, and $0.3n_1$ respectively, from left to right. The fraction of observations obtained is provided below each column. In this case, increasing noise variance results in a decrease in both the rank $r$ as well as the number $k$ of outliers that can be accurately identified.}
\label{fig:noise_sacos}
\vspace{-0.1in}
\end{figure}

We first investigate the performance of the ACOS method, following a similar experimental methodology as in  Section~\ref{sec:exp} to generate $\Lb$ and $\Cb$, except that now we renormalize each column of $(\Lb + \Cb)$ to have unit Euclidean norm (essentially to standardize the noise levels).  We consider three different noise levels ($\sigma = 0.001$, 0.0005 and 0.0001), three pairs of the row sampling parameter $m$ and the column sampling parameter $p$ ($m=0.1n_1,~p=0.1n_2$; \ $m=0.2n_1,~p=0.2n_2$; and \ $m=0.3n_1,~p=0.3n_2$) and for each we fix the column downsampling fraction to be $\gamma=0.2$; the corresponding sampling ratios are 2.1\%, 4.2\% and 6.3\%, respectively. We again perform 100 trials of Algorithm~\ref{alg:main} and record the success frequency for each. The results are given in Figure~\ref{fig:noisy}.

It can be observed from the results that increasing $m$ and $p$ promote accurate estimation of outlier column indices for increasing rank $r$ and numbers $k$ of outlier columns, which is exactly what we have seen in Figure~\ref{fig:sim1} for the noiseless case. However, the presence of noise degrades the estimation performance, albeit gracefully. This is reasonable, since in Step 2 of Algorithm~\ref{alg:main}, the measurements $\yb_2$ might be perturbed more seriously as the energy of noise increases, which results in more difficult recovery of true supports of $\cbb$. Under this scenario, we will require larger $p$ to enable better recovery of the underlying true supports. 
	
We also evaluate the SACOS procedure in noisy settings for three choices of $m$ ($m=0.1n_1$, $0.2n_1$ and  $0.3n_2$) and fixed column downsampling fraction $\gamma=0.2$.  Here, we again normalize columns of $(\Lb + \Cb)$, but consider three higher noise levels, corresponding to $\sigma=0.03$, $0.02$ and $0.01$. The results are presented in Figure~\ref{fig:noise_sacos}.  Here, we again observe a graceful performance degradation with noise. Notice, however, that higher level of variances of noise can be tolerated for SACOS compared with ACOS, which is an artifact of the difference between the second (inference) steps of the two procedures.

\subsection{Missing Data}
	
We also describe and demonstrate an extension of our SACOS method that is amenable to scenarios characterized by missing data.  Suppose that there exists some underlying matrix $\Mb$ that admits a decomposition of the form $\Mb = \Lb + \Cb$ with $\Lb$ and $\Cb$ as above, but we are only able to observe $\Mb$ at a subset of its locations.   Formally, we denote by $\bOmega \subseteq [n_1]\times[n_2]$ the set of indices corresponding to the available elements of $\Mb$, and let $\Pb_{\bOmega}(\cdot)$ be the operator that masks its argument at locations not in $\bOmega$.  Thus, rather than operate on $\Mb$ itself, we consider procedures that operate on the sampled data $\Pb_{\bOmega}(\Mb)$.

In this setting, we can modify our SACOS approach so that the observations obtained in Step 1 are of the form $\Yb_{(1)} = \bPhi \Pb_{\bOmega} (\Mb)\Sbb$, where (as before) $\Sbb$ is a column selection matrix but $\bPhi$ is now a \emph{row} subsampling matrix (i.e., it is comprised of a subset of rows of the $n_1\times n_1$ identity matrix) containing some $m$ rows.   The key insight here is that the composite operation of sampling elements of $\Mb$ followed by row subsampling can be expressed in terms of a related operation of subsampling elements of a row-subsampled version of $\Mb$.  Specifically, we have that $\bPhi \Pb_{\bOmega} (\Mb) = \Pb_{\bOmega_{\bPhi}}(\bPhi \Mb)$, where $\Pb_{\bOmega_{\bPhi}}(\cdot)$ masks the same elements as $\Pb_{\bOmega}(\cdot)$ in the rows selected by $\bPhi$. 

Now, given $\Yb_{(1)}$, we solve a variant of RMC \cite{Chen:11}
\begin{align*}
\{\widehat{\Lb}_{(1)},\widehat{\Cb}_{(1)}\} = \argmin_{\bL,\bC} &\|\bL\|_{*} + \lambda \|\bC\|_{1,2}\\
\mbox{ s.t. } & \Yb_{(1)} = \Pb_{\bOmega_{\bPhi}}( \bL + \bC)
\end{align*}
in an initial step, identifying (as before) an estimate $\widehat{\Lb}_{(1)}$ whose column span is an estimate of the subspace spanned by the low-rank component of $\bPhi\Mb$.  

Then (in a second step) we perform the ``missing data'' analog of the orthogonal projection operation on every column $j \in [n_2]$ of $\bPhi \Pb_{\bOmega} (\Mb)$, as follows.  For each $j\in[n_2]$, we let $\cI_j\in[m]$ denote the locations at which observations of column $j$ of $\bPhi \Pb_{\bOmega} (\Mb)$ are available, and let $(\bPhi\Pb_{\bOmega} (\Mb))_{\cI_j,j}$ be the sub vector of $(\bPhi \Pb_{\bOmega} (\Mb))_{:,j}$ containing only the elements indexed by $\cI_j$.  Similarly, let $(\widehat{\Lb}_{(1)})_{\cI_j,:}$ be the row submatrix of $\widehat{\Lb}_{(1)}$ formed by retaining rows indexed by $\cI_j$. Now, 
let $\Pb_{\widehat{\cL}_{(1)_{j}}}$ denote the orthogonal projection onto the subspace spanned by columns of $(\widehat{\Lb}_{(1)})_{\cI_j,:}$ and compute the residual energy of the $j$-th column as $\|(\Ib-\Pb_{\widehat{\cL}_{(1)_{j}}}) (\bPhi \Pb_{\bOmega} (\Mb))_{\cI_j,j}\|_2$. Overall, the orthogonal projection for the $j$-th column of $\bPhi \Pb_{\bOmega} (\Mb)$ is only computed over the nonzero entries of that column, an approach motivated by a recent effort examining subsampling methods in the context of matrix completion \cite{Krishnamurthy:14}.
	
\begin{figure}[!t]
\footnotesize
\centering
	\subfloat[3\%]{
		\includegraphics[width=0.28\linewidth]{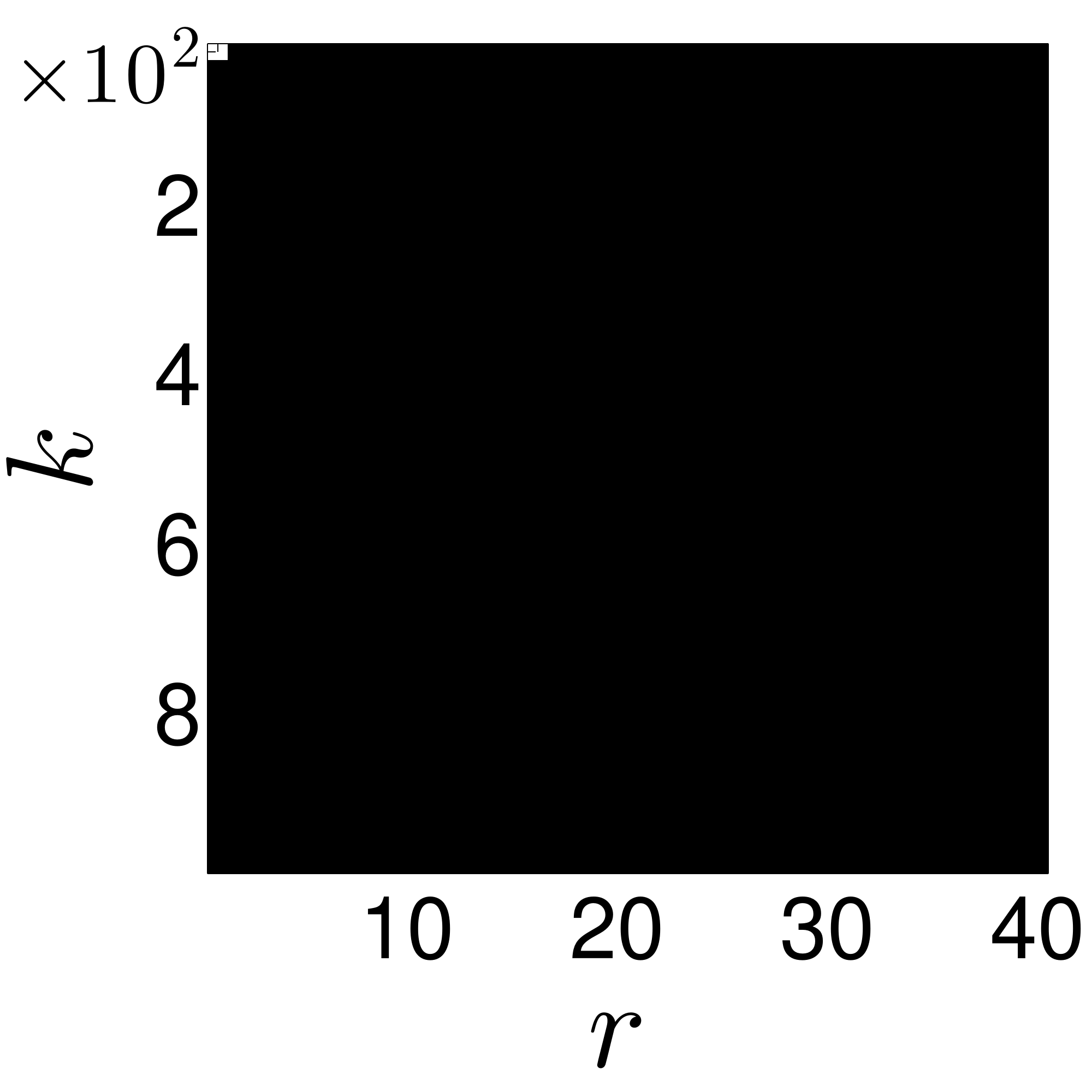}
	}
	\subfloat[5\%]{
		\includegraphics[width=0.28\linewidth]{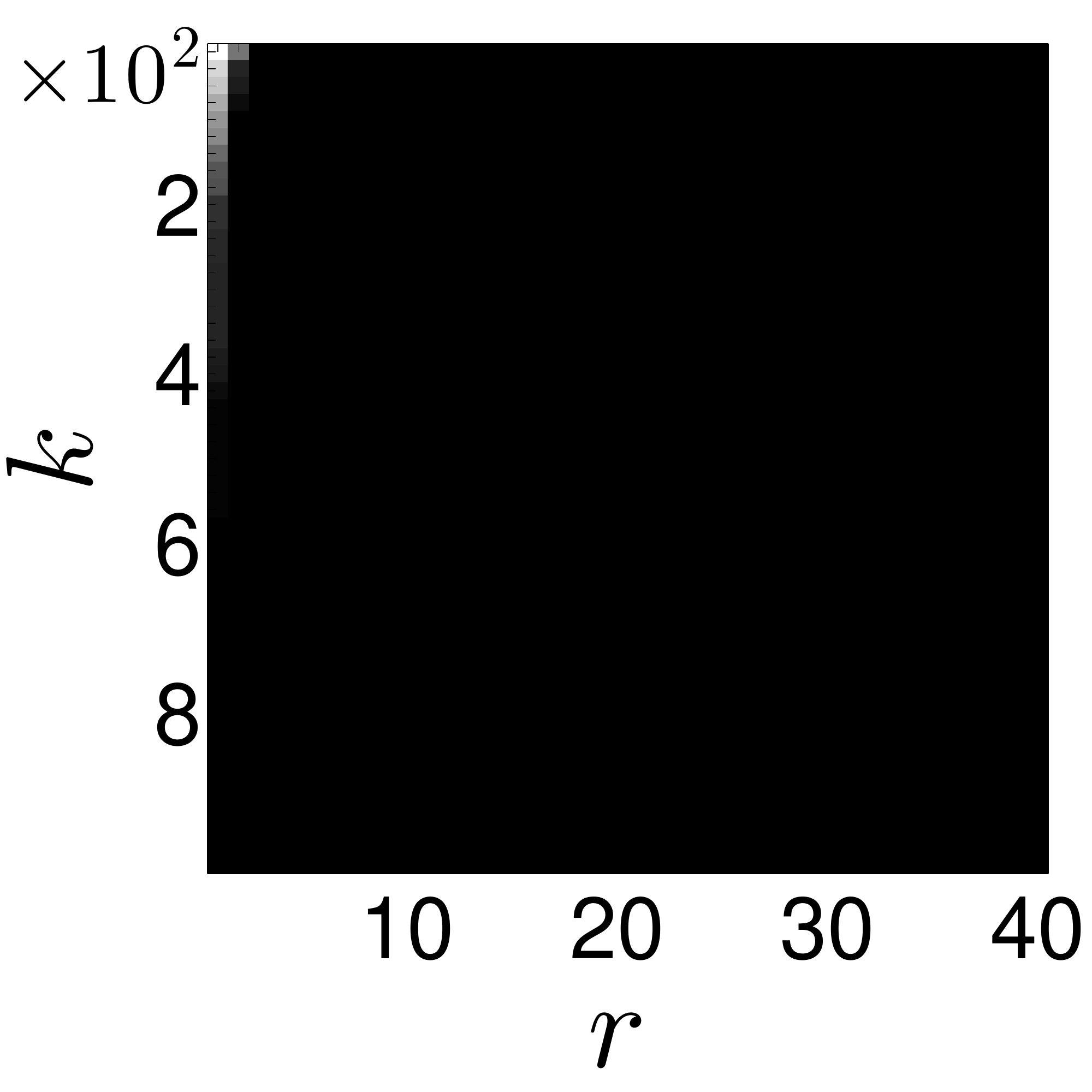}
	}
	\subfloat[7\%]{
		\includegraphics[width=0.28\linewidth]{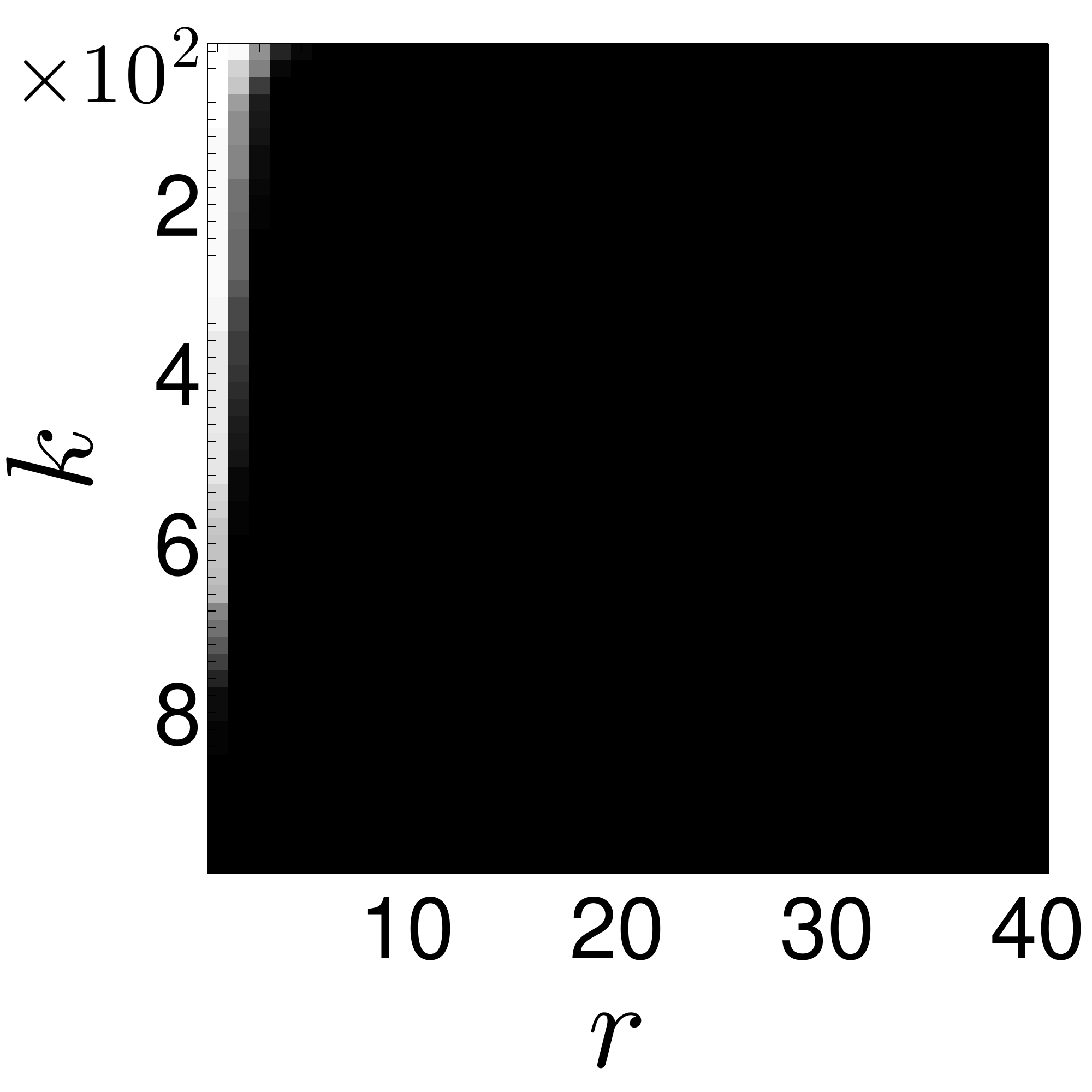}
	}\\
	\subfloat[6\%]{
		\includegraphics[width=0.28\linewidth]{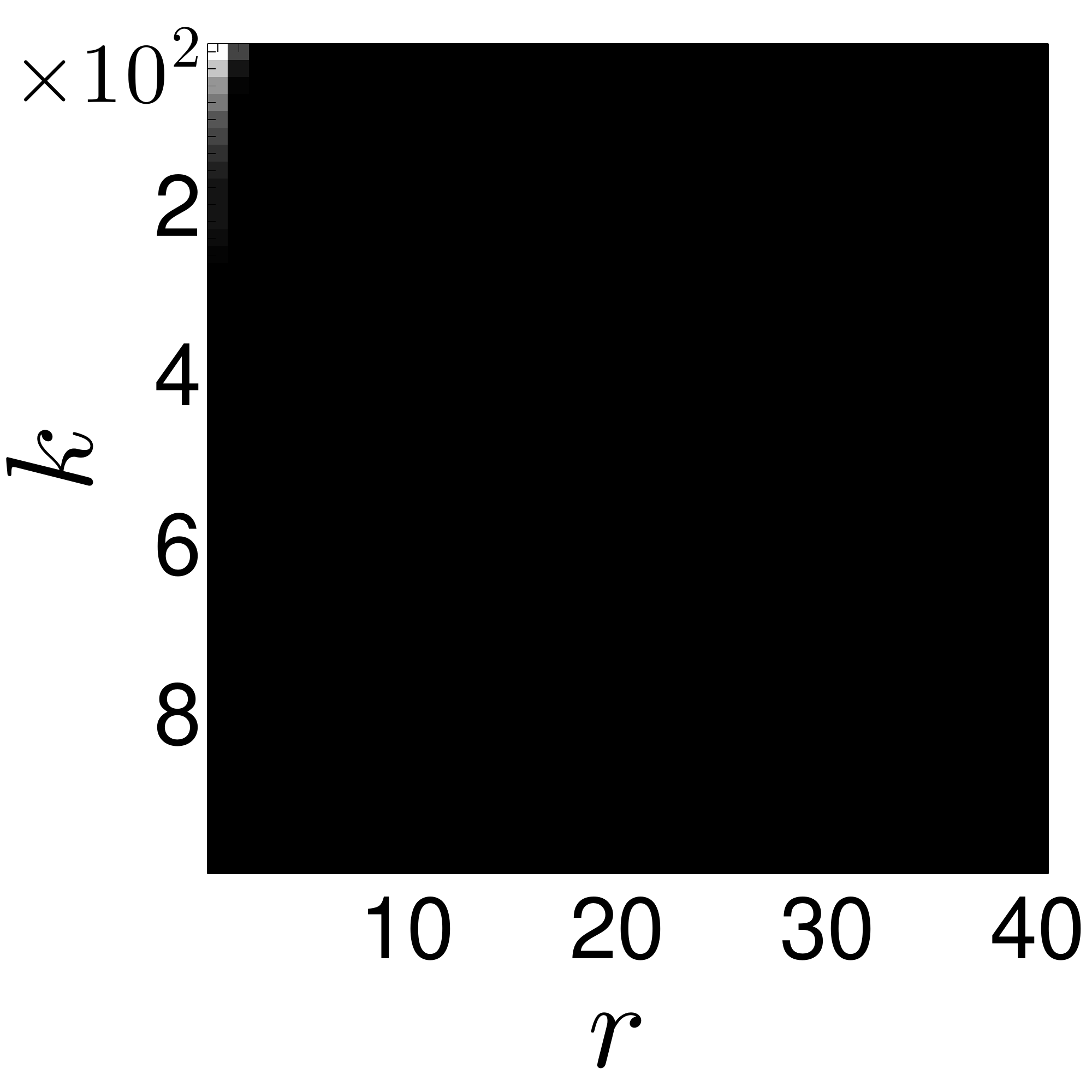}
	}
	\subfloat[10\%]{
		\includegraphics[width=0.28\linewidth]{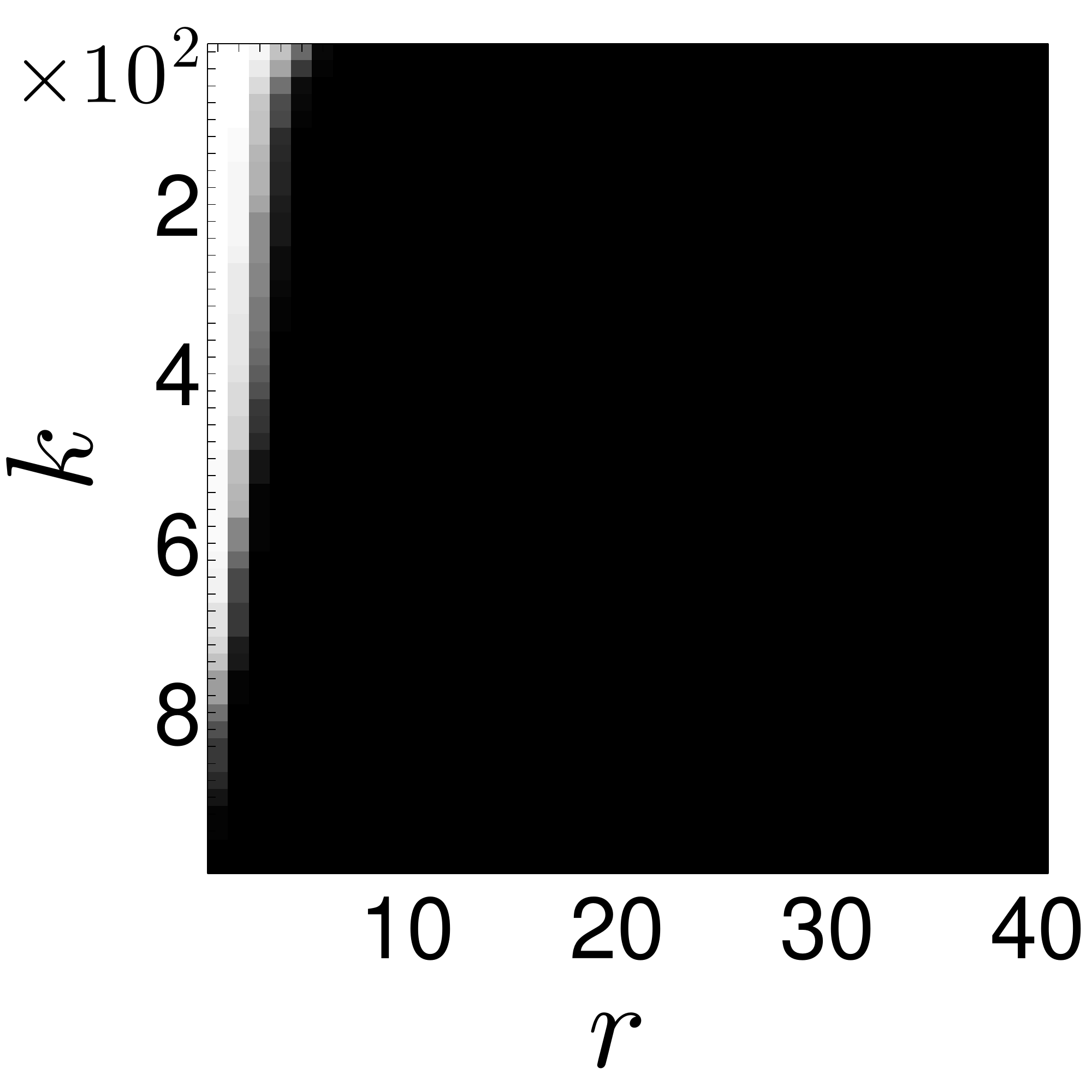}
	}
	\subfloat[14\%]{
		\includegraphics[width=0.28\linewidth]{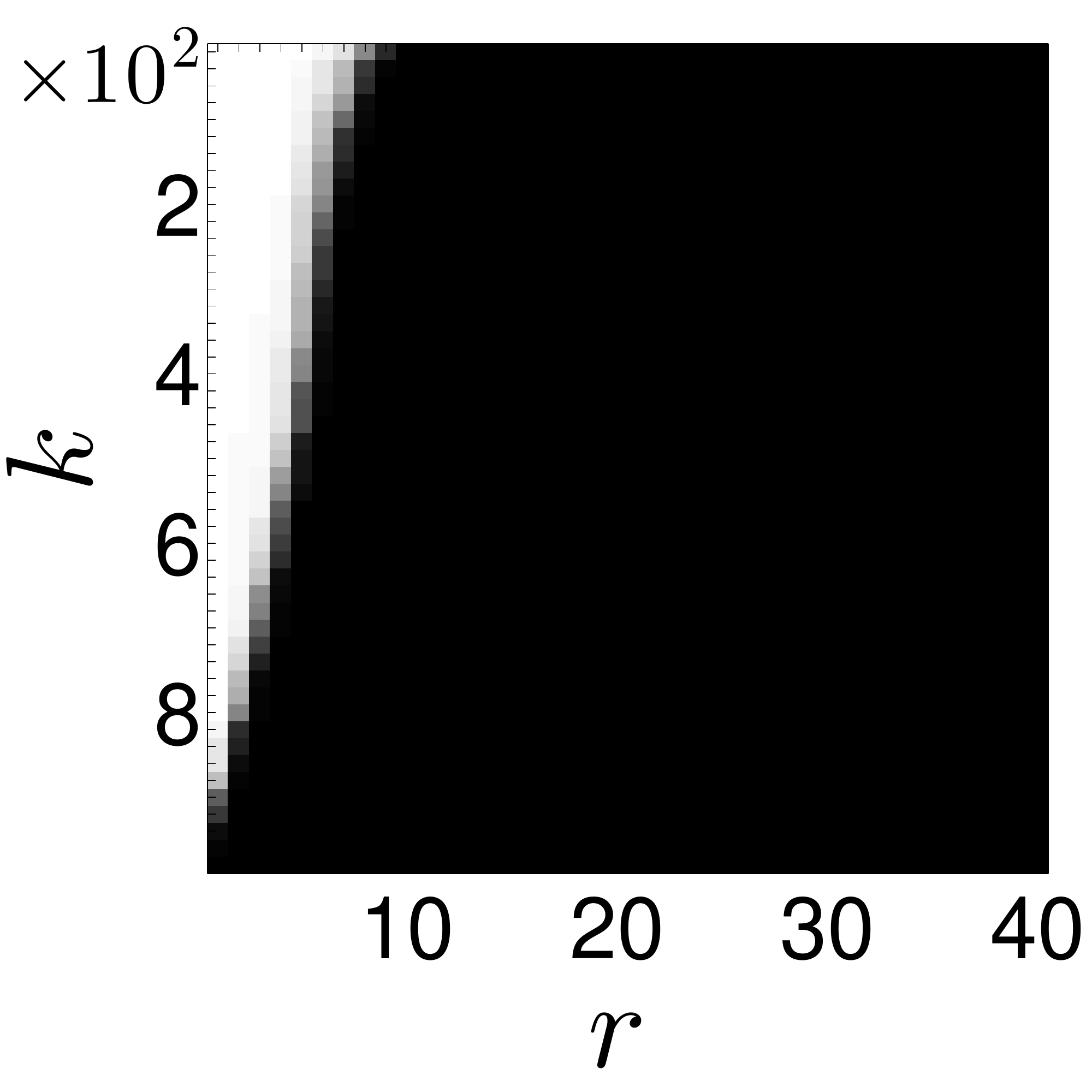}
	}\\
	\subfloat[9\%]{
		\includegraphics[width=0.28\linewidth]{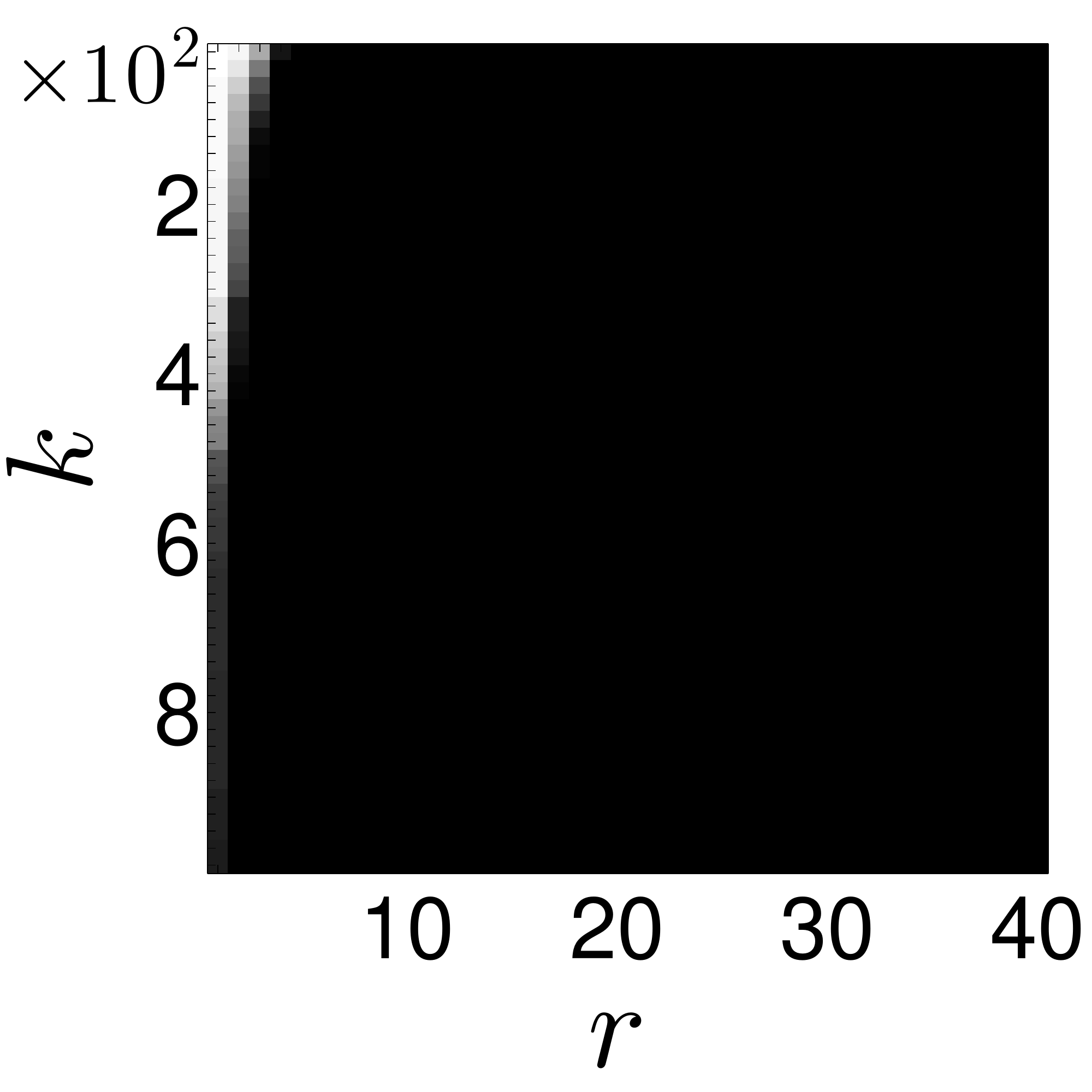}
	}
	\subfloat[15\%]{
		\includegraphics[width=0.28\linewidth]{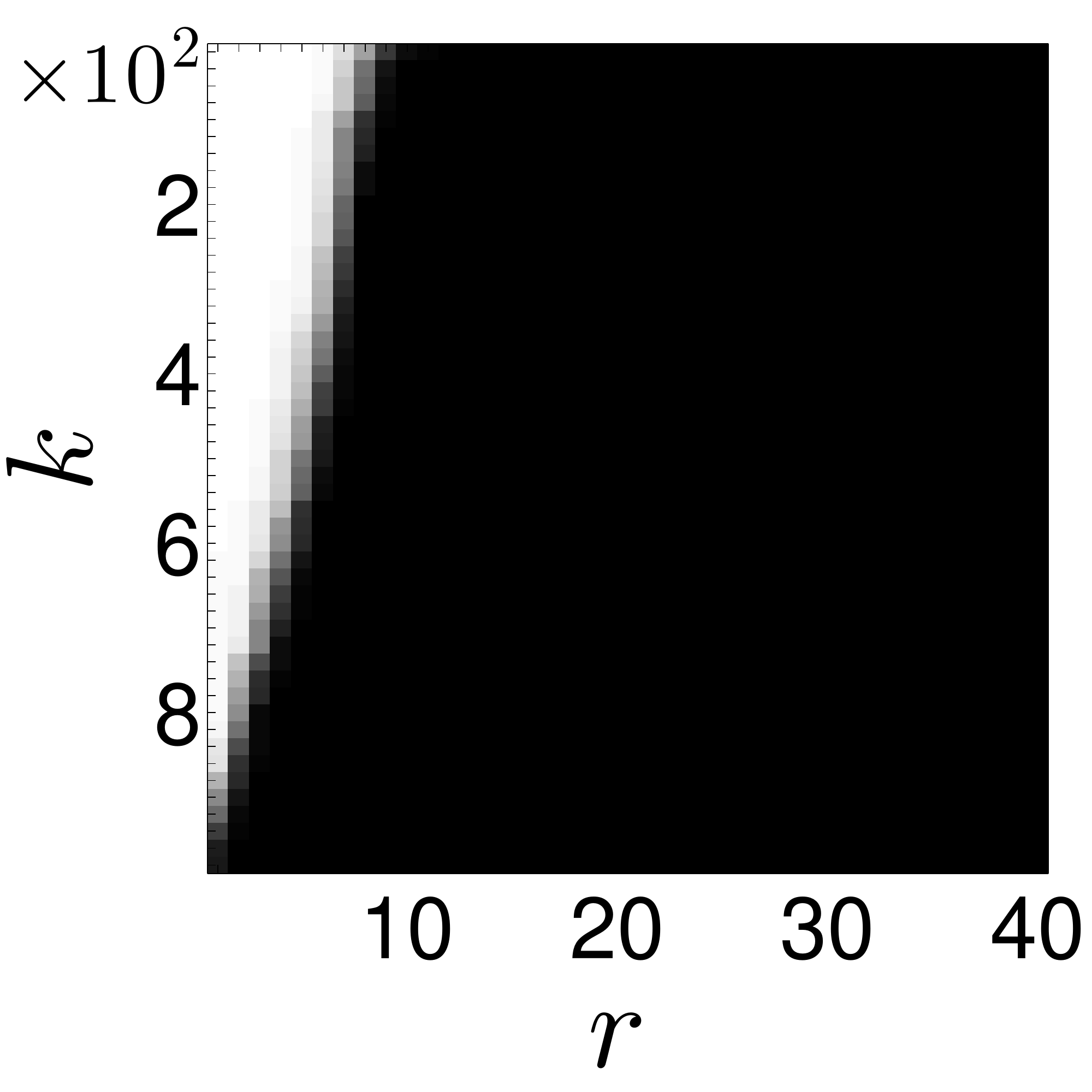}
	}
	\subfloat[21\%]{
		\includegraphics[width=0.28\linewidth]{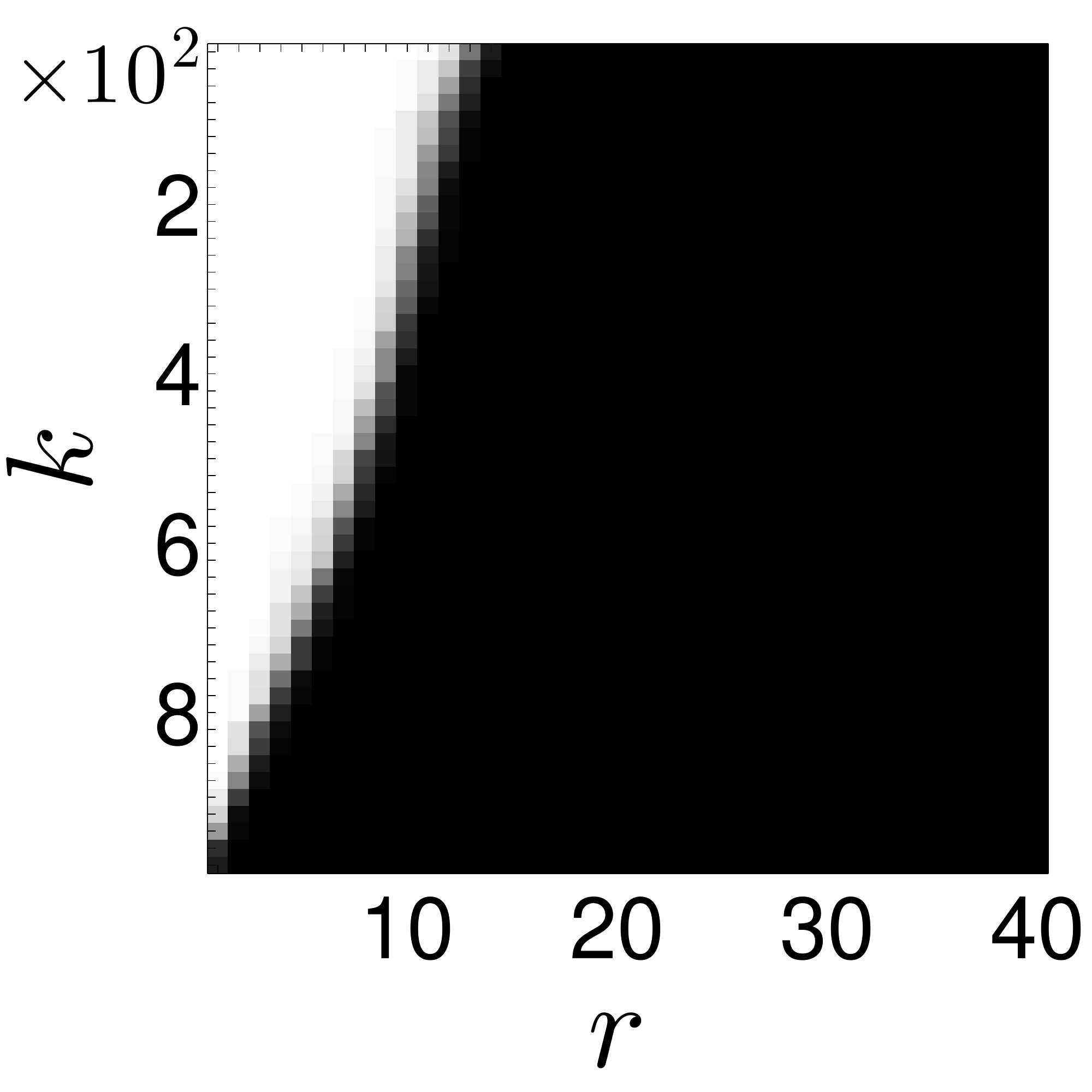}
	}\\
	%\vspace{-0.1in}
	\caption{Outlier recovery phase transitions plots for a ``missing data'' variant of the SACOS  method (white regions correspond to successful recovery). Rows correspond to available data fractions of $p_{\bOmega}$=0.3, 0.5 and 0.7 respectively, from top to bottom; columns corresponds to row sampling parameters $m=0.1n_1$, $0.2n_1$, and $0.3n_1$, respectively, from left to right. }
\label{fig:missing}
\vspace{-0.1in}
\end{figure}

We evaluate this approach empirically using the same data generation methods as above, and using an independent \emph{Bernoulli} model to describe the subsampling operation $\Pb_{\bOmega}(\cdot)$ (so that each $(i,j)\in\bOmega$ independently with probability $p_{\bOmega}$).  We consider noise-free settings, fix the column subsampling parameter $\gamma = 0.2$, and examine three different row-sampling scenarios ($m=0.1n_1$, $0.2n_1$ and $0.3n_1$) in each choosing subsets of $m$ rows uniformly at random from the collection of all $n_1 \choose m$ sets of cardinality $m$. The results are in Figure~\ref{fig:missing}.  Again, increasing $m$ and $p$ permits accurate estimation of outlier column indices for increasing rank $r$ and numbers $k$ of outlier columns.  Further, we do observe the performance degradation as the number of missing entries of $\Mb$ increases.

\section{Discussion and Future Directions}\label{sec:disc}

It is illustrative here to note a key difference between our approach and more conventional compressive sensing (CS) tasks.  Namely, the goal of the original CS works \cite{Candes:06:Freq, Donoho:06:CS, Candes:06:UES} and numerous follow-on efforts was to exactly recover or reconstruct a signal from compressive measurements, whereas the nature of our task here is somewhat simpler, amounting to a kind of multidimensional ``support recovery'' task (albeit in the presence of a low-rank ``background'').  Exactly recovering the low-rank and column-sparse components would be sufficient for the outlier identification task we consider here, but as our analysis shows it is not strictly necessary.  This is the insight that we exploit when operating on the ``compressed'' data $\bPhi\Mb$ instead of the original data matrix $\Mb$.  Ultimately, this allows us to successfully identify the locations of the outliers \emph{without first estimating the original (full size) low-rank matrix or the outliers themselves}.   For some regimes of $\mu_{\Lb}$, $r$ and $k$, we accomplish the outlier identification task using as few as $\cO\left(( r + \log k)(\mu_{\Lb} r\log r) + k\log(n_2/k) \right)$ observations. 

Along related lines, it is reasonable to conjecture that any procedure would require at least $r^2 + k$ measurements in order to identify $k$ outliers from an $r$-dimensional linear subspace.  Indeed, a necessary condition for the existence of outliers of a rank-$r$ subspace, as we have defined them, is that the number of rows of $\Mb$ be at least $r+1$.  Absent any additional structural conditions on the outliers and the subspace spanned by columns of the low-rank matrix, one would need to identify a collection of $r$ vectors that span the $r$-dimensional subspace containing the column vectors of the low-rank component (requiring specification of some $\cO(r^2)$ parameters) as well as the locations of the $k$ outliers (which would entail specifying another $k$ parameters).  In this sense, our approach may be operating near the sample complexity limit for this problem, at least for some regimes of $\mu_{\Lb}$, $r$ and $k$.

It would be interesting to see whether the dimensionality reduction insight that we exploit in our approach could be leveraged in the context of the Compressive Principal Component Pursuit (Compressive PCP) of \cite{Wright:13} in order to yield a procedure with comparable performance as ours, but which acquires only \emph{non-adaptive} linear measurements of $\Mb$.  Direct implementation of that approach in our experimental setting was somewhat computationally prohibitive (e.g., simulations at a $10\%$ sampling rate would require generation and storage of random matrices having $10^9$ elements). Alternatively, it is interesting to consider implementing the Compressive PCP method not on the full data $\Mb$, but on the a priori compressed data $\bPhi\Mb$. Our Lemma~\ref{lem:Mtilde} establishes that the row compression step preserves rank and column incoherence properties, so it is plausible that the Compressive PCP approach may succeed in recovering the components of the compressed matrix, which would suffice for the outlier identification task.  We defer this investigation along these lines to a future effort.

\begin{table}[!t]
\begin{center}
\caption{Computational complexities of outlier identification methods.  The stated results assume use of an accelerated first order method for all solvers (see text for additional details).}
\vspace{-0.04in}
\begin{tabular}{cc}
 \toprule
\hspace{-0.0in}  Method & Complexity\\
 \midrule
 \hspace{-0.00in}  OP & $\cO \left( {\rm IT} \cdot \left[ n_1 n_2\cdot \min\{n_1,n_2\} \right] \right)$\\
\hspace{-0.00in}  RMC & $\cO \left( {\rm IT} \cdot \left[ n_1 n_2\cdot \min\{n_1,n_2\} \right]  \right)$\\
\hspace{-0.00in}  ACOS & $\cO \left( {\rm IT}_1 \left[ m(\gamma n_2) \min\{m,\gamma n_2 \}\right] +  {\rm IT}_2\left[ p n_2  \right] \right)$\\
\hspace{-0.00in}  SACOS & $\cO \left( {\rm IT}_1 \left[ m(\gamma n_2) \min\{m,\gamma n_2 \}\right]  + m^2 n_2\right)$\\
 \bottomrule
 \end{tabular}\label{table:complex}
\end{center}
\vspace{-2em}
\end{table}

\sloppypar  We also comment briefly on the computational complexities of the methods we examined. We consider first the OP and RMC approaches, and assume that the solvers for each utilize an iterative accelerated first-order method (like those mentioned in the first part of Section~\ref{sec:exp}).  In this case, the computational complexity will be dominated by SVD steps in each iteration. Now, for an $n_1\times n_2$ matrix the computational complexity of the SVD is $\cO(n_1 n_2 \cdot \min\{n_1,n_2\})$; with this, and assuming some ${\rm IT}$ iterations are used, we have that the complexities of both OP and RMC scale as $\cO \left( {\rm IT} \cdot \left[ n_1 n_2\cdot \min\{n_1,n_2\} \right] \right)$.  By a similar analysis, we can conclude that the complexity of Step 1 of the ACOS and SACOS methods scales like $\cO \left( {\rm IT}_1 \cdot \left[ m (\gamma n_2)\cdot \min\{m,\gamma n_2\} \right] \right)$, where ${\rm IT}_1$ denotes the number of iterations for the solver in Step 1.  If we further assume an iterative accelerated first-order method for the LASSO in Step 2 of the ACOS approach, and that ${\rm IT}_2$ iterations are used, then the second step of the ACOS approach would have overall computational complexity $\cO \left( {\rm IT}_2 \cdot \left[ p n_2] \right) \right)$.  Along similar lines, Step 2 of SACOS would entail ${\cal O}(m^2 n_2 + mn_2) = {\cal O}(m^2 n_2)$ operations to compute the orthogonal projections and their $\ell_2$ norms. We summarize the overall complexity results in Table~\ref{table:complex}.  Since we will typically have $\gamma$ small, $m \ll n_1$, and $p\ll n_2$ in our approaches,  the computational complexity of our approaches can be much less than methods that operate on the full data or require intermediate SVD's of matrices of the same size as $\Mb$.

Note that we have not included here the complexity of acquiring or forming the observations in any of the methods.  For the ACOS method, this would comprise up to an additional $\cO \left(mn_1(\gamma n_2)\right)$ operations for Step 1 and $\cO(m^2 + mn_1 + n_1n_2 + n_2p) = \cO(mn_1 + n_1n_2 + n_2p)$ operations for Step 2, where the complexity for the second step is achieved by iteratively multiplying together the left-most two factors in the overall product, and using the fact that $m\leq n_1$.  Similarly, observations obtained via the SACOS approach could require up $\cO(m n_1 n_2)$ operations. On the other hand, depending on the implementation platform, forming the observations themselves could also have a negligible computational effect e.g., in our imaging example when linear observations are formed ``implicitly'' using a spatial light modulator or single pixel camera \cite{Duarte:08:SPC}.  Finally, we note that further reductions in the overall complexity of our approach may be achieved using fast or sparse JL embeddings along the lines of \cite{Ailon:06, Dasgupta:10}.

Finally, it is worth noting\footnote{Thanks to David B. Dunson and Alfred O. Hero for these suggestions.} that the performance in our our visual saliency application could likely be improved using an additional assumption that the salient regions be spatially clustered.  This could be implemented here using \emph{group sparse} regularization (e.g. \cite{Yuan:06}) in Step 2 of ACOS, or (more simply) by directly identifying groups of nonzero elements in Step 2 of SACOS. We defer investigations along these lines to a future effort.

\appendix
\subsection{Proof of Lemma~\ref{lem:Mtilde}}\label{a:lem1}

We proceed using the formalism of \emph{stable embeddings} that has emerged from the dimensionality reduction and compressive sensing literature (see, e.g., \cite{Davenport:10}). 
\begin{definition}[Stable Embedding]
For $\epsilon\in[0,1]$ and $\cU,\cV\subseteq \RR^n$, we say $\bPhi$ is an $\epsilon$-\emph{stable embedding} of $(\cU,\cV)$ if
\begin{equation}
(1-\epsilon) \|\ub-\vb\|_2^2 \leq \|\bPhi\ub-\bPhi\vb\|_2^2 \leq  (1+\epsilon) \|\ub-\vb\|_2^2
\end{equation}
for all $\ub\in\cU$ and $\vb\in\cV$. 
\end{definition}
Our proof approach is comprised of two parts. First, we show that each of the four claims in the lemma follow when $\bPhi$ is an $\epsilon$-stable embedding of 
\begin{equation}\label{eqn:epsdes}
\left(\cL, \cup_{i\in\cI_{\Cb}} \{\bC_{:,i}\} \cup  \{\bZero\}\right)
\end{equation}for any choice of $\epsilon<1/2$. Second, we show that for any $\delta\in(0,1)$, generating $\bPhi$ as a random matrix as specified in the lemma ensures it will be a $\sqrt{2}/4$-stable embedding of \eqref{eqn:epsdes} with probability at least $1-\delta$. The choice of $\sqrt{2}/4$ in the last step is somewhat arbitrary -- we choose this fixed value for concreteness here, but note that the structural conclusions of the lemma follow using any choice of $\epsilon<1/2$ (albeit with slightly different conditions on $m$). 

\subsubsection{Part 1}
Throughout this portion of the proof we assume that $\bPhi$ is an $\epsilon$-stable embedding of \eqref{eqn:epsdes} for some $\epsilon<1/2$, and establish each of the four claims in turn.  First, to establish that $\rk(\bPhi \Lb)=r=\rk(\Lb)$, we utilize an intermediate result of \cite{Gilbert:12}, stated here as a lemma (without proof) and formulated in the language of stable embeddings.
\begin{lemmai}[Adapted from \cite{Gilbert:12}, Theorem 1]\label{lem:Sketched}
Let $\Lb$ be an $n_1\times n_2$ matrix of rank $r$, and let $\cL$ denote the column space of $\Lb$, which is an $r$-dimensional linear subspace of $\RR^{n_1}$. If for some $\epsilon\in(0,1)$, $\bPhi$ is an $\epsilon$-stable embedding of $(\cL, \{\bZero\})$ then $\rk(\bPhi \Lb)=r=\rk(\Lb)$.
\end{lemmai}
Here, since $\bPhi$ being an $\epsilon$-stable embedding of \eqref{eqn:epsdes} implies it is also an $\epsilon$-stable embedding of $(\cL,\{\bZero\})$, the first claim (of Lemma~\ref{lem:Mtilde}) follows from Lemma~\ref{lem:Sketched}.

Next we show that $\bPhi\Lb$ has $n_{\Lb}$ nonzero columns.  Since $\bPhi$ is a stable embedding of $(\cL,\{\bZero\})$, it follows that for each of the $n_{\Lb}$ nonzero columns $\Lb_{:,i}$ of $\Lb$ we have $\|\bPhi\Lb_{:,i}\|_2^2 > (1-\epsilon) \|\Lb_{:,i}\|_2^2 > 0$, while for each of the remaining $n_2-n_{\Lb}$ columns $\Lb_{:,j}$ of $\Lb$ that are identically zero we have $\|\bPhi\Lb_{:,j}\|_2^2 = 0$ so that $\bPhi\Lb_{:,j}=0$.

Continuing, we show next that $\bPhi \Lb$ satisfies the column incoherence property with parameter $\mu_{\Lb}$.  Recall from above that we write the compact SVD of $\Lb$ as $\Lb = \Ub \bSigma \Vb^*$, where $\Ub$ is $n_1\times r$, $\Vb$ is $n_2\times r$, and $\bSigma$ is an $r\times r$ nonnegative diagonal matrix of singular values (all of which are strictly positive).  The incoherence condition on $\Lb$ is stated in terms of column norms of the matrix $\Vb^*$ whose rows form an orthonormal basis for the row space of $\Lb$.  Now, when the rank of $\bPhi\Lb$ is the same as that of $\Lb$, which is true here on account of Lemma~\ref{lem:Sketched}, the row space of $\bPhi\Lb$ is \emph{identical} to that of $\Lb$, since each are $r$-dimensional subspaces of $\RR^{n_2}$ spanned by linear combinations of the columns of the $\Vb^*$. Thus since the rank and number of nonzero columns of $\bPhi\Lb$ are the same as for $\Lb$, the coherence parameter of $\bPhi\Lb$ is just $\mu_{\Lb}$, and the third claim is established.

Finally, we establish the last claim, that the set of salient columns of $\bPhi\Cb$ is the same as for $\Cb$.  Recall that the condition that a column $\Cb_{:,i}$ be salient was equivalent to the condition that $\|\Pb_{\cL^{\perp}}\Cb_{:,i}\|_2 > 0$, where $\Pb_{\cL^{\perp}}$ is the orthogonal projection operator onto the orthogonal complement of $\cL$ in $\RR^{n_1}$. Here, our aim is to show that an analogous result holds in the ``projected'' space -- that for all $i\in\cI_{\Cb}$ we have $\|\Pb_{{(\bPhi\cL)}^{\perp}} \bPhi\Cb_{:,i}\|_2 > 0$, where $\bPhi\cL$ is the linear subspace spanned by the columns of $\bPhi\Lb$. For this we utilize an intermediate result of \cite{Davenport:10} formulated there in terms of a ``compressive interference cancellation'' method. We state an adapted version of that result here as a lemma (without proof).
\begin{lemmai}[Adapted from \cite{Davenport:10}, Theorem 5]\label{lem:IntCanc}
Let $\cV_1$ be an $r$-dimensional linear subspace of $\RR^{n}$ with $r<n$, let $\cV_2$ be any subset of $\RR^{n}$, and let $\check{\cV}_2 = \ \{\Pb_{\cV_1^{\perp}} \vb: \vb\in\cV_2\}$, where $\Pb_{\cV_1^{\perp}}$ is the orthogonal projection operator onto the orthogonal complement of $\cV_1$ in $\RR^{n}$.   If $\bPhi$ is an $\epsilon$-stable embedding of $(\cV_1, \check{\cV}_2 \cup \{\bZero\})$, then for all $\check{\vb} \in \check{\cV}_2$
\begin{equation}
\|\Pb_{(\bPhi\cV_1)^{\perp}} (\bPhi \check{\vb}) \|_2^2 \geq \left(1-\frac{\epsilon}{1-\epsilon}\right) \|\check{\vb} \|_2^2,
\end{equation} 
where $\Pb_{(\bPhi\cV_1)^{\perp}}$ is the orthogonal projection operator onto the orthogonal complement of the subspace of $\RR^{n}$ spanned by the elements of $\bPhi \cV_1 = \{\bPhi \vb: \vb\in\cV_1\} $.  
\end{lemmai}
Before applying this result we first note a useful fact, that $\bPhi$ being an $\epsilon$-stable embedding of $(\cV_1, \check{\cV}_2 \cup \{\bZero\})$ is equivalent to $\bPhi$ being an $\epsilon$-stable embedding of $(\cV_1, \cV_2 \cup \{\bZero\})$, which follows directly from the definition of stable embeddings and the (easy to verify) fact that $\left\{\vb_1 - \check{\vb}_2: \vb_1\in\cV_1, \check{\vb}_2\in\check{\cV}_2\cup \{\bZero\} \right\}=\left\{\vb_1 - \vb_2: \vb_1\in\cV_1, \vb_2\in\cV_2 \cup \{\bZero\}\right\}$. Now, to apply Lemma~\ref{lem:IntCanc} here, we let $\cV_1 = \cL$, $\cV_2 = \cup_{i\in\cI_{\Cb}} \{\Cb_{:,i}\}$, and $\check{\cV}_2 = \cup_{i\in\cI_{\Cb}} \{\Pb_{\cL^{\perp}}\Cb_{:,i}\}$.  Since $\bPhi$ is an $\epsilon$-stable embedding of \eqref{eqn:epsdes}, we have that for all $i\in\cI_{\Cb_{:,i}}$, $\|\Pb_{(\bPhi\cL)^{\perp}} (\bPhi \Cb)_{:,i} \|_2^2 \geq \left(1-\frac{\epsilon}{1-\epsilon}\right) \|\Pb_{\cL^{\perp}}\Cb_{:,i}\|_2^2$. Since $\epsilon<1/2$, the above result implies $\|\Pb_{(\bPhi\cL)^{\perp}} \bPhi \Cb_{:,i} \|_2>0$ for all $i\in\cI_{\Cb}$, while for all $j\notin\cI_{\Cb}$ we have $\Cb_{:,j}=\bZero$, implying that $\bPhi\Cb_{:,j}=\bZero$ and hence $\|\Pb_{(\bPhi\cL)^{\perp}} \bPhi \Cb_{:,j} \|_2=0$. Using this, and the fact that the nonzero columns of $\bPhi \Lb$ coincide with the nonzero columns of $\Lb$, we conclude that  $\cI_{\bPhi\Cb} = \{i:\|\Pb_{(\bPhi\cL)^{\perp}} \bPhi \Cb_{:,j} \|_2 > 0, (\bPhi\Lb)_{:,i} = \mathbf{0}\}$ is the same as $\cI_{\Cb}$.

\subsubsection{Part 2}

Given the structural result established in the previous step, the last part of the proof entails establishing that a random matrix $\bPhi$ generated as specified in the statement of Lemma~\ref{lem:Mtilde} is an $\sqrt{2}/4$-stable embedding of \eqref{eqn:epsdes}.   Our approach here begins with a brief geometric discussion, and a bit of ``stable embedding algebra.''  Appealing to the definition of stable embeddings, we see that $\bPhi$ being an $\epsilon$-stable embedding of \eqref{eqn:epsdes} is equivalent to $\bPhi$ being such that
\begin{equation}\label{eqn:JLcond}
(1-\epsilon) \|\vb\|_2^2 \leq \|\bPhi \vb\|_2^2 \leq (1+\epsilon) \|\vb\|_2^2
\end{equation} 
holds for all $\vb \in \cL \cup \bigcup_{i\in\cI_{\Cb}} \cL-\Cb_{:,i}$, where $\cL-\Cb_{:,i}$ denotes the $r$-dimensional \emph{affine} subspace of $\RR^{n_1}$ comprised of all elements taking the form of a sum between a vector in $\cL$ and the fixed vector $\Cb_{:,i}$. Thus, in words, establishing our claim here entails showing that a random $\bPhi$ (generated as specified in the lemma, with appropriate dimensions) approximately preserves the lengths of all vectors in a \emph{union of subspaces} comprised of one $r$-dimensional linear subspace and some $|\cI_{\Cb}|=k$, $r$-dimensional affine subspaces.  

Stable embeddings of linear subspaces using random matrices is, by now, well-studied (see, e.g., \cite{Baraniuk:08, Davenport:10, Gilbert:12}, as well as a slightly weaker result \cite[Lemma $10$]{Sarlos:06}), though stable embeddings of \emph{affine} subspaces has received less attention in the literature. Fortunately, using a straightforward argument we may leverage results for the former in order to establish the latter.  Recall the discussion above, and suppose that rather than establishing that \eqref{eqn:JLcond} holds for all $\vb \in \cL \cup \bigcup_{i\in\cI_{\Cb}} \cL-\Cb_{:,i}$ we instead establish a slightly stronger result, that \eqref{eqn:JLcond} holds for all $\vb \in \cL \cup \bigcup_{i\in\cI_{\Cb}} \cL^i$, where for each $i\in\cI_{\Cb}$, $\cL^i$ denotes the $(r+1)$-dimensional \emph{linear} subspace of $\RR^{n_1}$ spanned by the columns of the matrix $[\Lb \ \Cb_{:,i}]$. (That the dimension of each $\cL^i$ be $r+1$ follows from the assumption that columns $\Cb_{:,i}$ for $i\in\cI_{\Cb}$ be outliers.)  Clearly, if for some $i\in\cI_{\Cb}$ the condition \eqref{eqn:JLcond} holds for all $\vb\in\cL^i$, then it holds for all vectors formed as linear combinations of $[\Lb \ \Cb_{:,i}]$, so it holds in particular for all vectors in the $r$ dimensional affine subspace denoted by $\cL-\Cb_{:,i}$.  Further, that \eqref{eqn:JLcond} holds for any $i\in\cI_{\Cb}$ implies it holds for linear combinations that use a weight of zero on the component $\Cb_{:,i}$, so in this case \eqref{eqn:JLcond} holds also for all $\vb\in\cL$.  

Based on the above discussion, we see that a sufficient condition to establish that $\bPhi$ be an $\epsilon$-stable embedding of \eqref{eqn:epsdes} is that \eqref{eqn:JLcond} hold for all $\vb \in \bigcup_{i\in\cI_{\Cb}} \cL^i$; in other words, that $\bPhi$ preserve (up to multiplicative $(1\pm \epsilon)$ factors) the squared lengths of all vectors in a union of (up to) $k$ unique $(r+1)$-dimensional linear subspaces of $\RR^{n_1}$. To this end we make use of another result adapted from \cite{Gilbert:12}, and based on the union of subspaces embedding approach utilized  in \cite{Baraniuk:08}.
\begin{lemmai}[Adapted from \cite{Gilbert:12}, Lemma 1]\label{lem:JLSubsp}
Let $\bigcup_{i=1}^k \cV^i $ denote a union of $k$ linear subspaces of $\RR^{n}$, each of dimension at most $d$.  For fixed $\epsilon\in(0,1)$ and $\delta\in(0,1)$, suppose $\bPhi$ is an $m\times n$ matrix satisfying the distributional JL property with
\begin{equation}
m \geq \frac{d \log(42/\epsilon) + \log(k) + \log(2/\delta)}{f(\epsilon/\sqrt{2})}
\end{equation}
Then $(1-\epsilon) \|\vb\|_2^2 \leq \|\bPhi \vb\|_2^2 \leq (1+\epsilon) \|\vb\|_2^2$ holds simultaneously for all $\vb\in \bigcup_{i=1}^k \cV^i$ with probability at least $1-\delta$.
\end{lemmai}
Applying this lemma here with $d=r+1$ and $\epsilon=\sqrt{2}/4$, and using the fact that $\log(84\sqrt{2})< 5$ yields the final result.

\subsection{Proof of Lemma~\ref{lem:Step1}}\label{a:lem2}

Our approach is comprised of two parts.  In the first, we show that the two claims of Lemma~\ref{lem:Step1} follow directly when the following five conditions are satisfied
\begin{itemize}
\item[(\textbf{a1})] $\Sbb$ has $(1/2) \gamma n_2 \leq |\cS| \leq (3/2) \gamma n_2$ columns,
\item[(\textbf{a2})] $\widetilde{\Lb}\Sbb$ has at most $(3/2) \gamma n_{\Lb}$ nonzero columns,
\item[(\textbf{a3})] $\widetilde{\Cb}\Sbb$ has at most $k$ nonzero columns,
\item[(\textbf{a4})] $\sigma^2_1(\widetilde{\Vb}^*\Sbb)\leq (3/2) \gamma$, and 
\item[(\textbf{a5})] $\sigma^2_r(\widetilde{\Vb}^*\Sbb)\geq (1/2) \gamma$,
\end{itemize}
where the matrix $\widetilde{\Vb}^*$ that arises in (\textbf{a4})-(\textbf{a5}) is the matrix of right singular vectors from the compact SVD $\widetilde{\Lb}=\widetilde{\Ub}\widetilde{\bSigma}\widetilde{\Vb}^*$ of $\widetilde{\Lb}$, and $\sigma_i(\widetilde{\Vb}^*\Sbb)$ denotes the $i$-th largest singular value of $\widetilde{\Vb}^*\Sbb$.  Then, in the second part of the proof we show that (\textbf{a1})-(\textbf{a5}) hold with high probability when $\Sbb$ is a random subsampling matrix generated with parameter $\gamma$ in the specified range.  

We briefly note that parameters $(1/2)$ and $(3/2)$ arising in the conditions (\textbf{a1})-(\textbf{a5}) are somewhat arbitrary, and are fixed to these values here for ease of exposition.  Analogous results to that of Lemma~\ref{lem:Step1} could be established by replacing $(1/2)$ with any constant in $(0,1)$ and $(3/2)$ with any constant larger than $1$, albeit with slightly different conditions on $\gamma$.

\subsubsection{Part 1}

Throughout this portion of the proof, we assume that conditions (\textbf{a1})-(\textbf{a5}) hold.  Central to our analysis is a main result of \cite{Xu:12}, which we state as a lemma (without proof). 
\begin{lemmai}[Outlier Pursuit, adapted from \cite{Xu:12}]\label{lem:OP}
Let $\check{\Mb} = \check{\Lb} + \check{\Cb}$ be an $\check{n}_1\times \check{n}_2$ matrix whose components $\check{\Lb}$ and $\check{\Cb}$ satisfy the structural conditions
\begin{itemize}
\item[($\check{\cbb}$\textbf{1})] $\rk(\check{\Lb}) = \check{r}$,
\item[($\check{\cbb}$\textbf{2})] $\check{\Lb}$ has $n_{\check{\Lb}}$ nonzero columns,
\item[($\check{\cbb}$\textbf{3})] $\check{\Lb}$ satisfies the \emph{column incoherence property} with parameter $\mu_{\check{\Lb}}$, and
\item[($\check{\cbb}$\textbf{4})] $|\cI_{\check{\Cb}}|=\{i: \|\Pb_{\check{\cL}^{\perp}} \check{\Cb}_{:,i}\|_2 > 0, \check{\Lb}_{:,i} = \mathbf{0}\} = \check{k}$, where $\check{\cL}$ denotes the linear subspace spanned by columns of $\check{\Lb}$ and $\Pb_{\check{\cL}^{\perp}}$ is the orthogonal projection operator onto the orthogonal complement of $\check{\cL}$ in $\RR^{\check{n}_1}$,
\end{itemize} 
with
\begin{equation}
\check{k} \leq \left(\frac{1}{1+(121/9)\ \check{r}\mu_{\check{\Lb}}}\right)\check{n}_2.
\end{equation}  
For any upper bound $\check{k}_{\rm ub} \geq \check{k}$ and  $\lambda = \frac{3}{7 \sqrt{\check{k}_{\rm ub}}}$ any solutions of the \emph{outlier pursuit} procedure
\begin{equation}\label{eqn:OP2}
\{\widehat{\check{\Lb}},\widehat{\check{\Cb}}\} = \argmin_{\bL,\bC} \|\bL\|_* + \lambda \|\bC\|_{1,2} \ \mbox{ s.t. } \ \check{\Mb} = \bL + \bC,
\end{equation}
are such that the columns of $\widehat{\check{\Lb}}$ span the same linear subspace as the columns of $\check{\Lb}$, and the set of nonzero columns of $\widehat{\check{\Cb}}$ is the same as the set of locations of the nonzero columns of $\check{\Cb}$. 
\end{lemmai}
Introducing the shorthand notation $\check{\Lb}=\widetilde{\Lb}\Sbb$, $\check{\Cb}=\widetilde{\Cb}\Sbb$, and $\check{n}_2=|\cS|$, our approach will be to show that conditions (\textbf{a1})-(\textbf{a5}) along with the assumptions on $\widetilde{\Mb}$ ensure that ($\check{\cbb}$\textbf{1})-($\check{\cbb}$\textbf{4}) in Lemma~\ref{lem:OP} are satisfied for some appropriate parameters $\check{r}$, $n_{\check{\Lb}}$, $\mu_{\check{\Lb}}$, and $\check{k}$ that depend on analogous parameters of $\widetilde{\Mb}$.

First, note that (\textbf{a5}) implies that the matrix $\widetilde{\Vb}^*\Sbb$ has rank $r$, which in turn implies that $\check{\Lb}$ has rank $r$.  Thus, ($\check{\cbb}$\textbf{1}) is satisfied with $\check{r}=r$.  The condition ($\check{\cbb}$\textbf{2}) is also satisfied here for $n_{\check{\Lb}}$ no larger than $(3/2)\gamma n_{\Lb}$; this is a restatement of (\textbf{a2}). 

We next establish ($\check{\cbb}$\textbf{3}).  To this end, note that since $\check{\Lb}$ has rank $r$, it follows that the $r$-dimensional linear subspace spanned by the rows of $\check{\Lb}=\widetilde{\Ub}\widetilde{\bSigma}\widetilde{\Vb}^*\Sbb$ is the same as that spanned by the rows of $\widetilde{\Vb}^*\Sbb$.  Now, let $\Sbb^{T}\widetilde{\cV}$ denote the $r$-dimensional linear subspace of $\RR^{\check{n}_2}$ spanned by the columns of $\Sbb^{T}\widetilde{\Vb}$ and let $\Pb_{\Sbb^{T}\widetilde{\cV}}$ denote the orthogonal projection operator onto $\Sbb^{T}\widetilde{\cV}$.  Then, bounding the column incoherence parameter of $\check{\Lb}$ entails establishing an upper bound on $\max_{i\in[\check{n}_2]} \|\Pb_{\Sbb^{T}\widetilde{\cV}}\eb_i\|_2^2$, where $\eb_i$ is the $i$-th canonical basis vector of $\RR^{\check{n}_2}$.  Directly constructing the orthogonal projection operator (and using that $\widetilde{\Vb}^*\Sbb$ is a rank $r$ matrix) we have that
\begin{eqnarray}
\nonumber \lefteqn{\max_{i\in[\check{n}_2]} \|\Pb_{\Sbb^{T}\widetilde{\cV}}\eb_i\|_2^2 = \max_{i\in[\check{n}_2]} \left\|\Sbb^{T} \widetilde{\Vb}\left(\widetilde{\Vb}^*\Sbb \Sbb^{T} \widetilde{\Vb}\right)^{-1} \widetilde{\Vb}^*\Sbb \eb_i\right\|_2^2}\hspace{4em}&&\\
\nonumber &\stackrel{(a)}{\leq}&\max_{j\in[n_2]} \left\|\Sbb^{T} \widetilde{\Vb}\left(\widetilde{\Vb}^*\Sbb \Sbb^{T} \widetilde{\Vb}\right)^{-1} \widetilde{\Vb}^*\eb_j\right\|_2^2\\
\nonumber &\stackrel{(b)}{\leq}& \left(\frac{\sigma_1(\widetilde{\Vb}^*\Sbb)}{\sigma^2_r(\widetilde{\Vb}^*\Sbb)}\right)^2 \mu_{\Lb}\frac{r}{n_{\Lb}}\\
&\stackrel{(c)}{\leq}& \left(\frac{6}{\gamma}\right) \mu_{\Lb} \frac{r}{n_{\Lb}},
\end{eqnarray}
where $(a)$ follows from the fact that for any $i\in[\check{n}_2]$ the vector $\Sbb \eb_j$ is either the zero vector or one of the canonical basis vectors for $\RR^{n_2}$, $(b)$ follows from straightforward linear algebraic bounding ideas and the column incoherence assumption on $\widetilde{\Lb}$, and $(c)$ follows from (\textbf{a4})-(\textbf{a5}).  Now, we let $n_{\check{\Lb}}$ denote the number of nonzero columns of $\check{\Lb}$, and write 
\begin{equation}
\max_{i\in[\check{n}_2]} \|\Pb_{\Sbb^{T}\widetilde{\cV}}\eb_i\|_2^2 \leq \left(\frac{6}{\gamma}\right) \mu_{\Lb} \frac{r}{n_{\Lb}} \left(\frac{n_{\check{\Lb}}}{n_{\check{\Lb}}}\right)
\leq 9\mu_{\Lb} \frac{r}{n_{\check{\Lb}}},
\end{equation}
where the last inequality uses (\textbf{a2}).  Thus ($\check{\cbb}$\textbf{3}) holds with 
\begin{equation}\label{eqn:mucheck}
\mu_{\check{\Lb}}= 9\mu_{\Lb}.
\end{equation}

Next, we establish ($\check{\cbb}$\textbf{4}).  Recall from above that $\check{\Lb}$ has rank $r$, and is comprised of columns of $\widetilde{\Lb}$; it follows that the subspace $\check{\cL}$ spanned by columns of $\check{\Lb}$ is the same as the subspace $\widetilde{\cL}$ spanned by columns of $\widetilde{\Lb}$.  Thus, $\|\Pb_{\check{\cL}^{\perp}} \check{\Cb}_{:,i}\|_2 = \|\Pb_{\widetilde{\cL}^{\perp}} \check{\Cb}_{:,i}\|_2$, so to obtain an upper bound on $\check{k}$ we can simply count the number $\check{k}$ of nonzero columns of $\check{\Cb}=\widetilde{\Cb}\Sbb$.  By (\textbf{a3}) and \eqref{eqn:kmin},
\begin{eqnarray}
\nonumber \check{k}
&\leq& \left(\frac{1}{20(1+ 121 r \mu_{\Lb})}\right)\ \left(\frac{1}{2}\right)n_2\\
\nonumber &\stackrel{(a)}{\leq}& \left(\frac{1}{1+ 121 r \mu_{\Lb}}\right)\  \left(\frac{1}{2}\right) \gamma n_2\\
&\stackrel{(b)}{\leq}& \left(\frac{1}{1+ (121/9) \check{r} \mu_{\check{\Lb}}}\right)\  \check{n}_2,
\end{eqnarray}
where $(a)$ follows from the assumption that $\gamma\geq 1/20$, and $(b)$ follows from (\textbf{a1}) and \eqref{eqn:mucheck} as well as the fact that $\check{r}=r$.

Finally, we show that the two claims of Lemma~\ref{lem:Step1} hold.  The first follows directly from (\textbf{a1}).  For the second, note that for any $k_{\rm ub}\geq k$ we have that $\check{k}_{\rm ub} \triangleq k_{\rm ub} \geq \check{k}$.  Thus, since $\lambda =\frac{3}{7\sqrt{k_{\rm ub}}} = \frac{3}{7\sqrt{\check{k}_{\rm ub}}}$ and ($\check{\cbb}$\textbf{1})-($\check{\cbb}$\textbf{4}) hold, it follows from Lemma~\ref{lem:OP} that the optimization \eqref{eqn:OP2} produces an estimate $\widehat{\check{\Lb}}$ whose columns span the same linear subspace as that of $\check{\Lb}$.  But, since $\check{\Lb}$ has rank $r$ and its columns are just a subset of columns of the rank-$r$ matrix $\widetilde{\Lb}$, the subspace spanned by the columns of $\check{\Lb}$ is the same as that spanned by columns of $\widetilde{\Lb}$. 

\subsubsection{Part 2}

The last part of our proof entails showing (\textbf{a1})-(\textbf{a5}) hold with high probability when $\Sbb$ is randomly generated as specified.  Let $\cE_1,\dots,\cE_5$ denote the events that conditions (\textbf{a1})-(\textbf{a5}), respectively, hold.  Then  $\Prob\left( \ \left\{\bigcap_{i=1}^5 \cE_i\right\}^c \ \right)\leq \sum_{i=1}^5 \Prob(\cE_i^c)$, and we consider each term in the sum in turn.  

\sloppypar First, since $|\cS|$ is a Binomial($n_2,\gamma$) random variable, we may bound its tails using \cite[Theorem 2.3 (b-c)]{McDiarmid:98}. This gives that  $\Prob\left(|\cS| > 3\gamma n_2/2\right) \leq \exp\left(-3\gamma n_2/28\right)$  and  $\Prob\left(|\cS| < \gamma n_2/2\right) \leq\exp\left(-\gamma n_2/8\right).$  By union bound, we obtain that $\Prob(\cE_1^c) \leq \exp\left(-3\gamma n_2/28\right)  + \exp\left(-\gamma n_2/8\right).$

Next, observe that conditionally on $|\cS|=s$, the number of nonzero columns present in the matrix $\widetilde{\Lb} \Sbb$ is a hypergeometric random variable parameterized by a population of size $n_2$ with $n_{\Lb}$ positive elements and $s$ draws. Denoting this hypergeometric distribution here by ${\rm hyp}(n_2,n_{\Lb},s)$ and letting $H_{|\cS|}\sim {\rm hyp}(n_2,n_{\Lb},|\cS|)$, we have that $\Prob(\cE_2^c)=\Prob\left(H_{|\cS|}  > \left(\frac{3}{2}\right) \gamma n_{\Lb} \right)$. Using a simple conditioning argument, $\Prob(\cE_2^c) \leq \sum_{s=\lceil (2/3)\gamma n_2\rceil}^{\lfloor (4/3)\gamma n_2\rfloor} \Prob\left(H_{s}  > \left(\frac{3}{2}\right) \gamma n_{\Lb} \right)\Prob(|\cS| = s) + \Prob\left(\left| |\cS| -\gamma n_2 \right| > \left(\frac{1}{3}\right)\gamma n_2\right)$, and our next step is to simplify the terms in the sum.  Note that for any $s$ in the range of summation, we have $\Prob\left(H_{s}  > \left(\frac{3}{2}\right) \gamma n_{\Lb} \right) = \Prob\left(H_{s}  >  \left(\frac{3}{2}\right) \gamma n_{\Lb}\left(\frac{s n_2}{s n_2}\right) \right)$, and thus
\begin{eqnarray}
\nonumber \Prob\left(H_{s}  > \left(\frac{3}{2}\right) \gamma n_{\Lb} \right) 
&\stackrel{(a)}{\leq}& \Prob\left(H_{s} > \left(\frac{9}{8}\right) s\left(\frac{n_{\Lb}}{n_2}\right)\right)\\
\nonumber &\stackrel{(b)}{\leq}& \exp\left(-\frac{3s(n_{\Lb}/n_2)}{400}\right)\\
&\stackrel{(c)}{\leq}& \exp\left(-\frac{\gamma n_{\Lb}}{200}\right),  
\end{eqnarray}
where $(a)$ utilizes the largest value of $s$ to bound the term $\gamma n_2/s$, $(b)$ follows from an application of Lemma~\ref{lem:hyptail} in Appendix~\ref{a:hyp}, and $(c)$ results from using the smallest value of $s$ (within the range of summation) to bound the error term.  Assembling these results, we have that $\Prob(\cE_2^c) \leq \exp\left(-\gamma n_{\Lb}/200\right) + \exp\left(-\gamma n_2/24\right) +  \exp\left(-\gamma n_2/18\right)$, where we use the fact that the probability mass function of $|\cS|$ sums to one, and another application of \cite[Theorem 2.3(b,c)]{McDiarmid:98}. 

Bounding $\Prob(\cE^c_3)$ is trivial. Since $\widetilde{\Cb}$ itself has $k$ nonzero columns, the subsampled matrix $\widetilde{\Cb}\Sbb$ can have at most $k$ nonzero columns too.  Thus, $\Prob(\cE^c_3) = 0$.

Finally, we can obtain bounds on the largest and smallest singular values of $\widetilde{\Vb}^*\Sbb$ using the Matrix Chernoff inequalities of \cite{Tropp:12}.  Namely, letting $\Zb=\widetilde{\Vb}^*\Sbb$ we note that the matrix $\Zb\Zb^*$ may be expressed as a sum of independent positive semidefinite rank-one $r\times r$ Hermitian matrices, as $\Zb\Zb^* = \widetilde{\Vb}^*\Sbb\Sbb^{T}\widetilde{\Vb} = \sum_{i=1}^{n_2} S_i (\widetilde{\Vb}^*_{:,i})(\widetilde{\Vb}^*_{:,i})^*$, where the $\{S_i\}_{i=1}^{n_2}$ are i.i.d. Bernoulli($\gamma$) random variables as in the statement of Algorithm~\ref{alg:main} (and, $S_i^2=S_i$).  To instantiate the result of \cite{Tropp:12}, we note that $\lambda_{\rm max}(S_i (\widetilde{\Vb}^*_{:,i})(\widetilde{\Vb}^*_{:,i})^*) \leq \|\widetilde{\Vb}^*_{:,i}\|_2^2 \leq \mu_{\Lb} r/n_{\Lb}\triangleq R$ almost surely for all $i$, where the last inequality follows from the incoherence assumption ($\widetilde{\cbb}$\textbf{3}) (as well as ($\widetilde{\cbb}$\textbf{1})-($\widetilde{\cbb}$\textbf{2})). Further, direct calculation yields $\mu_{\rm min} \triangleq \lambda_{\rm min}\left(\EE\left[\Zb\Zb^*\right]\right) = \lambda_{\rm min}(\gamma \Ib) = \gamma$ and $\mu_{\rm max} \triangleq \lambda_{\rm max}\left(\EE\left[\Zb\Zb^*\right]\right) = \lambda_{\rm max}(\gamma \Ib) = \gamma$, where the identity matrices in each case are of size $r\times r$. Thus, applying \cite[Corollary 5.2]{Tropp:12} (with $\delta=1/2$ in that formulation) we obtain that $\Prob(\cE_4^c) = \Prob\left(\sigma^2_{1}\left(\widetilde{\Vb}^*\Sbb\right) \geq 3\gamma/2 \right) \leq r \cdot\left(9/10\right)^{\frac{\gamma n_{\Lb}}{r \mu_{\Lb}}}$, and $\Prob(\cE_5^c) = \Prob\left(\sigma^2_{r}\left(\widetilde{\Vb}^*\Sbb\right) \leq \gamma/2 \right)\leq r \cdot\left(9/10\right)^{\frac{\gamma n_{\Lb}}{r \mu_{\Lb}}}$.

Putting the results together, and using a further bound on $\Prob(\cE^c_1)$, we have $\Prob\left( \ \left\{\bigcap_{i=1}^5 \cE_i\right\}^c \ \right) \leq \exp\left(-\frac{\gamma n_{\Lb}}{200}\right) + 2\exp\left(-\frac{\gamma n_2}{24}\right) +  2\exp\left(-\frac{\gamma n_2}{18}\right) + r \cdot\left(\frac{9}{10}\right)^{\frac{\gamma n_{\Lb}}{r \mu_{\Lb}}} + r \cdot\left(\frac{9}{10}\right)^{\frac{\gamma n_{\Lb}}{r \mu_{\Lb}}}$, which is no larger than $\delta$ given that $\gamma$ satisfies \eqref{eqn:gammamin} (in particular, this ensures each term in the sum is no larger than $\delta/5$).

\subsection{Proof of Lemma~\ref{lem:Step2}}\label{a:lem3}

First, note that since $\widehat{\cL}_{(1)} = \widetilde{\cL}$, we have that $\|\Pb_{\widehat{\cL}_{(1)}^{\perp}} \widetilde{\Mb}_{:,i}\|_2 > 0$ for all $i\in\cI_{\widetilde{\Cb}}$, and $\|\Pb_{\widehat{\cL}_{(1)}^{\perp}} \widetilde{\Mb}_{:,i}\|_2 = 0$ otherwise.  This, along with the fact that the entries of $\bphi$ be i.i.d. realizations of a \emph{continuous} random variable, imply that with probability one the $1\times n_2$ vector $\xb^{T} \triangleq \bphi \Pb_{\widehat{\cL}_{(1)}^{\perp}} \widetilde{\Mb}$ is nonzero at every $i\in\cI_{\widetilde{\Cb}}$ and zero otherwise. Indeed, since for each $i\in\cI_{\widetilde{\Cb}}$ the distribution of ${\rm x}_i = \bphi \Pb_{\widehat{\cL}_{(1)}^{\perp}} \widetilde{\Mb}_{:,i}$ is a continuous random variable with nonzero variance, it takes the value zero with probability zero.  On the other hand, for $j\notin \cI_{\widetilde{\Cb}}$, ${\rm x}_j = \bphi \Pb_{\widehat{\cL}_{(1)}^{\perp}} \widetilde{\Mb}_{:,j}=0$ with probability one.   With this, we see that exact identification of $\cI_{\widetilde{\Cb}}$ can be accomplished if we can identify the support of $\xb$ from linear measurements of the form $\yb = (\yb_{(2)})^{T} = \Ab\xb $.  

To proceed, we appeal to (now, well-known) results from the compressive sensing literature.  We recall one representative result of \cite{Candes:08:RIP} that is germane to our effort below.  Here, we cast the result in the context of the stable embedding formalism introduced above, and state it as a lemma without proof.
\begin{lemmai}[Adapted from Theorem 1.2 of \cite{Candes:08:RIP}]\label{lem:RIP}
Let $\xb\in\RR^n$ and $\zb = \Ab \xb$.  If $\Ab$ is an $\epsilon$-stable embedding of $(\cU_{\binom{n}{2k}},\{\bZero\})$ for some $\epsilon < \sqrt{2}-1$ where $\cU_{\binom{n}{2k}}$ denotes the union of all $\binom{n}{2k}$ unique $2k$-dimensional linear subspaces of $\RR^n$ spanned by canonical basis vectors, and $\xb$ has at most $k$ nonzero elements, then the solution $\widehat{\xb}$ of
\begin{equation}
\argmin_{\bx} \|\bx\|_1 \ \mbox{ s.t. } \zb = \Ab \bx.
\end{equation}
is equal to $\xb$.  
\end{lemmai}

Now, a straightforward application of Lemma~\ref{lem:JLSubsp} above provides that for any $\delta\in(0,1)$, if 
\begin{equation}
p\geq \frac{2k\log(42/\epsilon) + \log\binom{n}{2k} + \log(2/\delta)}{f(\epsilon/\sqrt{2})}
\end{equation}
then the randomly generated $p\times n_2$ matrix $\bA$ will be an $\epsilon$-stable embedding of $(\cU_{\binom{n}{2k}},\{\bZero\})$ with probability at least $1-\delta$.  This, along with the well-known bound $\binom{n}{2k}\leq \left(\frac{en}{2k}\right)^{2k}$ and some straightforward simplifications, imply that the condition that $p$ satisfy \eqref{eqn:pmin}  is sufficient to ensure that with probability at least $1-\delta$, $\bA$ is a $(\sqrt{2}/4)$-stable embedding of $(\cU_{\binom{n}{2k}},\{\bZero\})$.  Since $\sqrt{2}/4 < \sqrt{2}-1$,  the result follows.

\subsection{An Upper Tail Bound for the Hypergeometric Distribution}\label{a:hyp}

Let ${\rm hyp}(N,M,n)$ denote the hypergeometric distribution parameterized by a population of size $N$ with $M$ positive elements and $n$ draws, so $H\sim{\rm hyp}(N,M,n)$ is a random variable whose value corresponds to the number of positive elements acquired from $n$ draws (without replacement). The probability mass function of $H\sim{\rm hyp}(N,M,n)$ is $\Prob(H=k) = \binom{M}{k}\binom{N-M}{n-k}/\binom{N}{n}$ for $k\in\{\max\{0,n+M-N\},\dots,\min\{M,n\}\}$, and its mean value is $\EE[H]=nM/N$.

It is well-known that the tails of the hypergeometric distribution are similar to those of the binomial distribution for $n$ trials and success probability $p=M/N$.  For example, \cite{Chvatal:79} established that for all $t\geq 0$, $\Prob(H -np \geq nt) \leq e^{-2t^2 n}$, a result that follows directly from Hoeffding's work \cite{Hoeffding:63}, and exhibits the same tail behavior as predicted by the Hoeffding Inequality for a Binomial($n,p$) random variable (see, e.g., \cite{McDiarmid:98}).  Below we provide a lemma that yields tighter bounds on the upper tail of $H$ when the fraction of positive elements in the population is near $0$ or $1$.  Our result is somewhat analogous to \cite[Theorem 2.3(b)]{McDiarmid:98} for the Binomial case.
\begin{lemmai}\label{lem:hyptail}
Let $H\sim{\rm hyp}(N,M,n)$, and set $p=M/N$.  For any $\epsilon\geq 0$,
\begin{equation}
\Prob(H \geq (1+\epsilon)np) \leq e^{-\frac{\epsilon^2 np}{2(1+\epsilon/3)}}.
\end{equation}
\end{lemmai}
\begin{IEEEproof}
We begin with an intermediate result of \cite{Chvatal:79}, that for any $t\geq 0$ and $h\geq 1$,
\begin{equation}
\Prob(H-pn \geq tn) \leq \left(h^{-(p+t)}(1-p+hp)\right)^n.
\end{equation}
Now, for the specific choices $t=\epsilon p$ and $h=1+\epsilon$ we have
\begin{eqnarray}
\nonumber \Prob(H-np \geq \epsilon np) &\leq& \left((1+\epsilon)^{-(1+\epsilon)p}(1+\epsilon p)\right)^n\\
\nonumber &\stackrel{(a)}\leq& \left((1+\epsilon)^{-(1+\epsilon)}e^{\epsilon}\right)^{np}\\
&\stackrel{(b)}\leq& e^{-\frac{\epsilon^2 np}{2(1+\epsilon/3)}},
\end{eqnarray}
where $(a)$ follows from the inequality $1+x\leq e^x$ (with $x=\epsilon p$), and $(b)$ follows directly from \cite[Lemma 2.4]{McDiarmid:98}.
\end{IEEEproof}

\bibliographystyle{IEEEbib}
\bibliography{salbib_jdh}

\begin{IEEEbiography}{Xingguo Li}
received the B.E. degree in 2010 in Communications Engineering from Beijing University of Posts and Telecommunications, and M.S. degree in 2013 with honor in Applied and Computational Mathematics from University of Minnesota Duluth. In 2010, he held a visiting research appointment in the Robotics Institute of School of Computer Science, Carnegie Mellon University. He is currently a Ph.D. student in the Department of Electrical and Computer Engineering, University of Minnesota, under the supervision of Professor Jarvis Haupt. His current research interest focuses on statistical signal processing, high-dimensional sparse regression and optimization with applications in image processing, computer vision and machine learning.
\end{IEEEbiography}

\begin{IEEEbiography}{Jarvis Haupt} 
(S'05--M'09) received the B.S., M.S. and Ph.D. degrees in electrical engineering from the University of Wisconsin-Madison in 2002, 2003, and 2009, respectively. From 2009-2010 he was a Postdoctoral Research Associate in the Dept.~of Electrical and Computer Engineering at Rice
University in Houston, Texas. He is currently an Assistant Professor in the Dept.~of Electrical and Computer Engineering at the University of
Minnesota. 

Professor Haupt is the recipient of several academic awards, including the Wisconsin Academic Excellence Scholarship, the Ford Motor Company
Scholarship, the Consolidated Papers and Mead Witter Foundation Tuition Scholarships, the Frank D. Cady Mathematics Scholarship, and the Claude and Dora Richardson Distinguished Fellowship. He received the DARPA Young Faculty Award in 2014. His research interests generally include high-dimensional statistical inference, adaptive sampling techniques, and statistical signal processing and learning theory, with applications in the biological sciences, communications, imaging, and networks.
\end{IEEEbiography}

\end{document}